\documentclass[11pt]{article}

\usepackage{amsfonts, amsthm}
\usepackage{amssymb}
\usepackage{amsmath}  
\usepackage{graphicx, comment, xcolor}
\usepackage{authblk}
\usepackage{mathrsfs}
\usepackage[colorlinks=true, allcolors=blue]{hyperref}
\usepackage{cleveref}
\usepackage{mathtools}

\usepackage[T1]{fontenc}
\allowdisplaybreaks
\usepackage[shortlabels]{enumitem}
\usepackage{tikz}
\usetikzlibrary{positioning,calc}
\usetikzlibrary{intersections}
\tikzset{every picture/.style={line width=0.75pt}} 

\usepackage[left=1.15in,right=1.15in,top=1.2in,bottom=1.2in]{geometry}

\newtheorem{theorem}{Theorem}
\newtheorem{lemma}[theorem]{Lemma}
\newtheorem{corollary}[theorem]{Corollary}
\newtheorem{proposition}[theorem]{Proposition}
\newtheorem{definition}[theorem]{Definition}
\newtheorem{conjecture}[theorem]{Conjecture}

\newcommand{\tr}{\mathrm{tr}}
\newcommand{\id}{\mathrm{id}}

\newcommand{\ac}{\mathbf{a}}
\newcommand{\bc}{\mathbf{b}}
\newcommand{\R}{\mathbb{R}}
\newcommand{\C}{\mathbb{C}}

\newcommand{\dd}{\text{\rm{d}}}

\newcommand{\ket}[1]{|#1\rangle}
\newcommand{\bra}[1]{\langle#1|}
\newcommand{\ketbra}[2]{|#1\rangle\!\langle#2|}
\newcommand{\bfa}{\mathbf{a}}
\newcommand{\Ntot}{N_{\mathrm{tot}}}

\title{From Qubit Tensorization to Gaussian Optimizers in\\ Bosonic Systems}

\author{Salman Beigi$^{1,2}$, Hami Mehrabi$^3$}
\affil{\it \footnotesize $^1$School of Mathematics, Institute for Research in Fundamental Sciences (IPM), P.O. Box 19395-5746, Tehran, Iran \\
\it \footnotesize $^2$Centre for Quantum Technologies, National University of Singapore, Singapore 117543, Singapore \\
\it \footnotesize $^3$School of Electrical and Computer Engineering, Cornell University, Ithaca, New York 14850, USA\\}

\date{}

\begin{document}
\maketitle

\begin{abstract}

We develop a method for deriving Gaussian optimality results
for bosonic quantum systems from inequalities for qubit systems.
Inspired by Gross's derivation of the Gaussian logarithmic
Sobolev inequality from an inequality on the two-point space,
we use qubits as quantum two-point spaces and lift bosonic
Weyl operators to pseudo-Weyl operators on large qubit systems.
By controlling weighted norms and relative entropy in the
fluctuation limit, we establish two transfer theorems that
convert tensorized qubit inequalities into inequalities for
bosonic Gaussian channels. As a first application, we prove
that the sharp $2$-logarithmic Sobolev inequality for the
bosonic Ornstein--Uhlenbeck semigroup tensorizes without
loss in the constant. We also establish a sharp output entropy
inequality for tensor powers of a class of qubit channels.
This includes a quantum Mrs.~Gerber's lemma determining
the minimum output entropy of tensor powers of the qubit
depolarizing channel at fixed input entropy. Applying our
relative entropy transfer theorem, we obtain multimode
constrained minimum output entropy bounds for Gaussian
additive-noise channels and thermal attenuators in the
heating range.
\end{abstract}

{\footnotesize
\tableofcontents
}

\section{Introduction}\label{sec:introduction}

Gaussian distributions play a central role in probability theory, analysis, and
information theory. Among distributions with a fixed covariance matrix,
they maximize differential entropy~\cite{LiebLoss, shannon1948mathematical}. They also arise as extremizers of fundamental functional and information-theoretic inequalities, making Gaussian optimality a recurring theme across these fields~\cite{LB, Lieb, stam1959some}.
Their appearance as optimizers is accompanied by a
familiar limiting phenomenon. Sums of independent random variables become
approximately Gaussian, and the same distribution emerges when one looks
at a fixed number of coordinates of a uniformly distributed point in a
Euclidean ball of radius $\sqrt d$ as its dimension $d$ tends to
infinity~\cite{DiaconisFreedman1987}. These observations suggest that, in
suitable settings, a sharp Gaussian inequality might be understood by
passing to the limit from a simpler finite-dimensional problem.

Gross's proof of the Gaussian logarithmic Sobolev inequality provides a
particularly useful realization of this idea~\cite{Gross1975}. For the
standard Gaussian measure $\gamma_d$ on $\R^d$, the inequality states that
\begin{align}\label{eq:intro:gross}
 \operatorname{Ent}_{\gamma_d}(f^2)
 &\le 2\int_{\R^d}|\nabla f|^2\,\dd\gamma_d,\qquad \quad 
 \operatorname{Ent}_{\gamma_d}(f^2)
 :=\int_{\R^d}f^2\ln\!\left(
       \frac{f^2}{\int f^2\,\dd\gamma_d}\right)\dd\gamma_d,
 \end{align}
for sufficiently regular real-valued functions $f$. The proof starts with an inequality on a two-point
space, tensorizes it over independent bits, and then uses the central
limit theorem to pass to the Gaussian measure by letting the number of bits tend to infinity. The Gaussian inequality is
therefore obtained from a discrete inequality whose constant survives
arbitrarily many independent copies. This argument connects the central
limit theorem with tensorization. It gives a strategy for deriving a
sharp Gaussian inequality from a simpler discrete one: the constant
must first survive tensorization, and the quantities in the inequality
must then survive the central limit.

In this paper, we develop a corresponding passage from qubits to bosonic
quantum systems. Qubits replace the two-point space, while collective
fluctuations of many qubits give rise to bosonic observables in the limit.
We use this correspondence to transfer sharp inequalities for qubit tensor
powers to inequalities for bosonic Gaussian channels. Two issues make
this passage substantially different from the classical argument.
First, a quantum inequality for one qubit need not retain its sharp
constant on entangled inputs to many qubits. Second,
convergence of fluctuation correlations does not by itself justify
passing to the limit in the nonlinear norm and entropy functionals that
appear in Gaussian optimization problems. Our construction addresses
this second issue and makes precise which tensorized qubit inequalities
are sufficient for the bosonic applications.

To introduce the optimization problems to which we apply this limiting procedure, we first describe the basic noise models for bosonic quantum systems. Gaussian
channels describe attenuation, amplification, and the addition of
Gaussian noise. Their Gaussian input states are natural candidates for optimizing
quantities such as output entropy and Schatten norms~\cite{GaussianOptimizersSurvey}. For the
phase-covariant channels considered here, the minimum
output entropy problem has been solved: the vacuum (and all coherent states) minimize the output
entropy of the channel and its tensor powers~\cite{GiovannettiHolevoGarciaPatron}. A more demanding question
arises when the input entropy is prescribed. Then, a pure input is
generally unavailable, and the proposed optimizer is a product of
identical thermal states. This is the constrained minimum output entropy (CMOE)
problem.

To formulate this question, let $\tau_\mu$ denote the one-mode thermal
state with mean photon number $\mu\ge0$. Its von Neumann entropy $S(\tau_\mu)=-\tr(\tau_\mu\ln\tau_\mu)$ is
\[
 S(\tau_\mu)=g(\mu),\qquad
 g(\mu)=(\mu+1)\ln(\mu+1)-\mu\ln\mu.
\] 
By definition, for an $m$-mode state $\rho$, the quantity $g^{-1}(S(\rho)/m)$ is the mean photon
number of a thermal state with the same entropy per mode. Suppose a
one-mode phase-covariant Gaussian channel $\Phi$ maps
$\tau_\mu$ to $\tau_{\mu'}=\tau_{\lambda\mu+w}$, where $\lambda\ge0$ is its gain and
$w\ge\max\{0,\lambda-1\}$ is the mean photon number of its vacuum
output. The CMOE conjecture is
\begin{equation}\label{eq:intro:cmoe}
 S\bigl(\Phi^{\otimes m}(\rho)\bigr)
 \ge m\,g\!\left(\lambda g^{-1}\!\left(\frac{S(\rho)}m\right)+w\right).
\end{equation}
We note that equality is attained by
$\rho=\tau_{\mu}^{\otimes m}$ where $\mu=g^{-1}(S(\rho)/m)$.

De Palma, Trevisan, and Giovannetti established this thermal optimality
for one-mode (i.e., $m=1$) phase-covariant Gaussian
channels~\cite{DePalmaCMOE}, using sharp norm estimates for the
quantum-limited amplifier together with the earlier attenuator result.
Their related work establishes a reduction to passive
states~\cite{DePalmaTrevisanGiovannetti2016Passive} and Gaussian optimality
for the $p\to q$ norms of one-mode quantum-limited attenuators and
amplifiers~\cite{DePalmaNorms}. In the multimode setting, CMOE is known
for inputs diagonal in a product basis, as well as, without this input
restriction, for entanglement-breaking attenuators and
amplifiers~\cite{DePalma2017ProductDiagonal,DePalma2018EntanglementBreaking}.
The passage from one mode to an arbitrary number of modes is the common
difficulty behind these questions. Even though the channel acts
independently on each mode, its input may be entangled across all of
them. A proof must therefore control these difficulties without
introducing a loss that grows with the number of modes.

A second family of optimization problems concerns logarithmic Sobolev
inequalities and hypercontractivity for the quantum Ornstein--Uhlenbeck
semigroup consisting of phase-covariant Gaussian channels. Let $\Phi_t=e^{t\mathcal L}$ be such a semigroup, with faithful
thermal invariant state $\tau_\mu$, where $\mu>0$. For a faithful
reference state $\xi$, define the weighted Schatten norms
\begin{equation}\label{eq:intro:weighted-norm}
 \|A\|_{p,\xi}
 :=\bigl\|\xi^{1/(2p)}A\xi^{1/(2p)}\bigr\|_p,
 \qquad
 \|T\|_p :=(\tr|T|^p)^{1/p}.
\end{equation}
Hypercontractivity asks how long the semigroup must act before it
contracts a weighted $p$-norm into a weighted $q$-norm with $q>p$.
For $m$ modes, the desired inequality is
\begin{equation}\label{eq:intro:hypercontractivity}
 \bigl\|(\Phi_t^\dagger)^{\otimes m}(B)\bigr\|_{q,\tau_\mu^{\otimes m}}
 \le \|B\|_{p,\tau_\mu^{\otimes m}},
 \qquad 1<p<q<\infty,
\end{equation}
for every bounded observable $B$. The sharp problem is to determine
the least admissible time $t$ and whether it remains unchanged as the
number of modes increases.

The corresponding infinitesimal question is expressed by the
$2$-logarithmic Sobolev inequality~\cite{Gross1975,CarboneSasso2008,OZ99}
\begin{equation}\label{eq:intro:2lsi}
 \mathcal E_{2,\widehat{\mathcal L},\tau_\mu^{\otimes m}}(\rho)
 \ge \alpha\,D(\rho\Vert\tau_\mu^{\otimes m}),
\end{equation}
where $D(\rho\Vert\xi)=\tr[\rho(\ln\rho-\ln\xi)]$ is quantum relative
entropy, $\widehat{\mathcal L}$ is the sum of the generators acting
on the individual modes, and $\mathcal E_2$ is the corresponding quadratic
Dirichlet form.
The largest admissible $\alpha$ is the $2$-logarithmic Sobolev
constant. With our normalization, the associated hypercontractive
time bound is
\[
 t\ge \frac{1}{4\alpha}\ln\frac{q-1}{p-1}.
\]
Beigi and Rahimi-Keshari determined the sharp one-mode
$p$-logarithmic Sobolev constants for $1\le p\le2$
\cite{Beigi-RahimiKeshari2023}.
The tensorization question is whether the sharp one-mode
constants remain valid in \eqref{eq:intro:2lsi} for every $m$ and every
finite-energy state $\rho$. Thus, here too, the central issue is extending a sharp
one-mode inequality to the setting of an arbitrary number of modes.

\subsection{Pseudo-Weyl operators}

We approach these optimization problems by transferring inequalities from qubit systems to bosonic Gaussian channels. The strategy is to
represent each bosonic mode through the collective fluctuations of
many qubits, apply an inequality that holds uniformly over qubit
tensor powers, and pass to the fluctuation limit. To implement this
strategy, we need a common language for qubit fluctuations and
bosonic observables, together with a limiting procedure that preserves
the norms and entropies appearing in the inequalities.

Weyl operators provide this common language. They are the unitary exponentials of the bosonic canonical observables. Moreover, Gaussian channels act on them in a particularly simple way: in the Heisenberg picture, a Gaussian channel maps an $m$-mode Weyl operator $W(z)$ to $f(z) W(L(z))$, where $f$ is a Gaussian function and $L$ is a linear map. This description in terms of bounded observables is well suited to a finite-dimensional approximation, where the bosonic canonical commutation relations cannot hold exactly.

For each bosonic mode, we take a block of $M$ qubits in a product
thermal reference state $\sigma_\mu^{\otimes M}$. Exponentials of suitably normalized collective
qubit observables then provide operators $W_M(z)$ corresponding to
$W(z)$. We call them \emph{pseudo-Weyl operators} since as finite-dimensional operators they cannot
satisfy the exact bosonic canonical commutation relations. Nevertheless,
products of the pseudo-Weyl operators have the same limiting expectations against the product qubit reference as the original Weyl operators against the thermal reference state when $M$ tends to infinity~\cite{goderis1989central,jakvsic2010quantum}. The relevant commutation relations are thus recovered in the asymptotic limit. This observation suggests associating to any \emph{Weyl expression} a qubit operator in terms of pseudo-Weyl operators that has the same asymptotic thermal correlations. We call this assignment \emph{Weyl lifting}.

For the intended applications, the convergence of these expectation values is not enough and 
these correlation limits must be
strengthened, particularly to control the weighted norms introduced
in \eqref{eq:intro:weighted-norm}. Convergence of thermal correlations
only controls fixed products of fluctuation observables. In contrast,
the weighted norms involve reference-state powers inserted around an
observable, followed by an absolute value and a possibly noninteger
power. Passing to the limit in these operations requires additional
analysis, especially because both the reference states and the Hilbert
spaces vary with $M$. We prove that Weyl lifting preserves the weighted
norms for every finite $p\ge1$ in the limit.  
These nonlinear limits allow us to transfer finite-time
hypercontractive inequalities from qubits to bosonic systems and,
by the hypercontractivity to logarithmic Sobolev implication, obtain
logarithmic Sobolev bounds.

Weyl lifting is also compatible with the action of channels: applying
a suitable qubit channel to a lifted observable reproduces, in the
fluctuation limit, the action of the corresponding bosonic Gaussian
channel. Combining this compatibility with the nonlinear convergence results allows us to transfer two classes of qubit inequalities to bosonic systems.
Fix a Gaussian channel $\Phi$ and a qubit counterpart $\Psi$, with
faithful input and output thermal reference states $\tau_\mu,\tau_{\mu'}$ and
$\sigma_\mu,\sigma_{\mu'}$, respectively. We show that 
for $1\le p,q<\infty$, a bound
\begin{align}\label{eq:intro-qubit-hc}
 \|(\Psi^\dagger)^{\otimes n}(B)\|_{q,\sigma_\mu^{\otimes n}}
 \le\|B\|_{p,\sigma_{\mu'}^{\otimes n}}
\end{align}
valid for every $n$ and every $n$-qubit operator $B$ implies the same
bound for $(\Phi^\dagger)^{\otimes m}$ at the corresponding thermal
references, for every $m$ and every bounded bosonic operator. Moreover, writing
$D(\cdot\|\cdot)$ for quantum relative entropy, we prove that for $c>0$ an inequality
\begin{align}\label{eq:intro-qubit-relative entropy-c}
 D(\Psi^{\otimes n}(\omega)\Vert\sigma_{\mu'}^{\otimes n})
 \le cD(\omega\Vert\sigma_\mu^{\otimes n})
\end{align}
valid for every $n$ and every state implies the corresponding inequality for $\Phi^{\otimes m}$ for every $m$, with the same coefficient. These are
Theorem~\ref{sandwich:norm-transfer} and Theorem~\ref{prop:transfer}.

\subsection{Applications}

Our first application resolves the sharp $2$-logarithmic Sobolev
tensorization problem for the quantum Ornstein--Uhlenbeck semigroup
described above. Although the sharp one-mode constant was determined
by Beigi and Rahimi-Keshari, their multimode lower bound decreases
to zero as the number of modes grows
\cite{Beigi-RahimiKeshari2023}. The remaining question is whether the
one-mode constant remains valid for arbitrary correlated multimode
states. In finite-dimensional quantum systems, exact tensorization is known
for several qubit semigroups, including the depolarizing semigroup
and, more generally, self-adjoint unital qubit semigroups
\cite{MontanaroOsborne2010,King2014}. Recent work of Dong, Gao, Ou,
Yao, and Zhou also establishes dimension-independent approximate
tensorization for a class of qudit semigroups and exact tensorization
for its qubit members~\cite{DongGaoOuYaoZhou2026}. On the bosonic side, however, to the best of our knowledge, no strictly positive lower bound independent of the number
of modes is known for the $2$-logarithmic Sobolev constants of tensor powers of this semigroup. Using Weyl lifting, we prove that the sharp one-mode constant is preserved under every tensor power.

\begin{theorem}[Sharp $2$-logarithmic Sobolev tensorization, informal]
\label{intro:2-LS}
For a quantum Ornstein--Uhlenbeck semigroup with a faithful thermal
invariant state, the sharp $2$-logarithmic Sobolev constant of every
finite tensor power equals the sharp one-mode constant. Displaced
copies of the product invariant thermal state attain equality.
\end{theorem}

The proof of this theorem starts from an improved nonlinear logarithmic Sobolev inequality
for the qubit depolarizing channel~\cite{Beigi2021}. Its nonlinear
dependence on entropy contains more information than a fixed linear
logarithmic Sobolev bound. Then, a supporting line to this nonlinear entropy term
produces a new sharp logarithmic Sobolev inequality at each biased thermal qubit reference.
We express the resulting inequality as a finite-time hypercontractivity
bound as in~\eqref{eq:intro-qubit-hc}, and then transfer this bound to the bosonic semigroup as stated above. The hypercontractivity to logarithmic Sobolev implication of~\cite{OZ99,CarboneSasso2008} then gives the desired result. 

To address CMOE through Weyl lifting, we need corresponding constrained
minimum output entropy bounds on the qubit side. The sharp quadratic
logarithmic Sobolev inequality alone does not provide the nonlinear
entropy profile required for this purpose. We therefore prove an
entropy inequality for a two-parameter family of qubit channels formed
by composing partial Pauli pinchings. This inequality gives a sharp
lower bound on the output entropy at fixed input entropy, valid for
arbitrary tensor powers of the channel. A distinguished member
of this family is the depolarizing channel, for which the bound takes
the form of a quantum Mrs.~Gerber's lemma.\footnote{Here this term
refers to a minimum output entropy statement for the depolarizing
channel. It is distinct from the conditional information-combining
question studied under the same name in~\cite{HircheReeb2018}.}
These qubit inequalities are optimizer results in their own right
and supply the finite-dimensional input for our bosonic CMOE bounds.

Let
\[
 \mathcal D_\lambda(\omega)=\lambda\omega+(1-\lambda)I/2
 ,
\]
be the qubit depolarizing channel. Also, let $h(x)=-x\ln x-(1-x)\ln(1-x)$ be the binary entropy function with
$h^{-1}$ being its inverse restricted to $[0,1/2]$.

\begin{theorem}[Quantum Mrs.~Gerber's lemma for depolarizing channels]
\label{intro:MGL}
For every $n$-qubit state $\omega$ and every $0\le\lambda\le1$,
\begin{equation}\label{eq:intro:mgl}
 S\bigl(\mathcal D_\lambda^{\otimes n}(\omega)\bigr)
 \ge n\,h\!\left(\frac{1-\lambda}{2}
          +\lambda h^{-1}\!\left(\frac{S(\omega)}n\right)\right).
\end{equation}
Products of qubit states each having entropy $S(\omega)/n$ attain the
bound.
\end{theorem}

This theorem is a special case of a more general entropy inequality
for a two-parameter family of qubit channels consisting of compositions of partial Pauli pinchings
(Theorem~\ref{thm:composed-tensor}). The general result also yields
a sharp relative entropy contraction bound with respect to thermal
reference states, which we transfer to bosonic Gaussian channels
to obtain our CMOE bounds.

The input $\omega$ in this theorem may be entangled across all $n$ qubits.
To handle these difficulties, we first separate one qubit from the
remaining system and keep information about the bias of its binary
measurement outcomes. A partial-pinching inequality bounds the local
entropy change. To preserve this local estimate while noise acts on
the remaining qubits, we use a projective-measurement optimized conditional entropy,
familiar from the definition of quantum discord~\cite{OllivierZurek2002}.
These local estimates combine through a convexity argument and yield
the tensor-power inequality by induction. The stronger two-parameter
statement provides the relative entropy estimate used in the bosonic
application, while the depolarizing case is also a finite-dimensional
optimizer result in its own right.

Transferring the relative entropy consequence of this qubit inequality to
bosonic systems gives CMOE for the additive Gaussian noise channel. Combining it
with attenuation about a fixed thermal reference gives an additional
attenuator range. Write $\Phi^{\mathrm{add}}_\gamma$ for the single-mode channel
that adds $\gamma/2$ photons, and
$\Phi^{\mathrm{att}}_{\lambda,N}$ for the single-mode attenuator with
transmissivity $\lambda$ and a thermal environment of mean photon
number $N$. Their action on a thermal input is respectively
$\tau_\mu\mapsto\tau_{\mu+\gamma/2}$ and
$\tau_\mu\mapsto\tau_{\lambda\mu+(1-\lambda)N}$.

\begin{theorem}[Multimode CMOE]
\label{intro:CMOE}
Let $\rho$ be a finite-energy $m$-mode state and put
$\mu=g^{-1}(S(\rho)/m)$. For every $\gamma\ge0$,
\[
 S\bigl((\Phi^{\mathrm{add}}_\gamma)^{\otimes m}(\rho)\bigr)
 \ge m\,g(\mu+\gamma/2).
\]
For $0\le\lambda\le1$ and $N\ge\mu$,
\[
 S\bigl((\Phi^{\mathrm{att}}_{\lambda,N})^{\otimes m}(\rho)\bigr)
 \ge m\,g\bigl(\lambda\mu+(1-\lambda)N\bigr).
\]
Both bounds are attained by $\rho=\tau_\mu^{\otimes m}$.
\end{theorem}

The attenuator condition $N\ge\mu$ means that the input entropy per
mode does not exceed the entropy of one environment mode:
$S(\rho)/m\le g(N)$. The proof uses the decomposition of such an
attenuator into attenuation preserving $\tau_\mu$, followed by
additive noise. Thus the same relative entropy bound applies after the thermal-reference attenuation step.

To extend this method to pure-loss and amplifier channels, we formulate
a sharp relative entropy contraction conjecture for tensor powers of
the qubit amplitude-damping and amplitude-raising channels. Through Weyl lifting, the
positive-polarization regime of this conjecture would imply CMOE for
pure-loss channels. The negative-polarization regime, related by
bit-flip conjugation to amplitude raising, would yield the corresponding
amplifier bound. The conjecture thus provides a finite-dimensional
target for extending our bosonic results.

\subsection{Related work}

The appearance of bosonic observables in collective qubit fluctuations
belongs to the theory of quantum central limit
theorems~\cite{goderis1989central,jakvsic2010quantum}.
This theory identifies the canonical commutation relations and Gaussian
states emerging in the fluctuation limit and controls expectations of
suitable products of exponentials of fluctuation observables.
Our construction takes this correspondence as its starting point.
As discussed above, however, these correlation limits alone do not
justify the passage to weighted norms and relative entropies.
For this purpose, we must also control expressions in which fractional
powers of thermal reference states are interspersed with observables,
followed by nonlinear operations such as taking absolute values,
noninteger powers, and logarithms.

There is also a connection with the lifting constructions used in our
work on the bosonic central limit theorem and entropy under quantum
convolution~\cite{beigi2023towards,BeigiMehrabi2025Monotonicity}.
Those constructions use exact representations of the Weyl--Heisenberg
group to relate a bosonic system to a larger bosonic system. In the
present setting, the lift takes values in a finite qubit algebra, so
an exact representation of the bosonic commutation relations is
unavailable. The relations needed by the proof must instead be
recovered asymptotically. This change in the destination of the lift
is the reason for introducing pseudo-Weyl operators and for proving
the additional nonlinear convergence results.

Local asymptotic normality gives another connection between many-copy
qubit models and bosonic Gaussian systems~\cite{GutaKahn2006}.
It constructs channels that compare local statistical models in the
two descriptions. Such a comparison suggests an alternative route to transferring qubit inequalities to bosonic systems, provided the weighted norm and entropy quantities
are controlled as well. We work directly with Weyl observables because
their simple channel action is suited to the inequalities considered
here. The construction therefore does not require the Schur--Weyl
decomposition used in the qubit local asymptotic normality approach.

Recent preprints provide further connections with finite-dimensional work.
Gao and Zhao study relative entropy contraction for generalized
quantum depolarization, including conditional-expectation and dephasing
examples~\cite{GaoZhao2026}. The polarized pinching result here retains
both input entropy and polarization and yields a moving-reference
bound to product Gibbs references for arbitrary correlated inputs. Ahmed and collaborators obtain a bosonic Gaussian
channel from qubit depolarization on symmetric subspaces and use that
limit for quantum-capacity codes~\cite{AhmedEtAl2026}. Their coding
problem concerns a different limit construction; our results
control nonlinear weighted norms and relative entropy in the thermal
fluctuation representation.

\paragraph{Added note.} In the final stages of preparing this manuscript, we learned of the
OpenAI manuscript \cite{OpenAIEPNI}, which claims the entropy
photon-number inequality (EPNI). As stated there, choosing one input to be a product thermal state
gives multimode CMOE for every thermal attenuator. It also implies
additive-noise CMOE through an appropriate limit, although this is not stated explicitly in
that manuscript. Thus EPNI implies both CMOE conclusions
of Theorem~\ref{intro:CMOE}. Nevertheless, the beam-splitter EPNI does not directly give the CMOE bound for amplifier channels. 

Our framework applies beyond CMOE and provides a general method for transferring qubit inequalities to bosonic systems. It isolates
finite-dimensional targets for further Gaussian optimizer problems. Focusing on entropy inequalities,  the cited EPNI statement does not supply our quantum MGL nor our
conjectured amplitude-damping contraction, which, if proved, would imply CMOE for all phase-covariant Gaussian channels including amplification channels.

\paragraph{Organization of the paper.} Section~\ref{sec:preliminaries} introduces Gaussian channels, their qubit counterparts, and the pseudo-Weyl operators that connect them. Section~\ref{sec:norm-lifting} develops Weyl lifting and uses it to transfer weighted superoperator norm inequalities from qubits to bosonic systems. Section~\ref{sec:LS} applies this result to establish sharp $2$-logarithmic Sobolev tensorization and hypercontractivity for the bosonic Ornstein--Uhlenbeck semigroup. Section~\ref{sec:entropy-transfer} constructs qubit states through Weyl lifting and uses them to transfer relative entropy inequalities to bosonic systems, and derives CMOE bounds for additive Gaussian noise and thermal attenuation in the heating range. The required qubit contraction bound is stated there; Section~\ref{sec:qubit-results} states and proves the stronger qubit entropy inequalities and derives that bound. Section~\ref{sec:amp-damp} formulates the amplitude-damping contraction conjecture and shows how it would imply the full CMOE bound for the phase-covariant Gaussian channels considered here. We conclude with some remarks and collect additional technical details in the appendices.

\section{Gaussian channels and their qubit counterparts}
\label{sec:preliminaries}

To identify the Gaussian channels arising from qubit fluctuations, we
first describe the two systems using compatible reference states and
observables. We fix the notation for bosonic states, Weyl operators,
and phase-covariant Gaussian channels, and then introduce their qubit
counterparts.

\subsection{Bosonic states and Weyl operators}

An $m$-mode bosonic system has Hilbert space
\[
 \mathcal H_m=\bigotimes_{j=1}^m\ell^2(\mathbb N_0).
\]
Its number basis consists of the vectors
$\ket{k_1,\ldots,k_m}$, where $k_j$ is the number of photons in mode
$j$. The annihilation and creation operators act on this basis as
\[
 \begin{aligned}
 \ac_j\ket{k_1,\ldots,k_j,\ldots,k_m}
 &=\sqrt{k_j}\ket{k_1,\ldots,k_j-1,\ldots,k_m},\\
 \ac_j^\dagger\ket{k_1,\ldots,k_j,\ldots,k_m}
 &=\sqrt{k_j+1}\ket{k_1,\ldots,k_j+1,\ldots,k_m}.
 \end{aligned}
\]
with the first expression understood to be zero when $k_j=0$.
On the linear span of the number basis, they satisfy the canonical
commutation relations
\begin{equation}\label{eq:bosonic-CCR}
 [\ac_j,\ac_k^\dagger]=\delta_{jk}I,
 \qquad [\ac_j,\ac_k]=[\ac_j^\dagger,\ac_k^\dagger]=0.
\end{equation}
The number operators $\ac_j^\dagger\ac_j$ have the number basis as
their joint eigenbasis. We write
$\Ntot=\sum_{j=1}^m\ac_j^\dagger\ac_j$ for the total photon number.

We denote the Banach space of trace-class operators on $\mathcal H_m$
by $\mathcal S_1(\mathcal H_m)$, with norm
$\|T\|_1=\tr|T|$. A state is a positive operator
$\rho\in\mathcal S_1(\mathcal H_m)$ with $\tr\rho=1$. Unless otherwise stated, all states in this paper are assumed to have finite energy, meaning that 
$\tr(\rho\Ntot)<\infty$.\footnote{For an unbounded positive operator
$B$ and a spectral decomposition $\rho=\sum_r p_r\ketbra{\psi_r}{\psi_r}$,
we understand $\tr(\rho B)$ as the extended quadratic-form expectation
$\sum_r p_r\|B^{1/2}\psi_r\|^2$. It is $+\infty$ if an occupied
vector is outside the form domain or the sum diverges. Finite energy
does not require the product $\rho\Ntot$ to be trace class.}
This ensures that the von Neumann entropy
$S(\rho)=-\tr(\rho\ln\rho)$ is finite. More precisely, if the total
mean photon number is $E$, then $S(\rho)\le m\,g(E/m)$, where $g(\cdot)$
is the thermal entropy function
\begin{equation}\label{eq:g-function}
 g(\mu)=(\mu+1)\ln(\mu+1)-\mu\ln \mu.
\end{equation}

Creation and annihilation operators are unbounded, so to compare with qubit operators, it is more convenient
to use their unitary exponentials. For $z\in\C^m$, the \emph{Weyl
operator}, also called the displacement operator with the displacement vector $z$, is defined as
\begin{equation}\label{eq:bosonic-Weyl}
 W(z)=\exp\!\left[\sum_{j=1}^m
             (z_j\ac_j^\dagger-\bar z_j\ac_j)\right].
\end{equation}
Translating the canonical commutation relations~\eqref{eq:bosonic-CCR}, these unitaries satisfy
\begin{equation}\label{eq:Weyl-relations}
 W(z)W(w)=e^{(z^\top\bar w-\bar z^\top w)/2}W(z+w),
 \qquad W(z)^\dagger=W(-z),
\end{equation}
for every $z, w \in \C^m$. 

For a given trace-class operator $T$,
its characteristic function is $\chi_T(z)=\tr(TW(z))$. This function determines $T$ uniquely. A state $\rho$ is \emph{Gaussian} if its characteristic function is a Gaussian function. A Gaussian state is fully specified by its first and second moments. The centered Gaussian states that will serve as our references are products of one-mode thermal states:
\begin{equation}\label{eq:notation}
 \tau_\mu= (1-e^{-\beta})e^{-\beta\ac^\dagger\ac} = \frac1{\mu+1}\sum_{k=0}^\infty
             \left(\frac\mu{\mu+1}\right)^k\ketbra{k}{k},
 \qquad \mu\ge0, \, \beta=\beta(\mu)=\ln\frac{\mu+1}{\mu}.
\end{equation}
Here $\tr(\tau_\mu\ac^\dagger\ac)=\mu$, so $\mu$ is the mean
photon number of the reference. Its characteristic function and entropy are $\chi_{\tau_\mu^{\otimes m}}(z)=e^{-(\mu+1/2)|z|^2}$ and $S(\tau_\mu)=g(\mu)$, where $|z|^2=\sum_j|z_j|^2$. At $\mu=0$, the thermal state is the vacuum $\ketbra00$.
The lifting construction below uses $\mu>0$, for which the reference
is faithful. Vacuum endpoints in our applications are treated by limits.

\subsection{Phase-covariant Gaussian channels}

A quantum channel on the bosonic system is a bounded linear,
completely positive, trace-preserving map
$\Phi:\mathcal S_1(\mathcal H_m)\to\mathcal S_1(\mathcal H_m)$.
We will use both its action on states and its action on observables.
The dual space of $\mathcal S_1(\mathcal H_m)$ is the Banach space
$\mathcal B(\mathcal H_m)$ of bounded operators, equipped with the
operator norm $\|A\|_\infty$ and with pairing
$(T,B)\mapsto\tr(TB)$. Accordingly, the \emph{dual channel}
$\Phi^\dagger$ is defined by
\begin{equation}\label{eq:channel-duality}
 \tr(\Phi(T)B)=\tr(T\Phi^\dagger(B)),
 \qquad T\in\mathcal S_1(\mathcal H_m),\quad
 B\in\mathcal B(\mathcal H_m).
\end{equation}
It is a normal, completely positive, unital map on bounded operators.
This is the Heisenberg description of the channel; $\Phi$ itself is
its Schr\"odinger description.

A quantum channel is called Gaussian if it maps Gaussian states to Gaussian states. The action of a Gaussian channel is determined by linear maps that describe the transformation of the first and second moments of a Gaussian state. We study the family of one-mode phase-covariant Gaussian channels and
their tensor powers. A one-mode channel is called phase-covariant if it commutes with conjugation by unitaries $U_\theta = \exp(i\theta \bfa^\dagger \bfa)$ for all $\theta\in \mathbb R$.

After removing a fixed phase rotation, the one-mode channels in this
family are specified by a gain $\lambda$ and a noise parameter
$\gamma$ through~\cite[Appendix A]{Beigi-RahimiKeshari2023}
\begin{equation}\label{eq:gaussian-general}
 \chi_{\Phi_{\lambda,\gamma}(T)}(z)
 =e^{-\gamma|z|^2/2}\chi_T(\sqrt\lambda\,z),
 \qquad \lambda\ge0,\quad \gamma\ge|1-\lambda|.
\end{equation}
The restriction on $\gamma$ is precisely the complete-positivity
condition. Equivalently, the dual of the $m$-fold product acts as
\begin{equation}\label{eq:Gaussian-dual-Weyl}
 (\Phi_{\lambda,\gamma}^\dagger)^{\otimes m}(W(z))
 =e^{-\gamma|z|^2/2}W(\sqrt\lambda\,z).
\end{equation}
This simple action is the property we will reproduce using qubits.
It separates the contraction or expansion of the displacement argument
from the Gaussian noise added by the channel.

A thermal input remains thermal under the action of these channels. Setting
$w=(\gamma+\lambda-1)/2$, we obtain
\begin{equation}\label{eq:thermal-mean-map}
 \Phi_{\lambda,\gamma}(\tau_\mu)=\tau_{\lambda\mu+w}.
\end{equation}
Thus $w$ is the mean photon number at the output when the input is
the vacuum. The usual attenuator, amplifier, and additive-noise
channels are given by
\begin{equation}\label{eq:Gaussian-subclasses}
 \begin{aligned}
 \Phi^{\mathrm{att}}_{\lambda,N}
   &=\Phi_{\lambda,(1-\lambda)(2N+1)}, &&0\le\lambda<1,\quad N\ge0,\\
 \Phi^{\mathrm{amp}}_{\lambda,N}
   &=\Phi_{\lambda,(\lambda-1)(2N+1)}, &&\lambda>1,\quad N\ge0,\\
 \Phi^{\mathrm{add}}_\gamma&=\Phi_{1,\gamma}, &&\gamma\ge0.
 \end{aligned}
\end{equation}
Attenuation results from mixing the input with a thermal environment
of mean photon number $N$ at a beamsplitter. Amplification uses a
two-mode squeezing interaction with the same type of environment.
The cases $N=0$ are pure loss and quantum-limited amplification.
Additive noise increases the thermal mean by $\gamma/2$.
The identity channel is included at $\lambda=1$, $\gamma=0$.

The continuous-time evolutions in this family describe photon creation
and annihilation at fixed rates. For a jump operator $B$, write
\begin{equation}\label{eq:dissipator}
 \mathfrak D[B](T)=BTB^\dagger-\tfrac12\{B^\dagger B,T\},
 \qquad \{A,B\}=AB+BA.
\end{equation}
This term is called a dissipator because it describes an irreversible
jump process rather than Hamiltonian evolution. The jump term is
balanced by the anticommutator so that the trace is preserved.
We use the Schr\"odinger generator
\begin{equation}\label{eq:boson-generator}
 \mathcal L_{\nu_0,\nu_1}
 =\nu_0\mathfrak D[\ac^\dagger]+\nu_1\mathfrak D[\ac],
 \qquad \nu_0,\nu_1\ge0,
\end{equation}
with semigroup $\Phi_t=e^{t\mathcal L_{\nu_0,\nu_1}}$.
Since $\ac$ is unbounded, the displayed generator formula is initially
interpreted on finite matrices in the number basis. The corresponding
channel semigroup is specified on trace-class operators by
\eqref{eq:gaussian-general}, with parameters
\begin{equation}\label{eq:general-rates}
 \lambda_t=e^{-2c_\nu t},\qquad
 \gamma_t=
 \begin{cases}
  \dfrac{\nu_0+\nu_1}{2c_\nu}(1-e^{-2c_\nu t}),&c_\nu\ne0,\\[4pt]
  2\nu_0t,&c_\nu=0,
 \end{cases}
 \qquad c_\nu=\frac{\nu_1-\nu_0}{2}.
\end{equation}
Here $\nu_0$ is the creation rate and $\nu_1$ is the annihilation rate.
Along a thermal trajectory, the mean obeys
$\dot\mu_t=\nu_0-2c_\nu\mu_t$.
Positive $c_\nu$ gives attenuation towards the mean
$\nu_0/(\nu_1-\nu_0)$; zero $c_\nu$ gives additive noise; negative
$c_\nu$ gives amplification.

Every channel in \eqref{eq:gaussian-general} with $\lambda>0$ occurs
at time $t=1$ in one of these semigroups. Indeed, at time one take
$c_\nu=-\tfrac12\ln\lambda$ and
$\nu_0=2c_\nu w/(1-\lambda)$, $\nu_1=\nu_0+2c_\nu$ when
$\lambda\ne1$; for $\lambda=1$, take $\nu_0=\nu_1=\gamma/2$.
The complete-positivity condition guarantees nonnegative rates.
The endpoint $\lambda=0$ is a thermal replacement channel, obtained
as an infinite-time limit of attenuation.

For $m$ modes, let $\mathcal L_{\nu_0,\nu_1}^{(j)}$ be the generator acting on the $j$th mode and write
\begin{equation}\label{eq:product-generator}
 \widehat{\mathcal L}_{\nu_0,\nu_1}
 =\sum_{j=1}^m\mathcal L_{\nu_0,\nu_1}^{(j)},
 \qquad e^{t\widehat{\mathcal L}_{\nu_0,\nu_1}}
 =\Phi_t^{\otimes m}.
\end{equation}
Its finite-time dual action is
\begin{equation}\label{eq:OU-Weyl}
 (\Phi_t^\dagger)^{\otimes m}(W(z))
 =e^{-\gamma_t|z|^2/2}W(e^{-c_\nu t}z).
\end{equation}

When $0<\nu_0<\nu_1$, the semigroup is stationary and maps any input to the faithful thermal
state $\tau_\mu$ with
$\mu=\nu_0/(\nu_1-\nu_0)$ and $\beta=\ln(\nu_1/\nu_0)$ in the limit.
This is the quantum Ornstein--Uhlenbeck semigroup. Fixing $\beta$
determines the ratio of the two rates, while their common scale fixes
the speed of relaxation. We choose the normalization
\begin{equation}\label{eq:thermal-rates}
 \mathcal L_\beta=e^{-\beta/2}\mathfrak D[\ac^\dagger]
                         +e^{\beta/2}\mathfrak D[\ac],
 \qquad c_\beta=\sinh(\beta/2),\quad \beta>0.
\end{equation}
It preserves $\tau_\mu$, where $\mu=(e^\beta-1)^{-1}$, and has
$\lambda_t=e^{-2c_\beta t}$ and
$\gamma_t=(1-e^{-2c_\beta t})\coth(\beta/2)$.
The corresponding generator for any $0<\nu_0<\nu_1$ satisfying $\beta=\ln(\nu_1/\nu_0)$
is $(c_\nu/c_\beta)\mathcal L_\beta$, so we can recover its constants
by rescaling time. Pure loss has the vacuum as its fixed point and corresponds to
the nonfaithful endpoint $\nu_0=0<\nu_1$.

\subsection{Thermal qubit references and phase-covariant channels}
\label{sec:qubit-semigroups}

The qubit that represents one fluctuation variable has Hilbert space
$\C^2$ and computational basis $\{\ket0,\ket1\}$.
These two vectors label qubit levels; they are distinct from the
infinite number basis of a bosonic mode. Define Pauli operators
\begin{equation}\label{eq:def-qubitPauli}
    \bc=\ketbra01,\qquad
 X=\bc+\bc^\dagger,\qquad
 Y=i(\bc^\dagger-\bc),\qquad
 Z=\ketbra00-\ketbra11=[\bc,\bc^\dagger].
\end{equation}
The operators $\mathbf b, \mathbf b^\dagger$ are called qubit annihilation and creation operators. 
The last identity shows the basic difference from an oscillator:
their commutator is $Z$ rather than $I$. Nevertheless, as will be shown below, their thermal expectations converge to bosonic ones.

To match a bosonic thermal state, we choose a qubit thermal state at the
same inverse temperature. Thus, for $\mu>0$ and
$\beta=\beta(\mu)$, set
\begin{align}\label{eq:qubit-thermal-notation}
 &\sigma_\mu
 =\frac{1}{1+e^{-\beta}}\big(\ketbra00+e^{-\beta}\ketbra11\big)
 =\operatorname{diag}\!\left(\frac{\mu+1}{2\mu+1},
                             \frac\mu{2\mu+1}\right),\\
                        &\kappa_\mu:=\tr(\sigma_\mu Z)
 =\frac1{2\mu+1}=\tanh(\beta/2).
\end{align}
The parameter $\mu$ remains the mean photon number of the bosonic
reference thermal state. It is not the mean excitation of the qubit, which is
$\mu/(2\mu+1)$. What the two thermal states share is the ratio
$e^{-\beta}$ of successive level weights. The positive polarization
$\kappa_\mu$ then gives the expected commutator of the qubit annihilation and creation operators.

A qubit phase rotation is conjugation by $V_\theta=e^{i\theta\bc^\dagger\bc}$. A qubit channel $\Psi$ is phase-covariant if it commutes with conjugation by $V_\theta$ for all $\theta\in \mathbb R$.
Writing $\omega=(I+xX+yY+zZ)/2$, and removing a fixed output phase
rotation, such a channel has the Bloch action
\begin{equation}\label{eq:qubit-standard-Bloch}
 (x,y,z)\longmapsto(sx,sy,\ell z+h).
\end{equation}
The parameters $s$, $\ell$, and $h$ are real. Here $s$ describes the
transverse contraction, while $\ell$ and $h$ describe the longitudinal
contraction and translation. The output phase can be chosen so that
$s\ge0$; with this convention, complete positivity is equivalent
to~\cite[Eqs.~(4)--(6)]{FilippovGlinovLeppajarvi}
\begin{equation}\label{eq:qubit-shared-CP}
 1-\ell\ge|h|,\qquad
 1+\ell\ge\sqrt{h^2+4s^2}.
\end{equation}

Let us compute the action of the qubit channel~\eqref{eq:qubit-standard-Bloch} on the thermal
reference $\sigma_\mu$. Since $\sigma_\mu$ has Bloch vector
$(0,0,\kappa_\mu)$, its image under $\Psi$ has Bloch vector
$(0,0,\ell\kappa_\mu+h)$. Thus, provided
$0<\ell\kappa_\mu+h<1$, we have
$\Psi(\sigma_\mu)=\sigma_{\mu'}$, where
\[
 \kappa_{\mu'}=\ell\kappa_\mu+h,
 \qquad
 \mu'=\frac12\left(\frac1{\kappa_{\mu'}}-1\right).
\]
We next associate to $\Psi$ a bosonic Gaussian channel
$\Phi_{\lambda,\gamma}$ that satisfies $\Phi_{\lambda, \gamma}(\tau_\mu)=\tau_{\mu'}$.
Matching the parameters $\mu, \mu'$ of thermal references requires $\mu'=\lambda\mu+\frac{\gamma+\lambda-1}{2}$, or equivalently $\frac1{\kappa'}=\frac{\lambda}{\kappa}+\gamma$, where for simplicity we use $\kappa=\kappa_\mu$ and $\kappa'=\kappa_{\mu'}$.
To determine $\lambda$ and $\gamma$ separately, we also match the
transverse action of the qubit channel. Comparing $\Psi^{\dagger}(\mathbf b) = s\mathbf b$ and $\Phi_{\lambda,\gamma}^\dagger(\mathbf a)=\sqrt{\lambda}\,\mathbf a$, and taking into account the
fluctuation normalizations defined in~\eqref{eq:collective-qubit} at the input and output references, we find that 
\begin{equation}\label{eq:general-map}
 \lambda=s^2\frac{\kappa}{\kappa'},\qquad
 \gamma=\frac{1-s^2}{\kappa'}.
\end{equation}
These choices satisfy the thermal-reference matching condition above. We will see below that this choice of parameters $\lambda, \gamma$ reproduces the
full Gaussian action in the fluctuation limit and establishes the required correspondence between the qubit and bosonic channels.

These parameters always satisfy the Gaussian complete-positivity
condition. To see this, apply the Schwarz inequality for the unital
completely positive map $\Psi^\dagger$ to $\bc$ and $\bc^\dagger$, and then take the
expectation in $\sigma_\mu$. Since
$\Psi^\dagger(\bc)=s\bc$, we obtain $s^2(1-\kappa)\le1-\kappa'$ and $s^2(1+\kappa)\le1+\kappa'$. Using \eqref{eq:general-map}, these inequalities are respectively
$\gamma\ge1-\lambda$ and $\gamma\ge\lambda-1$.

Conversely, start with $\Phi_{\lambda,\gamma}$ and $\mu>0$ such
that $\mu'=\lambda\mu+(\gamma+\lambda-1)/2>0$. Then we can recover the qubit parameters by
\begin{equation}\label{eq:qubit-output-reference}
 s=\sqrt{\lambda\kappa'/\kappa},\qquad
 h=\kappa'-\ell\kappa.
\end{equation}
We denote any channel with these parameters satisfying
\eqref{eq:qubit-shared-CP} by
$\Psi_{\lambda,\gamma;\ell}^{(\mu)}$. Such a choice always exists:
put
\[
 r=\min\!\left\{\frac{1-\kappa'}{1-\kappa},
                 \frac{1+\kappa'}{1+\kappa}\right\},
 \qquad \ell=r,\qquad h=\kappa'-r\kappa.
\]
Then $0<r\le1$, $|h|=1-r$, and Gaussian complete positivity gives
$s^2\le r$, which verifies both conditions in
\eqref{eq:qubit-shared-CP}. Other choices of $\ell$ may also be
possible. The superscript $\mu$ specifies the reference used to label
the channel; it does not make the action of the channel depend on
the state to which it is applied.

For the logarithmic Sobolev application, the qubit channel $\Psi$ belongs to a semigroup and the reference is fixed by
its stationary state. In this case, $\kappa'=\kappa$ and
$h_t=(1-\ell_t)\kappa$. Moreover, the shared parameters are reduced to
\[
 \lambda_t=s_t^2,\qquad \gamma_t=(1-s_t^2)/\kappa.
\]

In the following we fix parameters of a semigroup of qubit channels that match those of the quantum Ornstein--Uhlenbeck semigroup with generator~\eqref{eq:thermal-rates}. 
Using the same notation for dissipators as in the bosonic case, we let
\begin{equation}\label{eq:Kq}
 \mathcal K_\beta
 =\tanh(\beta/4)\left(
   e^{\beta/2}\mathfrak D[\bc]
  +e^{-\beta/2}\mathfrak D[\bc^\dagger]
  +\tfrac12\mathfrak D[Z]\right).
\end{equation}
Its semigroup $\Psi_t=e^{t\mathcal K_\beta}$ preserves $\sigma_\mu$ and has eigenoperators
\begin{equation}\label{eq:Kq-eigen}
 \begin{aligned}
 \mathcal K_\beta^\dagger(\bc)&=-c_\beta\bc,
 &\mathcal K_\beta^\dagger(\bc^\dagger)&=-c_\beta\bc^\dagger,\\
 \mathcal K_\beta^\dagger(Z-\kappa_\mu I)
 &=-d_\beta(Z-\kappa_\mu I),
 &d_\beta&=2\tanh(\beta/4)\cosh(\beta/2),
 \end{aligned}
\end{equation}
where as before $c_\beta=\sinh(\beta/2)$.
Thus $s_t=e^{-c_\beta t}$ and $\ell_t=e^{-d_\beta t}$.

\subsection{Pseudo-Weyl operators and thermal fluctuations}

Although finite-dimensional operators cannot satisfy the canonical
commutation relations~\eqref{eq:bosonic-CCR} exactly, the quantum central limit theorem
provides a way to recover them in the fluctuation limit. In the following we consider normalized sums of qubit creation and annihilation operators so that
their commutator has expectation one in the reference state, while its
fluctuations vanish as the number of qubits grows. These collective operators thus
play the roles of bosonic annihilation and creation operators in the
limit.

Consider an $m$-mode bosonic system and for each bosonic mode $1\leq j\leq m$, take a group of $M$ qubits. The full Hilbert space
is $\mathcal H_{m,M}^{\mathrm{qubit}}
 =\bigotimes_{j=1}^m(\C^2)^{\otimes M}$, and we consider $\sigma_\mu^{\otimes mM}$ as the corresponding reference state. 
Let $\bc_{j,k}$ and $Z_{j,k}$ be qubit operators defined in~\eqref{eq:def-qubitPauli} acting on qubit $1\leq k\leq M$ in group $j$ and
as the identity on all other qubits. Define the collective qubit operator
\begin{equation}\label{eq:collective-qubit}
 \ac_{j,M}^{\mathrm{qubit}}
 =\frac1{\sqrt{M\kappa_\mu}}\sum_{k=1}^M\bc_{j,k}.
\end{equation}
Operators corresponding to different groups $j, j'$ commute. Within one group, we have
\begin{equation}\label{eq:collective-commutator}
 \Big[\ac_{j,M}^{\mathrm{qubit}},(\ac_{j,M}^{\mathrm{qubit}})^\dagger\Big] =\frac1{M\kappa_\mu}\sum_{k=1}^M Z_{j,k}.
\end{equation}
From this equation we realize that the normalization in~\eqref{eq:collective-qubit} is chosen in such a way that $\tr\big(\sigma_\mu^{\otimes mM}\, \big[\ac_{j,M}^{\mathrm{qubit}},(\ac_{j,M}^{\mathrm{qubit}})^\dagger\big] \big)=1$. We also have
\begin{equation} \label{eq:qudratic-commut-PW}
\tr\!\left[\sigma_\mu^{\otimes mM}
 \left(\Big[\ac_{j,M}^{\mathrm{qubit}},
         (\ac_{j,M}^{\mathrm{qubit}})^\dagger\Big]-I\right)^2\right] =\tr\!\left[\sigma_\mu^{\otimes mM}
\left(\frac{1}{M\kappa_\mu}\sum_{k=1}^{M}Z_{j,k}
- I\right)^2\right]=\frac{1-\kappa_\mu^2}{M\kappa_\mu^2}.
\end{equation}
The commutator therefore tends to $I$ in thermal mean square. We emphasize that this
statement concerns only the chosen product reference state, and does not necessarily imply convergence in any other sense.

The \emph{pseudo-Weyl operators} are now defined in terms of the exponentials of
these operators. For any $z\in \mathbb C^m$ we define
\begin{equation}\label{eq:pseudo-Weyl}
 W_{M,\mu}(z)  \coloneqq\exp\!\left[\sum_{j=1}^m
  \left(z_j(\ac_{j,M}^{\mathrm{qubit}})^\dagger
                 -\bar z_j\ac_{j,M}^{\mathrm{qubit}}\right)\right] =\bigotimes_{j=1}^m
 \left[e^{(z_j\bc^\dagger-\bar z_j\bc)/\sqrt{M\kappa_\mu}}
 \right]^{\otimes M}.
\end{equation}
Pseudo-Weyl operators are unitary and satisfy $W_{M,\mu}(z)^\dagger=W_{M,\mu}(-z)$.
Their reference expectations have the following exact finite-$M$
expression and scalar limit as $M \to \infty$
\begin{equation}\label{eq:pseudo-thermal-characteristic}
 \tr(\sigma_\mu^{\otimes mM}W_{M,\mu}(z))
 =\prod_{j=1}^m
   \cos\!\left(\frac{|z_j|}{\sqrt{M\kappa_\mu}}\right)^M
 \longrightarrow e^{-|z|^2/(2\kappa_\mu)}
 =\chi_{\tau_\mu^{\otimes m}}(z).
\end{equation}
This is a quantum central limit theorem in its simplest form:
the centered fluctuations of independent qubits have a Gaussian limit.
The normalization chosen in \eqref{eq:collective-qubit} depends on $\mu$ and makes its
covariance agree with that of $\tau_\mu$. When it is clear from the context, we may drop the subscript $\mu$ from $W_{M, \mu}(z)$.

Matching a single characteristic function does not yet reveal a signature of the Weyl--Heisenberg group in the limiting system. For that, one must keep
products of pseudo-Weyl operators. For example, a second-order
expansion of the pseudo-Weyl operators gives
\begin{align*} W_{M,\mu}(z)W_{M,\mu}(w)=W_{M,\mu}(z+w)\times\exp\!\left[ \frac{1}{2M\kappa_\mu} \sum_{j=1}^{m}(z_j\bar w_j-\bar z_jw_j) \sum_{k=1}^{M}Z_{j,k} \right] +O(M^{-1/2}),
\end{align*}
where the error is in operator norm. Then, using~\eqref{eq:qudratic-commut-PW} we find that
\begin{equation}\label{eq:pseudo-two-word-limit}
 \begin{aligned}
 &\lim_{M\to\infty}\tr\!\big[
  \sigma_\mu^{\otimes mM}W_{M,\mu}(z)W_{M,\mu}(w)\big]
 =e^{(z^\top\bar w-\bar z^\top w)/2}
        e^{-|z+w|^2/(2\kappa_\mu)}
 =\tr\!\left[\tau_\mu^{\otimes m}W(z)W(w)\right].
 \end{aligned}
\end{equation}
The same expansion applies to every fixed
finite product. These are thermal correlation limits of the type
studied in quantum central limit theory~\cite{goderis1989central,jakvsic2010quantum}.
They recover the bosonic relations in expectations, while the
finite-$M$ operators need not obey those relations exactly.

\subsection{Channel correspondence}

We can now make precise what it means for a qubit channel to be a
counterpart of a Gaussian channel. The two act on different Hilbert
spaces, so they cannot be compared directly as maps on a common space. Instead, we compare their actions on the corresponding Weyl
observables using their reference states.

Let $\Psi=\Psi_{\lambda,\gamma;\ell}^{(\mu)}$ be a qubit channel and let
$\mu'>0$ be its output reference parameter: $\Psi(\sigma_\mu) = \sigma_{\mu'}$. For $z_j\ne0$, put
\[
 V_j=\frac{z_j\bc^\dagger-\bar z_j\bc}{|z_j|},
 \qquad \theta_{j,M}=\frac{|z_j|}{\sqrt{M\kappa_{\mu'}}}.
\]
Since $V_j^2=-I$ we have
\begin{equation}\label{eq:W-expand-I-b}
W_{M,\mu'}(z)
= \bigotimes_{j=1}^{m}
   \left( e^{\theta_{j,M} V_j} \right)^{\otimes M}= \bigotimes_{j=1}^{m}
   \big( \cos(\theta_{j,M}) I
   + \sin(\theta_{j,M}) V_j \big)^{\otimes M}.
\end{equation}
Then using $\Psi^\dagger(\bc)=s\bc$ and $\Psi^\dagger(\bc^\dagger)=s\bc^\dagger$, the exact dual action on pseudo-Weyl operators is computed as
\begin{equation}\label{eq:qubit-dual-pseudo-Weyl}
 (\Psi^\dagger)^{\otimes mM}(W_{M,\mu'}(z))
 =\bigotimes_{j=1}^m
   \big(\cos\theta_{j,M}\,I+s\sin\theta_{j,M}\,V_j\big)^{\otimes M}.
\end{equation}
We note that the reference state of the output of the channel $\Psi$ is $\sigma_{\mu'}$, so $\Psi^\dagger$ acts on pseudo-Weyl operators associated with parameter $\mu'$. 

\begin{proposition} \label{lem:adjoint-limit}
Fix $m$, $\mu,\mu'>0$, an admissible channel
$\Psi_{\lambda,\gamma;\ell}^{(\mu)}$, and $z\in\C^m$. Then
\begin{equation}\label{eq:qubit-dual-Gaussian-limit}
 \left\|(\Psi^\dagger)^{\otimes mM}(W_{M,\mu'}(z))
       -e^{-\gamma|z|^2/2}W_{M,\mu}(\sqrt\lambda\,z)\right\|_\infty
 =O(M^{-1/2}).
\end{equation}
\end{proposition}

\begin{proof}
For any $\theta$ with $|\theta| < \pi/2$, we can write
\[
 \cos\theta\,I+s\sin\theta\,V
 =r_\theta e^{\varphi_\theta V},\qquad
 r_\theta^2=1-(1-s^2)\sin^2\theta,\quad
 \varphi_\theta=\arctan(s\tan\theta).
\]
Indeed,
$e^{\varphi V}=\cos\varphi\,I+\sin\varphi\,V$,
$r_\theta\cos\varphi_\theta=\cos\theta$, and
$r_\theta\sin\varphi_\theta=s\sin\theta$.
Consequently,~\eqref{eq:qubit-dual-pseudo-Weyl} yields 
\begin{equation}\label{eq:exact}
 (\Psi^\dagger)^{\otimes mM}(W_{M,\mu'}(z))
 =
 \left(\prod_{j=1}^m r_{\theta_{j,M}}^M\right)
 \bigotimes_{j=1}^m
 \left[e^{\varphi_{\theta_{j,M}}V_j}\right]^{\otimes M}.
\end{equation}
The Taylor expansions
\[
 \ln r_\theta=-\tfrac12(1-s^2)\theta^2+O(\theta^4),
 \qquad
 \varphi_\theta=s\theta+O(\theta^3)
\]
and $\theta_{j,M}=O(M^{-1/2})$ give
\[
 M\ln r_{\theta_{j,M}}
 =-\frac{\gamma|z_j|^2}{2}+O(M^{-1}),
 \qquad
 |\varphi_{\theta_{j,M}}-s\theta_{j,M}|=O(M^{-3/2}).
\]
Since $m$ is fixed, summing the first expansion over the modes and
exponentiating yields
\begin{equation}\label{eq:scalar}
 \prod_{j=1}^m r_{\theta_{j,M}}^M
 =e^{-\gamma|z|^2/2}+O(M^{-1}).
\end{equation}

We next compare the unitary factors. For real $x, y$, the fundamental
theorem of calculus gives $e^{xV_j}-e^{yV_j}=\int_y^x V_j e^{tV_j}\,dt$. Since $e^{tV_j}$ is unitary and $\|V_j\|_\infty=1$, this implies
\begin{equation}\label{eq:single}
 \|e^{xV_j}-e^{yV_j}\|_\infty\le |x-y|.
\end{equation}
Next, the telescoping identity $A^{\otimes M}-B^{\otimes M} = \sum_{k=1}^M B^{\otimes(k-1)}\otimes(A-B)\otimes A^{\otimes(M-k)}$ and the triangle inequality give, for contractions $A$ and $B$,
$ \|A^{\otimes M}-B^{\otimes M}\|_\infty \le M\|A-B\|_\infty$. Therefore,~\eqref{eq:single} yields
\begin{equation}\label{eq:mode}
 \left\|
 [e^{\varphi_{\theta_{j,M}}V_j}]^{\otimes M}
 -[e^{s\theta_{j,M}V_j}]^{\otimes M}
 \right\|_\infty
 \le M|\varphi_{\theta_{j,M}}-s\theta_{j,M}|
 =O(M^{-1/2}),
\end{equation}
for any $1\leq j\leq m$.
The same argument applies across the modes: since $s\theta_{j,M}=\sqrt{\lambda}|z_j|/\sqrt{M\kappa_\mu}$ and $m$
 is a constant, we conclude that 
\begin{align}\label{eq:allmodes}
 \left\|
 \bigotimes_{j=1}^m\Big[e^{\varphi_{\theta_{j,M}}V_j}\Big]^{\otimes M}
 -
 W_{M, \mu}(\sqrt \lambda z)
 \right\|_\infty&=\left\|
 \bigotimes_{j=1}^m\Big[e^{\varphi_{\theta_{j,M}}V_j}\Big]^{\otimes M}
 -
 \bigotimes_{j=1}^m\Big[e^{s\theta_{j,M}V_j}\Big]^{\otimes M}
 \right\|_\infty\\
 &\le
 M\sum_{j=1}^m|\varphi_{\theta_{j,M}}-s\theta_{j,M}|
 =O(M^{-1/2}).
\end{align}
Finally, to take care of the scalars we use~\eqref{eq:exact} and~\eqref{eq:scalar} and the triangle inequality $\|xA-yB\|_{\infty}\leq |x-y|\|A\|_\infty+|y|\|A-B\|_\infty$
to conclude 
\begin{align*}
 &\left\|(\Psi^\dagger)^{\otimes mM}(W_{M,\mu'}(z))
 -e^{-\gamma|z|^2/2}W_{M,\mu}(\sqrt{\lambda}\,z)\right\|_\infty\\
 &\quad\le
 \left|\prod_{j=1}^m r_{\theta_{j,M}}^M-e^{-\gamma|z|^2/2}\right|
 +e^{-\gamma|z|^2/2}
 M\sum_{j=1}^m|\varphi_{\theta_{j,M}}-s\theta_{j,M}|\\
 &\quad=O(M^{-1})+O(M^{-1/2})=O(M^{-1/2}).
 \end{align*}

\end{proof}

Comparing this proposition with \eqref{eq:Gaussian-dual-Weyl} justifies 
the choice of shared parameters in \eqref{eq:general-map}. 
We also note that the longitudinal coefficient $\ell$ does not appear in
\eqref{eq:qubit-dual-pseudo-Weyl}. The point is that, as is clear from~\eqref{eq:W-expand-I-b}, the linear span of pseudo-Weyl operators excludes $Z$.
Thus two dual qubit channels can agree on every pseudo-Weyl operator
while acting differently on $Z$. Products of
pseudo-Weyl
operators generate the missing direction, but a channel dual
need not preserve products. In contrast, the full bosonic Weyl
family determines every trace-class operator via its characteristic function. This explains why the Gaussian
channel is determined by its Weyl action while a qubit counterpart
can retain an additional longitudinal parameter.

We also note that, for the fixed-reference semigroup $\Psi_t=e^{t\mathcal K_\beta}$,
\eqref{eq:Kq-eigen} gives $\lambda_t=e^{-2c_\beta t}$ and $\gamma_t=(1-e^{-2c_\beta t})/\kappa_\mu$. These are precisely the parameters of $e^{t\mathcal L_\beta}$.

\section{Weighted norm inequalities}
\label{sec:norm-lifting}

This section develops a method for deriving superoperator norm inequalities for bosonic Gaussian channels from corresponding inequalities for tensor powers of qubit channels. Its main results are Theorem~\ref{sandwich:norm-transfer}, which establishes this implication, and
Theorem~\ref{norm:weighted-lifting}, which provides the nonlinear convergence required for its proof. Section~\ref{sec:LS} then uses this method to prove sharp
$2$-logarithmic Sobolev tensorization for the bosonic
Ornstein--Uhlenbeck semigroup.

\subsection{Transferring weighted norm inequalities to bosonic systems}\label{sec:transfer-principles}
We denote the Hilbert--Schmidt inner product and its associated norm by
$\langle A,B\rangle_{\mathrm{HS}}=\tr(A^\dagger B)$ and
$\| A\|_{\mathrm{HS}} = \langle A,A\rangle_{\mathrm{HS}}^{1/2}$, respectively. The operators with finite Hilbert--Schmidt norm form a Hilbert space under this inner product, which we call the Hilbert--Schmidt space.
For $r\ge0$ and a faithful state $\xi$ define
\begin{align}\label{eq:def-Gamma}
\Gamma_\xi^r(A)=\xi^{r/2}A\xi^{r/2},
\end{align}
and for bounded $B$ and $0<p<\infty$, set
\begin{equation}\label{norm:weighted-norm}
 \|B\|_{p,\xi}
 =\Big\|\Gamma_\xi^{1/p}(B)\Big\|_p
 =\Big\|\xi^{1/(2p)}B\xi^{1/(2p)}\Big\|_p,
 \qquad \|T\|_p=\big(\tr|T|^p\big)^{1/p}.
\end{equation}
For $p\ge1$ these are weighted Schatten norms; for $0<p<1$ the
same expression is a quasi-norm. At $p=2$, the corresponding inner product is
\begin{equation}\label{eq:thermal-inner-product}
 \langle A,B\rangle_{2,\xi}
 =\tr\big(\xi^{1/2}A^\dagger\xi^{1/2}B\big).
\end{equation}

Fix $\mu,\mu'>0$ and an admissible qubit channel
$\Psi=\Psi_{\lambda,\gamma;\ell}^{(\mu)}$ from
Section~\ref{sec:qubit-semigroups}, satisfying
$\Psi(\sigma_\mu)=\sigma_{\mu'}$. Also, let $\Phi=\Phi_{\lambda,\gamma}$ be its corresponding Gaussian channel given by
\eqref{eq:general-map}, which satisfies 
$\Phi(\tau_\mu)=\tau_{\mu'}$. For a fixed number of modes $m$, the input and output bosonic
references are $\tau_\mu^{\otimes m}$ and $\tau_{\mu'}^{\otimes m}$,
while the corresponding qubit references are
$\sigma_\mu^{\otimes mM}$ and $\sigma_{\mu'}^{\otimes mM}$.
An observable at the output is measured using the output reference,
while its dual-channel image is measured using the input reference.
The following theorem respects this change of reference.

\begin{theorem}[Transferring weighted norm inequalities to bosonic systems]
\label{sandwich:norm-transfer}
Let $1\le p,q<\infty$. Suppose that for every $n\ge1$ and every
$n$-qubit operator $A$,
\begin{equation}\label{sandwich:qubit-norm-bound}
 \|(\Psi^\dagger)^{\otimes n}(A)\|_{q,\sigma_\mu^{\otimes n}}
 \le\|A\|_{p,\sigma_{\mu'}^{\otimes n}}.
\end{equation}
Then for every $m\ge1$ and every bounded $m$-mode operator $B$,
\begin{equation}\label{sandwich:boson-norm-bound}
 \|(\Phi^\dagger)^{\otimes m}(B)\|_{q,\tau_\mu^{\otimes m}}
 \le\|B\|_{p,\tau_{\mu'}^{\otimes m}}.
\end{equation}
The inequality also extends by continuity between the corresponding weighted
$L_p$ and $L_q$ completions.
\end{theorem}

We emphasize that the hypothesis on the qubit  channel must hold for every $n$, because in the proof each bosonic mode is
represented by a block of $M$ qubits and $M$ tends to infinity. The rest of this section is devoted to the proof of this theorem.

\subsection{Weyl expressions}
\label{sec:lift-and-limit}

To use the qubit inequalities, we assign a qubit observable to a
bosonic one through pseudo-Weyl operators. At finite $M$,
the pseudo-Weyl operators do not obey the exact bosonic relations. Therefore, in this assignment we keep an  expression for the bosonic observable in terms of Weyl operators fixed while $M$ grows,
and evaluate that same expression on the pseudo-Weyl operators. The following definition clarifies this point. 

\begin{definition}[Weyl lifting]\label{norm:lift-definition}
Choose finitely many displacement vectors $z_1,\ldots,z_L\in\C^m$ and let $P$ be a noncommutative polynomial with complex coefficients and
with adjoints allowed. Let
\[
 X=P\big(W(z_1),\ldots,W(z_L)\big),
\]
be a \emph{Weyl expression} for a bosonic observable $X$.
The qubit lift of this Weyl expression is defined as
\[
 X_{M,\mu}=P(W_{M,\mu}(z_1),\ldots,W_{M,\mu}(z_L)).
\]
In particular, the lift for elementary finite-sums is given by
\begin{equation}\label{norm:Weyl-sum-lift}
 X=\sum_{\ell=1}^L c_\ell W(z_\ell)
 \quad\longmapsto\quad
 X_{M,\mu}=\sum_{\ell=1}^L c_\ell W_{M,\mu}(z_\ell).
\end{equation}
When the reference parameter $\mu$ is known, we write $X_M=X_{M,\mu}$.
\end{definition}

We emphasize that a Weyl expression is not unique for a given operator $X$. In fact, due to the multiplication rule for Weyl operators, every operator admitting a Weyl expression is a finite sum of Weyl operators as in~\eqref{norm:Weyl-sum-lift}. Nevertheless, pseudo-Weyl operators do not satisfy the same multiplication rule. This is why in defining the Weyl lift it is crucial to first fix the Weyl expression. The following theorem shows that this dependence on the Weyl expression
disappears for the limiting norms used in the proof of Theorem~\ref{sandwich:norm-transfer}.

\begin{theorem}[Weighted norm convergence]
\label{norm:weighted-lifting}
For every specified finite Weyl expression for $X$, and
every real $1\le p<\infty$ we have
\begin{equation}\label{norm:all-p-limit}
 \lim_{M\to\infty}\|X_M\|_{p,\sigma_\mu^{\otimes mM}}=\|X\|_{p,{\tau_\mu^{\otimes m}}}.
\end{equation}
\end{theorem}

We first extend the correlation limit to products of Weyl operators with intervening powers of the thermal reference states. For positive observables, this already gives the desired convergence when $p$ is an integer. Our idea to handle noninteger exponents is to introduce a thermally rescaled positive functional whose $p$-th moment equals the $p$-th power of the weighted norm. The required integer moments converge by the thermal correlation limit. To establish convergence of the corresponding resolvent expressions, we represent them on Hilbert--Schmidt space and use a uniformly convergent Neumann expansion, whose coefficients are treated by bounded functional calculus for the modular operators. An integral representation of fractional powers then gives convergence for every positive exponent. For a general observable, we first apply Gaussian averaging along the modular group of the reference state. This makes the required conjugations by reference powers bounded and reduces the problem to positive trace-power convergence with exponent $p/2$. Finally, we remove the averaging using convergence in the weighted $2$-norm, together with weighted norm comparisons and interpolation, to finish the proof.

\subsection{Mixed thermal correlations}
\label{sec:mixed-thermal-words}

The correlation limit from Section~\ref{sec:preliminaries} compares
products of Weyl and pseudo-Weyl operators in the two reference states. A weighted
Schatten norm requires more. For instance, if $Y$
is positive and $p$ is an integer, then
\[
 \|Y\|_{p,{\tau_\mu^{\otimes m}}}^p
 =\tr\bigl(\bigl(\tau_\mu^{\otimes m}\bigr)^{1/(2p)}Y\bigl(\tau_\mu^{\otimes m}\bigr)^{1/(2p)}\bigr)^p.
\]
Here the reference powers occur between successive copies of $Y$.
They cannot be collected into a single reference factor unless $Y$
commutes with ${\tau_\mu^{\otimes m}}$. We therefore need to extend the correlation limit
to products with intervening thermal powers. 

Throughout this subsection, $m$ and $\mu>0$ are fixed, and
$\beta=\ln(1+1/\mu)$. The limit is always the scalar
limit as $M\to\infty$. For a faithful reference $\xi$ and a complex
number $\alpha$, the power $\xi^\alpha$ is defined by spectral calculus.
When $\operatorname{Re}\alpha\ge0$, it is bounded and if
$\operatorname{Re}\alpha>0$, then 
$\|\xi^\alpha\|_{1/\operatorname{Re}\alpha}=1$. For a purely
imaginary exponent, the power is unitary. These observations explain
both the condition on the exponents and the trace bounds in the next
lemma.

\begin{lemma}[Mixed thermal correlations]\label{norm:thermal-words}
Let $A_1,\ldots,A_k$ be specified Weyl expressions, and let
$A_{1,M},\ldots,A_{k,M}$ be their lifts as in
Definition~\ref{norm:lift-definition}. For complex numbers
$\alpha_1,\ldots,\alpha_k$ satisfying
$\operatorname{Re}\alpha_j\ge0$ and $\sum_j\alpha_j=1$, we have
\begin{equation}\label{norm:word-limit}
 \lim_{M\to\infty}
 \tr\Big(\bigl(\sigma_\mu^{\otimes mM}\bigr)^{\alpha_1}A_{1,M}\cdots
                 \bigl(\sigma_\mu^{\otimes mM}\bigr)^{\alpha_k}A_{k,M}\Big)
 =\tr\Big(\bigl(\tau_\mu^{\otimes m}\bigr)^{\alpha_1}A_1\cdots
                 \bigl(\tau_\mu^{\otimes m}\bigr)^{\alpha_k}A_k\Big).
\end{equation}
All the products in this formula are trace class. If the operator
norms of the corresponding factors are bounded by $C_1,\ldots,C_k$,
the absolute value of either trace is at most $\prod_j C_j$. 
\end{lemma}

\begin{proof}
By expanding the Weyl expressions, it suffices to prove the statement
in the special case where each $A_j$ is a product of  Weyl operators. Furthermore, by increasing the number of operators and adding dummy exponents equal to $0$, we can assume without loss of generality that each $A_j$ is a single Weyl operator.
On the other hand, since in this case the mode groups are independent and appear as tensor products, it suffices to prove the lemma in the case of $m=1$.

For displacement $z$, put
\[
 G_z=\frac{z\bc^\dagger-\bar z\bc}{\sqrt{\kappa_\mu}}.
\]
Then the one-mode qubit trace in~\eqref{norm:word-limit} is equal to
\[
 f_M^M=\Big[\tr\Big(\sigma_{\mu}^{\alpha_1}e^{G_{z_1}/\sqrt M}
                    \cdots
               \sigma_{\mu}^{\alpha_k}e^{G_{z_k}/\sqrt M}\Big)\Big]^M.
\]
Expanding each exponential in powers of $M^{-1/2}$, the constant term is
$\tr\sigma_\mu=1$ and the terms of order $M^{-1/2}$ vanish because
$G_z$ is off-diagonal. The contribution of the $j$th factor at order
$M^{-1}$ is $\frac1{2M}\tr(\sigma_\mu G_{z_j}^2)$.
For $i<j$, put $s_{ij}=\alpha_{i+1}+\cdots+\alpha_j$.
The mixed contribution is
$M^{-1}\tr(\sigma_\mu^{1-s_{ij}}G_{z_i}\sigma_\mu^{s_{ij}}G_{z_j})$.
Direct computation gives
\begin{equation}\label{norm:covariance}
 \begin{aligned}
 \frac12\tr\big(\sigma_{\mu} G_z^2\big)
   &=-\frac{|z|^2}{2\kappa_\mu},\\
 \tr\big(\sigma_{\mu}^{1-s}G_z\sigma_{\mu}^sG_w\big)
   &=-\frac{e^{-\beta(1-s)}z\bar w
                   +e^{-\beta s}\bar z w}{1-e^{-\beta}}.
 \end{aligned}
\end{equation}
Consequently,
\begin{equation}\label{norm:word-quadratic}
 f_M=1+\frac{Q}{M}+O(M^{-3/2}),\qquad
 Q=-\sum_{j=1}^k\frac{|z_j|^2}{2\kappa_\mu}
 -\sum_{i<j}
  \frac{e^{-\beta(1-s_{ij})}z_i\bar z_j
           +e^{-\beta s_{ij}}\bar z_i z_j}{1-e^{-\beta}}.
\end{equation}
Taking the $M$-th power therefore gives $f_M^M\to e^Q$ as $M\to \infty$.

We now identify $e^Q$ with the bosonic trace. The following
commutation and normal-ordering identities are initially interpreted
in number-basis matrix elements; the intermediate exponentials need
not be bounded. Absolute convergence, justified below, permits their
use inside the thermal trace. Using $\tau_\mu=(1-e^{-\beta})e^{-\beta\ac^\dagger\ac}$, we find that $\tau_\mu^{-r}\mathbf a\tau_\mu^r
=e^{-\beta r}\mathbf a$. Therefore, 
\begin{align}\label{eq:W-tau-commutator}
W(z)\tau_\mu^{\,r}=e^{-|z|^2/2}e^{z\ac^\dagger}e^{-\bar z\ac}\tau_\mu^{\, r} =e^{-|z|^2/2} \tau_\mu^{\, r} e^{e^{\beta r}z\ac^\dagger}e^{-e^{-\beta r}\bar z\ac}=\tau_\mu^{\,r} \exp\!\left( e^{\beta r}z\mathbf a^\dagger -e^{-\beta r}\bar z\mathbf a \right).
\end{align}
Repeated use of this as well as $
e^{u\mathbf a^\dagger-v\mathbf a}
=e^{-uv/2}e^{u\mathbf a^\dagger}e^{-v\mathbf a}$ and $
e^{-v\mathbf a}e^{u\mathbf a^\dagger}
=e^{-v u}
e^{u\mathbf a^\dagger}e^{-v\mathbf a}$ 
gives
\begin{align*}
\tr\!\left(
\tau_\mu^{\alpha_1}W(z_1)\cdots
\tau_\mu^{\alpha_k}W(z_k)\right)& =
\tr\!\left(
\tau_\mu\prod_{i=1}^k
e^{u_i\mathbf a^\dagger-v_i\mathbf a}\right) \\
& =  \exp\!\left(
-\frac12\sum_i u_i v_i-\sum_{i<j}v_i u_j
\right) \tr\!\left(
\tau_\mu 
e^{(\sum_i u_i)\mathbf a^\dagger}
e^{-(\sum_i v_i)\mathbf a}\right),
\end{align*}
where $r_i=\alpha_{i+1}+\cdots+\alpha_k$, and $u_j=e^{\beta r_j}z_j, v_j=e^{-\beta r_j}\bar z_j$. Computing the trace by expanding in the number basis and using $
\mu=\frac{e^{-\beta}}{1-e^{-\beta}}$, gives $\operatorname{tr}\!\left(
\tau_\mu e^{u\mathbf a^\dagger}e^{-v\mathbf a}\right) = e^{-\mu uv}$.
Therefore, 
\[
\begin{aligned}
&\operatorname{tr}\!\left(
\tau_\mu^{\alpha_1}W(z_1)\cdots
\tau_\mu^{\alpha_k}W(z_k)\right)\\
&\quad=
\exp\!\left[
-\frac12\sum_i u_i v_i
-\sum_{i<j}v_i u_j
-\mu\left(\sum_i u_i\right)\left(\sum_j v_j\right)
\right]\\
&\quad=
\exp\!\left[
-\left(\mu+\frac12\right)\sum_i u_i v_i
-\sum_{i<j}
\bigl(\mu u_i v_j+(\mu+1)v_i u_j\bigr)
\right].
\end{aligned}
\]
Finally, using $u_iv_i=|z_i|^2$, $r_i-r_j=s_{ij}$, 
$u_i v_j=e^{\beta s_{ij}}z_i\bar z_j$,
$v_i u_j=e^{-\beta s_{ij}}\bar z_i z_j$, and $
\mu+\frac12=\frac1{2\kappa_\mu}$
we arrive at
\begin{align*}\label{norm:bosonic-word-evaluation} &\tr\Big(\tau_\mu^{\alpha_1}W(z_1)\cdots
                   \tau_\mu^{\alpha_k}W(z_k)\Big)
 =
 \exp\left[-\left(\mu+\frac12\right)\sum_i u_i v_i
             -\sum_{i<j}\bigl(\mu u_i v_j+(\mu+1)v_i u_j\bigr)
      \right]
 =e^Q,
 \end{align*}
as desired. To justify the rearrangements, put
$u_i^+=e^{\beta\operatorname{Re}r_i}|z_i|$ and
$v_i^+=e^{-\beta\operatorname{Re}r_i}|z_i|$.
Expanding the normally ordered Weyl factors in number-basis matrix
elements and taking absolute values gives a nonnegative series bounded by
\[
 \exp\!\left[-\frac12\sum_i|z_i|^2
       +\sum_{i<j}v_i^+u_j^+
       +\mu\Big(\sum_i u_i^+\Big)\Big(\sum_i v_i^+\Big)\right]<\infty.
\]
Thus these rearrangements are absolutely convergent. Finally, Schatten
H\"older proves the trace-class assertion and the stated bound, since
the real parts of the thermal exponents sum to one.
\end{proof}

As a special case of~\eqref{eq:W-tau-commutator}, for any real $t$ we have
\begin{equation}\label{norm:modular-rotation}
 \bigl(\tau_\mu^{\otimes m}\bigr)^{it}W(z)\bigl(\tau_\mu^{\otimes m}\bigr)^{-it}=W\big(e^{-i\beta t}z\big).
\end{equation}
Interestingly, the same holds in the qubit case:
\begin{align}\label{norm:modular-rotation-2}
\bigl(\sigma_\mu^{\otimes mM}\bigr)^{it}W_{M,\mu}(z)\bigl(\sigma_\mu^{\otimes mM}\bigr)^{-it}
       =W_{M,\mu}\big(e^{-i\beta t}z\big).
\end{align}

\begin{definition}[Jointly admissible families]
\label{def:jointly-admissible}
Let $A_{j,M}$ be qubit operators paired with bounded bosonic
operators $A_j$, for $1\le j\le k$. Overloading the notation, we do not require
$A_{j,M}$ to necessarily be a Weyl lift of $A_j$.
The families $(A_{j, M}, A_j)_{j=1}^k$ are called \emph{jointly admissible} if
they have uniform operator bounds,
\[
 \sup_M \|A_{j,M}\|_\infty<\infty,
 \qquad 1\le j\le k,
\]
and the mixed thermal-correlation limit
\eqref{norm:word-limit} holds not only for them, but also for every finite word formed
from these paired families, their adjoints, finite Weyl
expressions paired with their lifts, and real modular
rotations of all these factors as in~\eqref{norm:modular-rotation} and~\eqref{norm:modular-rotation-2}.
\end{definition}

We emphasize that in the above definition we have an operator $A_{j, M}$ acting on $mM$ qubits, for every $M$. We take this convention in the following whenever we express $(A_M, A)$ as a pair.

Lemma~\ref{norm:thermal-words} and equations~\eqref{norm:modular-rotation} and~\eqref{norm:modular-rotation-2} show that finite
Weyl expressions, paired with their Weyl lifts, are jointly admissible. On the other hand, by definition, sums, products, and adjoints of jointly admissible families are again jointly admissible. 

There are two other crucial closure properties collected in the following lemma, whose proof is straightforward and is omitted.

\begin{lemma}[Closure properties of jointly admissible families]
\label{lem:admissible-closure}

Joint admissibility has the
following closure properties.
\begin{enumerate}[(i)]
\item \emph{Uniform operator-norm approximation.}
Suppose that 
for each integer $r$, the paired families
$\big(A_{j,M}^{(r)},A_j^{(r)}\big)_{j=1}^k$ are jointly admissible, and they satisfy
\[
 \lim_{r\to\infty}
 \left(\sup_M\big\|A_{j,M}-A_{j,M}^{(r)}\big\|_\infty
              +\big\|A_j-A_j^{(r)}\big\|_\infty\right)=0,
\]
for some operators $A_j, A_{j, M}$. Then, the limiting paired families
$(A_{j,M},A_j)$ are also jointly admissible.

\item \emph{Modular averaging.}
Suppose $(A_{j,M},A_j)_{j=1}^k$ are jointly admissible.
For $f_j\in L^1(\mathbb R)$, define the weak operator
integrals
\begin{align}\label{norm:modular-integrals}
 B_{j, M}&=\int_{\R}f_j(t)\bigl(\sigma_\mu^{\otimes mM}\bigr)^{it}A_{j, M}\bigl(\sigma_\mu^{\otimes mM}\bigr)^{-it}\,\dd t,\\
 B_j&=\int_{\R}f_j(t)\bigl(\tau_\mu^{\otimes m}\bigr)^{it}A_j\bigl(\tau_\mu^{\otimes m}\bigr)^{-it}\,\dd t,
 \end{align}
Then the pairs $(B_{j,M},B_j)_{j=1}^k$, together with
the original pairs $(A_{j,M},A_j)_{j=1}^k$, are jointly
admissible.
\end{enumerate}
\end{lemma}

\subsection{Weighted norm convergence for positive operators}
\label{sec:positive-trace-powers}

In this subsection we prove the following statement regarding the limit of weighted $p$-norms of jointly admissible families. 

\begin{proposition}
\label{norm:positive-powers}
Let $(Y_M,Y)$ be a jointly admissible family of operators as defined above. 
Suppose that $0\le Y_M\le bI$ and $0\le Y\le bI$ for a fixed $b<+\infty$.
Then, for every real $p>0$,
\begin{equation}\label{norm:positive-limit}
 \lim_{M\to\infty}\|Y_M\|_{p,{\sigma_\mu^{\otimes mM}}}
 =\|Y\|_{p,{\tau_\mu^{\otimes m}}}.
\end{equation}
\end{proposition}

\medskip

For a positive observable, the weighted norm is a trace power, $\|Y\|_{p,\xi}^p
 =\tr\bigl(\xi^{1/(2p)}Y\xi^{1/(2p)}\bigr)^p$. For integer $p$, expanding the power gives a mixed thermal word,
so Lemma~\ref{norm:thermal-words} already gives the desired convergence~\eqref{norm:positive-limit}. To handle noninteger $p$, we use an integral representation of fractional powers in terms of resolvents and prove a similar convergence for resolvents. 

Let $\xi = \sum_j r_j \ketbra{v_j}{v_j}$ be the spectral decomposition of a faithful state $\xi$. For a bounded
positive operator $Y$ set
\begin{equation}\label{norm:trace-functional}
 \mathcal F_{\xi,Y,p}(f)
 =\sum_j r_j\big\langle v_j,f(r_j^{-1/p}B)v_j\big\rangle, \qquad B=\xi^{1/(2p)}Y\xi^{1/(2p)}.
\end{equation}
This is a positive functional on the space of bounded Borel functions, with
$\mathcal F_{\xi,Y,p}(1)=\tr \xi=1$. It is independent of the choice of spectral decomposition of $\xi$, and extends to nonnegative functions
by the monotone convergence theorem. From the definition we clearly have 
\begin{equation}\label{norm:trace-functional-power}
 \mathcal F_{\xi,Y,p}(x^p)=\tr B^p=\|Y\|_{p,\xi}^p.
\end{equation}
Here $x$ denotes the scalar argument of the function $f$. This observation can be generalized for every integer $k\ge1$ satisfying
$1-(k-1)/p\ge0$:
\begin{equation}\label{norm:integer-moment}
 \mathcal F_{\xi,Y,p}(x^k)
 =\tr\left[\xi^{1-(k-1)/p}Y(\xi^{1/p}Y)^{k-1}\right].
\end{equation}
Indeed, in each diagonal matrix element defining the moment, the
two outer factors $\xi^{1/(2p)}$ can be absorbed into its scalar
weight. This gives the trace above using only nonnegative thermal
powers. Their exponents sum to one, so Schatten H\"older shows that
the product is trace class and, if $0\le Y\le bI$, that
$0\le\mathcal F_{\xi,Y,p}(x^k)\le b^k$.

\begin{lemma}
\label{norm:resolvent-lemma}
Suppose $Y_M,Y$ are jointly admissible, and
$aI\le Y_M\le bI$ and $aI\le Y\le bI$ for some $0<a\le b<\infty$.
For fixed $p>0$ let $\mathcal F_M=\mathcal F_{\sigma_\mu^{\otimes mM},Y_M,p}$ and $
 \mathcal F=\mathcal F_{\tau_\mu^{\otimes m},Y,p}$. Then, for every
$t>0$,
\begin{equation}\label{norm:resolvent-convergence}
 \mathcal F_M\!\left(\frac1{x+t}\right)
 \longrightarrow
 \mathcal F\!\left(\frac1{x+t}\right).
\end{equation}
\end{lemma}

We first deduce the proposition from this lemma, and then prove the
resolvent limit~\eqref{norm:resolvent-convergence} in the sequel.

\begin{proof}[Proof of Proposition~\ref{norm:positive-powers}]
We first assume $aI\le Y_M\le bI$ and $aI\le Y\le bI$ with $a>0$.
For integer $p$, the result follows from
Lemma~\ref{norm:thermal-words}. Otherwise write $p=n+\theta$, where
$n\ge0$ is an integer and $0<\theta<1$. For $0\le k\le n+1$, by the above discussion and the same lemma we have
\begin{equation}\label{norm:moment-convergence}
 \mathcal F_M(x^k)\longrightarrow\mathcal F(x^k),
 \qquad 0\le\mathcal F_M(x^k),\mathcal F(x^k)\le b^k.
\end{equation}
The scalar identity
\[x^\theta=\frac{\sin(\pi\theta)}{\pi}
 \int_0^\infty t^{\theta-1}\frac{x}{x+t}\,\dd t,
 \qquad x\ge0,
\]
implies
\[
 x^p=\frac{\sin(\pi\theta)}{\pi}
 \int_0^\infty t^{\theta-1}\frac{x^{n+1}}{x+t}\,\dd t,
 \qquad x\ge0.
\]
Then, the positivity of the functional gives
\begin{equation}\label{norm:fractional-integral}
 \|Y_M\|_{p,\sigma_\mu^{\otimes mM}}^p
 =\frac{\sin(\pi\theta)}{\pi}
 \int_0^\infty t^{\theta-1}
 \mathcal F_M\!\left(\frac{x^{n+1}}{x+t}\right)\,\dd t,
\end{equation}
and similarly for $\|Y\|_{p,\tau_\mu^{\otimes m}}^p$.
For fixed $t>0$, we have
\[
 \frac{x^{n+1}}{x+t}
 =\sum_{j=0}^n(-t)^j x^{n-j}
   +\frac{(-t)^{n+1}}{x+t}.
\]
Thus by \eqref{norm:moment-convergence}
and Lemma~\ref{norm:resolvent-lemma}, we have $\mathcal F_M\!\left(\frac{x^{n+1}}{x+t}\right)\to \mathcal F\!\left(\frac{x^{n+1}}{x+t}\right)$. On the other hand, the positivity of the functional yields
\begin{equation}\label{norm:fractional-domination}
 0\le\mathcal F_M\!\left(\frac{x^{n+1}}{x+t}\right)
 \le\min\{\mathcal F_M(x^n),\mathcal F_M(x^{n+1})/t\}
 \le\min\{b^n,b^{n+1}/t\}.
\end{equation}
Multiplication by $t^{\theta-1}$ gives an integrable bound,
independent of $M$. Then, the dominated convergence theorem can be applied to conclude the desired convergence~\eqref{norm:positive-limit}. 

Finally, replace $Y_M,Y$ by $Y_M+\varepsilon I,Y+\varepsilon I$, and note that they remain jointly admissible. For $p\ge1$, the weighted-norm
triangle inequality bounds the change in each norm by $\varepsilon$.
For $0<p<1$, eigenvalue monotonicity and Rotfel\'d's inequality~\cite{bhatia_matrix_1997} give
\[
 0\le\|Y_M+\varepsilon I\|_{p,\sigma_\mu^{\otimes mM}}^p-\|Y_M\|_{p,\sigma_\mu^{\otimes mM}}^p
 \le\varepsilon^p.
\]
The estimates are uniform in $M$. Taking $M\to\infty$ and then
$\varepsilon\downarrow0$ removes the strict lower bounds $Y_M\geq aI$ and $Y\geq aI$.

\end{proof}

It remains to prove Lemma~\ref{norm:resolvent-lemma}. Write
\[
 \xi_M=\sigma_\mu^{\otimes mM},\qquad
 \xi=\tau_\mu^{\otimes m},
\] 
Denote left and right multiplications by
$\mathcal L_A(U)=AU$ and $\mathcal R_A(U)=UA$ and define modular operators
\[
 \Delta_M=\mathcal L_{\xi_M}\mathcal R_{\xi_M^{-1}},
 \qquad
 \Delta=\mathcal L_{\xi}\mathcal R_{\xi^{-1}}.
\]
These superoperators are understood spectrally: given the spectral decomposition $\xi =\sum_j r_j \ketbra{v_j}{v_j}$, we have 
$\Delta\ketbra{v_i}{v_j}
 =\frac{r_i}{r_j}\ketbra{v_i}{v_j}.$
In particular, $\Delta_M\xi_M^{1/2}=\xi_M^{1/2}$ and $ \Delta\xi^{1/2}=\xi^{1/2}$. We also note that $\Delta$ is a positive operator on the Hilbert--Schmidt space.

We record the following lemma as the functional-calculus consequence of joint
admissibility needed for the resolvent expansion below. The proof of this lemma is based on the Kaplansky
density theorem and is presented in Appendix~\ref{app:modular-FC}. 

\begin{lemma}[Bounded modular functional calculus]
\label{norm:bounded-modular-calculus}
Let $A_{j,M},A_j$ be jointly admissible bounded families.
Fix functions $f_0, \dots, f_k\in C_b(\mathbb R)$ and indices $j_1, \dots, j_k$. Then, 
\begin{align}
&\lim_{M\to \infty}\Big\langle \xi_M^{1/2}, 
f_0(\ln\Delta_M)\,
\mathcal L_{A_{j_1,M}}\,
f_1(\ln\Delta_M)\cdots
\mathcal L_{A_{j_k,M}}\,
f_k(\ln\Delta_M) \xi_M^{1/2}\Big\rangle_{\mathrm{HS}} \nonumber\\
& \qquad\qquad \qquad \qquad   = \Big\langle \xi^{1/2}, 
f_0(\ln\Delta)\,
\mathcal L_{A_{j_1}}\,
f_1(\ln\Delta)\cdots
\mathcal L_{A_{j_k}}\,
f_k(\ln\Delta) \xi^{1/2}\Big\rangle_{\mathrm{HS}}.\label{eq:modular-FC-limit}
\end{align}
\end{lemma}

\medskip

\begin{proof}[Proof of Lemma~\ref{norm:resolvent-lemma}]
We give the construction for $\xi$; the same construction applies to $\xi_M$.
Set $B=\xi^{1/(2p)}Y\xi^{1/(2p)}$ and define the positive self-adjoint operator
\[
 \mathcal K=\mathcal L_B\mathcal R_{\xi^{-1/p}}
 =\Delta^{1/(2p)}\mathcal L_Y\Delta^{1/(2p)}.
\]
Here the first expression is the columnwise direct sum described
below, and the second equality is understood in the sense of quadratic
forms. Using $\xi=\sum_j r_j\ketbra{v_j}{v_j}$, we have 
\begin{align}\label{eq:action-of-K}
\mathcal K(X)\ket{v_j} = \big(r_j^{-1/p}\big) B X\ket{v_j},
\end{align}
for any operator $X$ in the domain of $\mathcal K$. Define an operator \(Z\) by its action on basis vectors:
\begin{align}
\label{eq:def-Z-K}
Z\ket{v_j}
=r_j^{1/2}\bigl(r_j^{-1/p}B+tI\bigr)^{-1}\ket{v_j}.
\end{align}
Because \(B\ge0\), we have $\big\|\bigl(r_j^{-1/p}B+tI\bigr)^{-1}\big\|_{\infty}\leq 1/t$. Therefore
\[
\|Z\|_{\mathrm{HS}}^2
=\sum_j r_j
\left\|\bigl(r_j^{-1/p}B+tI\bigr)^{-1}\ket{v_j}\right\|^2
\le\frac1{t^2}\sum_jr_j
=\frac1{t^2}.
\]
Thus $Z$ is a well-defined Hilbert--Schmidt operator. Moreover,
the bound $\|C(C+tI)^{-1}\|_\infty\le1$ for $C\ge0$ gives
\[
 \sum_j\|r_j^{-1/p}BZ\ket{v_j}\|^2
 =\sum_j r_j\|r_j^{-1/p}B(r_j^{-1/p}B+tI)^{-1}\ket{v_j}\|^2
 \le\sum_jr_j=1.
\]
This is the columnwise domain condition for $\mathcal K$, so
$Z\in\operatorname{Dom}(\mathcal K)$. Moreover,~\eqref{eq:action-of-K} and~\eqref{eq:def-Z-K} give
\[
\bigl((\mathcal K+t\id)Z\bigr)\ket{v_j}
=(r_j^{-1/p} B+tI)Z\ket{v_j}=r_j^{1/2}\ket{v_j}=\xi^{1/2}\ket{v_j}.\]
It follows that $
Z=(\mathcal K+t\id)^{-1}\xi^{1/2}$. Equivalently, this means $(\mathcal K+t\id)^{-1}\xi^{1/2}\ket{v_j} =r_j^{1/2} (r_j^{-1/p}B+tI)^{-1}\ket {v_j}$.
As a result
\begin{align}\label{l11:resolvent-representation}
 \mathcal F\left(\frac1{x+t}\right)
 =\sum_j r_j\big\langle v_j,(r_j^{-1/p}B+tI)^{-1}v_j\big\rangle=\left\langle\xi^{1/2},
       (\mathcal K+t\id)^{-1}\xi^{1/2}
   \right\rangle_{\mathrm{HS}}.
\end{align}

To expand this resolvent on the right-hand side, we compare $\mathcal K$ with
$c\Delta^{1/p}$, where $c=(a+b)/2$. Put $\mathcal S=(c\Delta^{1/p}+t\id)^{-1}$. Then, as quadratic forms,
\begin{equation}\label{l11:factorization}
\begin{aligned}
 \mathcal K+t\id
 &=c\Delta^{1/p}+t\id
   +\Delta^{1/(2p)}\mathcal L_{(Y-cI)}\Delta^{1/(2p)}\\
 &=\mathcal S^{-1/2}
   \Bigl[\id+
     \mathcal S^{1/2}\Delta^{1/(2p)}
     \mathcal L_{(Y-cI)}
     \Delta^{1/(2p)}\mathcal S^{1/2}\Bigr]
   \mathcal S^{-1/2}.
\end{aligned}
\end{equation}
The apparently unbounded factors inside the brackets occur only
through the bounded product
\[
 \mathcal Q:=\Delta^{1/(2p)}\mathcal S^{1/2}
 =\left(\Delta^{1/p}\big(c\Delta^{1/p}+t\id\big)^{-1}\right)^{1/2},
 \qquad \|\mathcal Q\|\le c^{-1/2}.
\]
Consequently,
\begin{equation}\label{l11:perturbation-bound}
 \|\mathcal Q\mathcal L_{(Y-cI)}\mathcal Q\|
 \le\frac{\|Y-cI\|_\infty}{c}
 \le\frac{b-a}{b+a}=:\eta<1.
\end{equation}
Here and below, norms of superoperators are their operator norms
on the Hilbert--Schmidt space. Inverting
\eqref{l11:factorization} therefore gives the norm-convergent
Neumann series
\begin{align}\label{l11:neumann-series}
 (\mathcal K+t\id)^{-1}
 =\mathcal S^{1/2}
   \sum_{\ell=0}^{\infty}
   \big(-\mathcal Q\mathcal L_{(Y-cI)}\mathcal Q\big)^\ell
   \mathcal S^{1/2}.
\end{align}
Since $\Delta\xi^{1/2}=\xi^{1/2}$, we have
$\mathcal S^{1/2}\xi^{1/2}=(c+t)^{-1/2}\xi^{1/2}$.
Combining \eqref{l11:resolvent-representation} and
\eqref{l11:neumann-series} yields
\begin{equation}\label{l11:scalar-series}
 \mathcal F\!\left(\frac1{x+t}\right)
 =\frac1{c+t}\sum_{\ell=0}^{\infty}(-1)^\ell
 \left\langle\xi^{1/2},
   (\mathcal Q\mathcal L_{Y-cI}\mathcal Q)^\ell\xi^{1/2}
 \right\rangle_{\mathrm{HS}}.
\end{equation}
By \eqref{l11:perturbation-bound}, the absolute value of the tail for $\ell> L$ is at most $\frac{\eta^{L+1}}{(c+t)(1-\eta)}$. The same bound holds for $\mathcal F_M$ uniformly in $M$.

It remains to compare each finite coefficient. We have
\[
 \mathcal Q=f_t(\ln\Delta),\qquad
 f_t(x)=\left(\frac{e^{x/p}}{ce^{x/p}+t}\right)^{1/2}
 \in C_b(\mathbb R).
\]
The pair $(Y_M-cI, Y-cI)$ is jointly admissible. Lemma~\ref{norm:bounded-modular-calculus} therefore gives, for each fixed $\ell$,
convergence of the corresponding inner products in~\eqref{l11:scalar-series}, with the corresponding reference states $\xi_M$ and $\xi$. Combining coefficient convergence
with the uniform geometric tail proves
\[
 \mathcal F_M
       \!\left(\frac1{x+t}\right)
 \longrightarrow
 \mathcal F
       \!\left(\frac1{x+t}\right),
\]
as required.
\end{proof}

\subsection{Weighted norm convergence for general operators}
\label{subsec:norm-arbitrary-observables}

The positive trace-power limit of Proposition~\ref{norm:positive-powers} is the main step in the proof
of Theorem~\ref{norm:weighted-lifting}. There is nevertheless one more
issue to resolve. For an arbitrary observable $X$, its weighted norm
involves the absolute value of
$\bigl(\tau_\mu^{\otimes m}\bigr)^{1/(2p)}X\bigl(\tau_\mu^{\otimes m}\bigr)^{1/(2p)}$, and is not covered by this proposition. We handle this placement of the weights by first
smoothing $X$ along the modular group of the reference state. The smoothed
observable admits a bounded continuation to complex modular times,
which allows us to move a thermal power from one side to the other.

\begin{proof}[Proof of Theorem~\ref{norm:weighted-lifting}]
Let $C_X=\max\{\sup_M\|X_M\|_\infty , \|X\|_\infty\}$. Define
\[
 \mathcal U_M(t)(A)=\bigl(\sigma_\mu^{\otimes mM}\bigr)^{it}A\bigl(\sigma_\mu^{\otimes mM}\bigr)^{-it},\qquad
 \mathcal U(t)(A)=\bigl(\tau_\mu^{\otimes m}\bigr)^{it}A\bigl(\tau_\mu^{\otimes m}\bigr)^{-it},\qquad t\in\R.
\]
For $\delta>0$, define
\begin{equation}\label{norm:modular-smoothing}
  g_\delta(t)=\frac{e^{-t^2/\delta^2}}{\sqrt\pi\,\delta},\qquad
  X_{\delta,M}=\int_\R g_\delta(t)\mathcal U_M(t)(X_M)\,\dd t,
  \qquad
  X_\delta=\int_\R g_\delta(t)\mathcal U(t)(X)\,\dd t.
\end{equation}
The integrals are weak operator integrals. Since the real modular
translations are unitary conjugations and $\int g_\delta=1$, both
smoothed observables $X_{\delta, M}, X_\delta$ have operator norm at most $C_X$.

The Gaussian kernel gives more than boundedness at real times. For
$\zeta\in\C$, we can define the continuation to the complex plane by
\[
 \mathcal U_M(\zeta)(X_{\delta,M})
   =\int_\R g_\delta(t-\zeta)\mathcal U_M(t)(X_M)\,\dd t,
 \qquad
 \big\|\mathcal U_M(\zeta)(X_{\delta,M})\big\|_\infty
   \le C_X e^{(\operatorname{Im}\zeta)^2/\delta^2},
\]
with the bound being a consequence of $\int_\R|g_\delta(t-\zeta)|\,\dd t
=e^{(\operatorname{Im}\zeta)^2/\delta^2}$. Moreover,  differentiation
of the kernel on compact sets shows that the continuation is entire.
In the bosonic setting, the same integral defines the bounded continuation;
it crucially avoids interpreting an unbounded inverse thermal power on an
arbitrary observable. We note that Lemma~\ref{lem:admissible-closure} shows that these continuations, their
adjoints, and their products are jointly admissible.

We apply Proposition~\ref{norm:positive-powers} 
to the following observables:
\[
 \begin{aligned}
 Z_{\delta,M}&=\mathcal U_M(-i/(2p))(X_{\delta,M}),&
 Y_{\delta,M}&=Z_{\delta,M}^\dagger Z_{\delta,M},\\
 Z_\delta&=\mathcal U(-i/(2p))(X_\delta),&
 Y_\delta&=Z_\delta^\dagger Z_\delta.
 \end{aligned}
\]
We note that $Y_{\delta, M}, Y_\delta$ are positive, and all the operators are uniformly
bounded for fixed $\delta$ and $p$.
For the qubit operators we have $Z_{\delta,M}=\bigl(\sigma_\mu^{\otimes mM}\bigr)^{1/(2p)}X_{\delta,M}\bigl(\sigma_\mu^{\otimes mM}\bigr)^{-1/(2p)}$,
so direct multiplication gives
\begin{align}\label{norm:smoothing-identity}
 &\bigl(\sigma_\mu^{\otimes mM}\bigr)^{1/p}Y_{\delta,M}\bigl(\sigma_\mu^{\otimes mM}\bigr)^{1/p}\nonumber\\
  &\qquad =\Big(\bigl(\sigma_\mu^{\otimes mM}\bigr)^{1/(2p)}X_{\delta,M}\bigl(\sigma_\mu^{\otimes mM}\bigr)^{1/(2p)}\Big)^\dagger \Big(\bigl(\sigma_\mu^{\otimes mM}\bigr)^{1/(2p)}X_{\delta,M}\bigl(\sigma_\mu^{\otimes mM}\bigr)^{1/(2p)}\Big).
\end{align}
The same bosonic identity follows on the finite number-basis core and then
by bounded extension, using the analytic continuation just defined.
Consequently,
Proposition~\ref{norm:positive-powers}, applied to $s=p/2$, gives
\begin{equation}\label{norm:smoothed-limit}
 \big\|X_{\delta,M}\big\|_{p,{\sigma_\mu^{\otimes mM}}}^p
   =\big\|Y_{\delta,M}\big\|_{s,{\sigma_\mu^{\otimes mM}}}^s
   \longrightarrow
   \big\|Y_\delta\big\|_{s,{\tau_\mu^{\otimes m}}}^s
   =\big\|X_\delta\big\|_{p,{\tau_\mu^{\otimes m}}}^p.
\end{equation}
Observe that when $1\le p<2$, the auxiliary exponent $s=p/2$ is below one.
This is why Proposition~\ref{norm:positive-powers} was proved for every
positive exponent, including those that define only a quasi-norm.

It remains to remove the smoothing. Put
$E_{\delta,M}=X_M-X_{\delta,M}$ and $E_\delta=X-X_\delta$. 
Direct computation shows that 
\begin{align}
\|E_{\delta,M}\|_{2,\xi_M}^{\,2}
=\operatorname{tr}\!\left[
\xi_M^{1/2}E_{\delta,M}^\dagger
\xi_M^{1/2}E_{\delta,M}
\right],
\end{align}
and similarly for $E_\delta$. Then, since $(E_{\delta, M}, E_\delta)$ is jointly admissible as verified above, we have  
\begin{equation}\label{norm:smoothing-error-two}
 \lim_{M\to\infty}\|E_{\delta,M}\|_{2,{\sigma_\mu^{\otimes mM}}}
      =\|E_\delta\|_{2,{\tau_\mu^{\otimes m}}}.
\end{equation}
Moreover, $\|E_\delta\|_{2,{\tau_\mu^{\otimes m}}}\to0$ as $\delta\downarrow0$.
To see this, the map
$A\mapsto\bigl(\tau_\mu^{\otimes m}\bigr)^{1/4}A\bigl(\tau_\mu^{\otimes m}\bigr)^{1/4}$ identifies the weighted $2$-norm
with the Hilbert--Schmidt norm, and under this identification
\[
 \bigl(\tau_\mu^{\otimes m}\bigr)^{1/4}\mathcal U(t)(X)\bigl(\tau_\mu^{\otimes m}\bigr)^{1/4}
 =\bigl(\tau_\mu^{\otimes m}\bigr)^{it}\Big(\bigl(\tau_\mu^{\otimes m}\bigr)^{1/4}X\bigl(\tau_\mu^{\otimes m}\bigr)^{1/4}\Big)\bigl(\tau_\mu^{\otimes m}\bigr)^{-it}.
\]
These conjugations form a strongly continuous unitary group on
Hilbert--Schmidt space. Averaging against the approximate identity
$g_\delta$ in~\eqref{norm:modular-smoothing} gives the convergence of the Hilbert--Schmidt operator $\bigl(\tau_\mu^{\otimes m}\bigr)^{1/4}X_\delta\bigl(\tau_\mu^{\otimes m}\bigr)^{1/4}$ to $\bigl(\tau_\mu^{\otimes m}\bigr)^{1/4}X\bigl(\tau_\mu^{\otimes m}\bigr)^{1/4}$.

The $2$-norm estimate also controls the smoothing error at every
finite $p$. For $1\le p<2$ the monotonicity of weighted $p$-norms in $p$ gives $\|E_\delta\|_{p,\tau_\mu^{\otimes m}}\le\|E_\delta\|_{2,\tau_\mu^{\otimes m}}$. For $2<p<\infty$, since weighted $p$-norms form an interpolation family (see e.g.~\cite{Jencova2018}), we have 
\begin{equation}\label{norm:smoothing-interpolation}
 \|E_\delta\|_{p,\tau_\mu^{\otimes m}}
 \le\|E_\delta\|_\infty^{1-2/p}\|E_\delta\|_{2,\tau_\mu^{\otimes m}}^{2/p}.
\end{equation}
By the triangle inequality we have $\|E_{\delta,M}\|_\infty\le2C_X$, so these estimates and
\eqref{norm:smoothing-error-two} imply
\[
 \lim_{\delta\downarrow0}\limsup_{M\to\infty}
        \|E_{\delta,M}\|_{p,{\sigma_\mu^{\otimes mM}}}=0,
 \qquad
 \lim_{\delta\downarrow0}\|E_\delta\|_{p,{\tau_\mu^{\otimes m}}}=0.
\]
Finally, the triangle inequality and~\eqref{norm:smoothed-limit} imply~\eqref{norm:all-p-limit} as desired.
\end{proof}

\subsection{Proof of Theorem~\ref{sandwich:norm-transfer}}
\label{sec:weighted-transfer}

We now combine weighted norm convergence with the channel correspondence from Section~\ref{sec:preliminaries} to prove Theorem~\ref{sandwich:norm-transfer}. 

Throughout the following argument, $\Psi$ and $\Phi$ are the channel
pair in Theorem~\ref{sandwich:norm-transfer}.
The output references are $\tau_{\mu'}^{\otimes m}$ and
$\sigma_{\mu'}^{\otimes mM}$, and the input references are
$\tau_\mu^{\otimes m}$ and $\sigma_\mu^{\otimes mM}$.
Proposition~\ref{lem:adjoint-limit}, applied to each term of a finite
Weyl sum $B$, gives
\begin{equation}\label{eq:adjoint-limit}
 \left\|(\Psi^\dagger)^{\otimes mM}(B_{M,\mu'})
       -\bigl[(\Phi^\dagger)^{\otimes m}(B)\bigr]_{M,\mu}\right\|_\infty
 \longrightarrow0.
\end{equation}
The expression in brackets is again a finite Weyl sum, by
\eqref{eq:Gaussian-dual-Weyl}, and its lift uses the input reference.
Thus both terms being compared act on the same qubit Hilbert space
for each $M$. This is the only channel convergence needed in the
weighted norm argument.

\begin{proof}[Proof of Theorem~\ref{sandwich:norm-transfer}]
First let $B$ be a finite Weyl sum. Apply the assumed qubit inequality
\eqref{sandwich:qubit-norm-bound} to $B_{M,\mu'}$, on $mM$ qubits. Once again by the monotonicity of weighted $p$-norm in $p$, the error in \eqref{eq:adjoint-limit} tends to zero in
the weighted $q$-norm as well. Then, two uses of Theorem~\ref{norm:weighted-lifting} at the two reference parameters
give
 \begin{align}
 \bigl\|(\Phi^\dagger)^{\otimes m}(B)\bigr\|_{q,{\tau_\mu^{\otimes m}}}
 &=\lim_{M\to\infty}
    \bigl\|(\Psi^\dagger)^{\otimes mM}(B_{M,\mu'})\bigr\|_{q,{\sigma_\mu^{\otimes mM}}}
 \nonumber \\
 &\le\lim_{M\to\infty}\|B_{M,\mu'}\|_{p,{\sigma_{\mu'}^{\otimes mM}}}
 =\|B\|_{p,{\tau_{\mu'}^{\otimes m}}}.
 \label{eq:B-Weyl-sum-limit-thm5}
 \end{align}
This proves \eqref{sandwich:boson-norm-bound} on finite Weyl sums.
We note that this argument applies even to not necessarily self-adjoint
observables; it does not require positivity of $B$.

We next extend the inequality to all bounded observables $B$. The Weyl
representation is irreducible, so the algebra of finite
Weyl sums is strongly dense in the space of bounded operators. The bicommutant
theorem and the Kaplansky density theorem provide a sequence of finite Weyl sums
$B_k$ such that both convergences $B_k\to B$ and $B_k^\dagger\to B^\dagger$ hold strongly, and there is a uniform operator norm
bound $\|B_k\|_\infty\le C$ for every $k$~\cite{Hall-Book,Takesaki}.

For a fixed density operator $\xi$, the sequence $B_k$
converges to $B$ in every weighted norm of finite order
$r\ge1$. To see this, put $E_k=B_k-B$ and increase $C$, if necessary, so that
$\|E_k\|_\infty\le C$ for every $k$.
Let $P_N$ be increasing finite-rank spectral projections
of $\xi$ such that $\tr\big(\xi(I-P_N)\big)\to0$, and write $\xi_r = \xi^{1/(2r)}$ and $\xi_{r, N} = P_N \xi_r=\xi_r P_N$. Then $\|\xi_r\|_{2r}=1$ and $\|\xi_{r, N}\|_{2r}\le1$, and as $N \to \infty$
\[
 \|\xi_r-\xi_{r, N}\|_{2r}
 =\bigl(\tr(\xi(I-P_N))\bigr)^{1/(2r)}
 \longrightarrow0.
\]
Decomposing the difference as $\xi_r E_k\xi_r -\xi_{r, N} E_k\xi_{r,N}
 =(\xi_r -\xi_{r, N})E_k\xi_{r}+\xi_{r, N}E_k(\xi_{r}-\xi_{r, N})$, and using the H\"older inequality gives
\[
 \|\xi_{r}E_k\xi_{r}-\xi_{r, N}E_k\xi_{r, N}\|_r
 \le 2C\|\xi_{r}-\xi_{r, N}\|_{2r}.
\]
Thus the truncation error tends to zero uniformly in $k$. Next, for fixed $N$, strong convergence $E_k\to0$ implies
$\|E_kP_N\|_\infty\to0$, because the range of $P_N$
is finite dimensional. Consequently,
\[
 \|\xi_{r,N}E_k\xi_{r,N}\|_r
 \le (\operatorname{rank}P_N)^{1/r}
      \|\xi_{r,N}\|_\infty^2\|E_kP_N\|_\infty
 \longrightarrow0,
\]
as $k$ tends to infinity. Combining the two estimates, we obtain
\[
 \limsup_{k\to\infty}\|B_k-B\|_{r,\xi}
 =\limsup_{k\to\infty}\|\xi_{r}E_k\xi_{r}\|_r
 \le 2C\bigl(\tr(\xi(I-P_N))\bigr)^{1/(2r)}.
\]
Letting $N\to\infty$ proves
$\|B_k-B\|_{r,\xi}\to0$.

The same approximation can be passed through
$(\Phi^\dagger)^{\otimes m}$. Indeed, $(\Phi^\dagger)^{\otimes m}$ is a normal
unital completely positive map, so the Schwarz inequality gives, for every vector $\ket v$,
\[
 \|(\Phi^\dagger)^{\otimes m}(E_k)\ket v\|^2
 \le \big\langle v,\, (\Phi^\dagger)^{\otimes m}(E_k^\dagger E_k)v\big\rangle = \tr\left[\Phi^{\otimes m}(\ketbra vv) E_k^\dagger E_k\right].
\]
Since $E_k\to0$ strongly as $k$ tends to infinity and the operators $E_k$ are
uniformly bounded, $E_k^\dagger E_k\to0$ strongly as well. 
Therefore the right-hand side tends to zero, proving that
$(\Phi^\dagger)^{\otimes m}(E_k) \to 0$ strongly. We similarly have $(\Phi^\dagger)^{\otimes m}(E_k^\dagger) \to 0$. On the other hand, unital
complete positivity gives
$\|(\Phi^\dagger)^{\otimes m}(B_k)\|_\infty\le\|B_k\|_\infty$, so these images
remain uniformly bounded. Therefore, the weighted-norm continuity
established in the preceding paragraph also applies
to $(\Phi^\dagger)^{\otimes m}(B_k)$. 

Now we can apply~\eqref{eq:B-Weyl-sum-limit-thm5} to any $B_k$, and take the limit $k\to \infty$ of both sides  by the above discussion to prove  the desired inequality
for $B$.

Finally, the weighted $p$-space is the completion of bounded
observables. Therefore, the desired inequality extends to the whole weighted $p$-space. 
\end{proof}

\section{Sharp logarithmic Sobolev tensorization}
\label{sec:LS}

In this section we apply Theorem~\ref{sandwich:norm-transfer} to the quantum Ornstein--Uhlenbeck
semigroup and prove that its sharp one-mode $2$-logarithmic Sobolev
constant survives arbitrary tensor powers. In the following subsection
after defining the Dirichlet forms and recalling the one-mode constants,
we state our combined tensorization and hypercontractivity theorem. Its proof
starts from the improved qubit logarithmic Sobolev inequality for the depolarizing channel~\cite{Beigi2021} and 
then uses Theorem~\ref{sandwich:norm-transfer} to transfer the resulting hypercontractivity inequalities to the bosonic side.

\subsection{Logarithmic Sobolev inequalities and hypercontractivity}

For a faithful reference state $\xi$, we use the weighted
norms and inner product of \eqref{norm:weighted-norm} and
\eqref{eq:thermal-inner-product}. Recall the bosonic Ornstein--Uhlenbeck  generator $\mathcal L_\beta$
and the qubit generator
$\mathcal K_\beta$, defined in~\eqref{eq:thermal-rates} and~\eqref{eq:Kq}:
\begin{align}
 &\mathcal L_\beta=e^{-\beta/2}\mathfrak D[\ac^\dagger]
                         +e^{\beta/2}\mathfrak D[\ac],\label{eq:generators-both-1}\\
    & \mathcal K_\beta
 =\tanh(\beta/4)\left(
   e^{\beta/2}\mathfrak D[\bc]
  +e^{-\beta/2}\mathfrak D[\bc^\dagger]
  +\tfrac12\mathfrak D[Z]\right).\label{eq:generators-both-2}
\end{align}
The dual generators $\mathcal L_\beta^\dagger$ and
$\mathcal K_\beta^\dagger$ are symmetric with respect to the inner product
$\langle A,B\rangle_{2,\xi}
 =\tr(\xi^{1/2}A^\dagger\xi^{1/2}B)$, with $\xi$ equal to
their respective invariant states $\tau_{\mu}$ and $\sigma_\mu$ where $\mu=(e^{\beta}-1)^{-1}$. 

A logarithmic Sobolev inequality bounds relative entropy by a
Dirichlet form measuring dissipation. To state it in the coordinates
used later, let $\mathcal G$ be a Schr\"odinger generator with faithful
invariant state $\xi$. Using $\Gamma_\xi(X) = \xi^{1/2}X\xi^{1/2}$, for $p=2$, the Dirichlet form is defined as
\begin{equation}\label{eq:state-2-form}
 \mathcal E_{2,\mathcal G,\xi}(\rho)
 =-\tr\!\left[
   \Gamma_\xi^{-1/2}(\sqrt\rho)\,
   \mathcal G\!\left(\Gamma_\xi^{1/2}(\sqrt\rho)\right)\right].
\end{equation}
In finite dimensions all reference powers are
well defined. In the bosonic setting inverse thermal powers are
unbounded, so the expression is first defined on finite number-basis
matrices and then interpreted through its quadratic form in $\sqrt\rho$.
For the stationary bosonic rates $0<\nu_0<\nu_1$ with the corresponding $m$-mode generator $\widehat{\mathcal L}_{\nu_0,\nu_1}
 =\sum_{j=1}^m\big(\nu_0\mathfrak D[\ac^\dagger_j]+\nu_1\mathfrak D[\ac_j]\big)$, this form has the explicit expression~\cite{Beigi-RahimiKeshari2023}
\begin{equation}\label{eq:bosonic-state-2-form}
 \mathcal E_{2,\widehat{\mathcal L}_{\nu_0,\nu_1},
                         \tau_\mu^{\otimes m}}(\rho)
 =\frac{2c_\nu}{1-e^{-\beta}}\sum_{j=1}^m
  \Big\|\ac_j\sqrt\rho-e^{-\beta/2}\sqrt\rho\,\ac_j\Big\|_{\mathrm{HS}}^2.
\end{equation}
Here, as before $\beta=\ln(\nu_1/\nu_0)$ and $c_\nu = \frac{\nu_1-\nu_0}{2}$. 
The right side is finite for finite-energy states and gives the
quadratic-form interpretation directly. We use the closed Dirichlet-form
realization and generator-core framework for the quantum Ornstein--Uhlenbeck
semigroup developed in~\cite{CiprianiFagnolaLindsay2000}, together with its
weighted $L_p$ formulation in~\cite{CarboneSasso2008}. For finitely many modes,
the product semigroup has the sum of the local closed forms. Under the
thermal sandwich, this is precisely \eqref{eq:bosonic-state-2-form}; finite-energy
states lie in its form domain.

For $1<p<\infty$, we write $p'=p/(p-1)$ for the H\"older conjugate and use the state convention
of~\cite{Beigi-RahimiKeshari2023}:
\begin{equation}\label{eq:state-p-form}
 \mathcal E_{p,\mathcal G,\xi}(\rho)
 =-\frac{pp'}4\tr\!\left[
   \Gamma_\xi^{-1/p'}(\rho^{1/p'})\,
   \mathcal G\!\left(\Gamma_\xi^{1/p'}(\rho^{1/p})\right)\right].
\end{equation}
At the entropy endpoint, $p=1$, Dirichlet form is
\begin{equation}\label{eq:state-1-form}
 \mathcal E_{1,\mathcal G,\xi}(\rho)
 =-\tfrac14\tr\big[(\ln\rho-\ln\xi)\mathcal G(\rho)\big].
\end{equation}
On a domain where differentiation is valid, this quantity is related to
the decay rate of 
$D(e^{t\mathcal G}(\rho)\Vert\xi)$ at $t=0$.

The $p$-logarithmic Sobolev inequality and its best constant are
\begin{equation}\label{eq:state-2lsi}
 \mathcal E_{p,\mathcal G,\xi}(\rho)
 \ge\alpha_p(\mathcal G)D(\rho\Vert\xi),\qquad
 \alpha_p(\mathcal G)
 =\inf_{\rho\ne\xi}
   \frac{\mathcal E_{p,\mathcal G,\xi}(\rho)}{D(\rho\Vert\xi)},
\end{equation}
where $D(\rho\Vert\xi)=\tr[\rho(\ln\rho-\ln\xi)]$ is the quantum relative entropy. 
For the qubit generators the infimum is over all states; for the
stationary bosonic generators it is over all finite-energy states,
using the preceding interpretation of the forms.
We note that the invariant state $\xi$ is understood in the notation $\alpha_p(\mathcal G)$.

\begin{theorem}[Sharp one-mode constants~\cite{Beigi-RahimiKeshari2023}]
\label{thm:one-mode}
For $0<\nu_0<\nu_1$, let $\beta=\ln(\nu_1/\nu_0)$ and
$c_\nu=(\nu_1-\nu_0)/2$. The sharp
one-mode constants for $1<p\le2$ are
\begin{equation}\label{eq:one-mode-p-constant}
 \alpha_p(\mathcal L_{\nu_0,\nu_1})
 =\frac{c_\nu}{\sinh(\beta/2)}
   \frac{pp'}{4\beta}e^{\beta/2}
      (1-e^{-\beta/p})(1-e^{-\beta/p'}).
\end{equation}
In particular,\footnote{The value of $\alpha_1(\mathcal L_{\nu_0, \nu_1})$ was first evaluated in~\cite{CarlenMass2}.}
\begin{equation}\label{eq:one-mode-endpoints}
 \alpha_2(\mathcal L_{\nu_0,\nu_1})
   =\frac{2c_\nu}{\beta}\tanh(\beta/4),
 \qquad \alpha_1(\mathcal L_{\nu_0,\nu_1})=\frac{c_\nu}{2}.
\end{equation}
\end{theorem}

For the normalized generator $\mathcal L_\beta$ corresponding to $\nu_0=e^{-\beta/2}$ and $\nu_1=e^{\beta/2}$, the $2$-logarithmic Sobolev constant from
\eqref{eq:one-mode-endpoints} reads $\alpha_\beta=4\sinh^2(\beta/4)/\beta$.
Our tensorization theorem stated in the following subsection asks whether this sharp one-mode value
remains valid for the generator $\widehat{\mathcal L}_\beta$ on arbitrarily many
modes.

A hypercontractivity inequality for the generator $\mathcal G$ with invariant state $\xi$ takes the form
\[
 \big\|e^{t\mathcal G^\dagger}(A)\big\|_{q,\xi}\le\|A\|_{p,\xi},
 \qquad 1<p\le q<\infty,
 \qquad t\ge\frac1{4\alpha}\ln\frac{q-1}{p-1}.
\]
For a positive semigroup that is symmetric on weighted $L_2$ and
contractive on weighted $L_p$, $1\le p\le\infty$, this estimate implies the
$2$-logarithmic Sobolev inequality with constant $\alpha$; see
\cite[Theorem~3.8]{OZ99} and its formulation for quantum
Ornstein--Uhlenbeck semigroups in~\cite[Theorem~4.1(a)]{CarboneSasso2008}.
This direction requires no additional $L_p$-regularity assumption on the
Dirichlet form. The converse holds under the appropriate reversibility
and regularity assumptions; for the strongly reversible qubit semigroups
below, we use~\cite[Corollary~17]{BeigiDattaRouze}.

\subsection{Sharp tensorization and hypercontractivity}

The one-mode logarithmic Sobolev constants bound the multimode one by considering tensor product input states $\rho$. Our main result in this section proves the matching lower bound
for arbitrary states $\rho$.

\begin{theorem}[Sharp $2$-logarithmic Sobolev tensorization and hypercontractivity]
\label{thm:2-tensor}
For $\beta>0$ recall that $\mu=(e^\beta-1)^{-1}$ and $c_\beta = \sinh(\beta/2)$, and set $\alpha_\beta=4\sinh^2(\beta/4)/\beta$. Then, 
for every finite number of modes $m$, we have
\begin{equation}\label{eq:2-tensor}
 \alpha_2(\widehat{\mathcal L}_{\beta})
 =\alpha_2(\mathcal L_{\beta})
 =\frac{2c_\beta}{\beta}\tanh(\beta/4).
\end{equation}
In particular, for every finite-energy $m$-mode bosonic state $\rho$,
\begin{equation}\label{eq:unweighted-2lsi}
 \mathcal E_{2,\widehat{\mathcal L}_{\beta},
                         \tau_\mu^{\otimes m}}(\rho)
 \ge \frac{2c_\beta}{\beta}\tanh(\beta/4)
                  D(\rho\Vert\tau_\mu^{\otimes m}).
\end{equation}
This inequality is tight for every nontrivial displacement of the
product thermal reference $\tau_\mu^{\otimes m}$. Moreover, 
for the semigroup $\Phi_t=e^{t\mathcal L_\beta}$, if $1<p\le q<\infty$ and
\begin{equation}\label{eq:hypercontractive-time}
 t\ge\frac1{4\alpha_\beta}\ln\frac{q-1}{p-1},
\end{equation}
then, for every bounded $m$-mode observable $B$,
\begin{equation}\label{eq:bosonic-hypercontractivity}
 \| (\Phi_t^\dagger)^{\otimes m}(B)\|_{q,{\tau_\mu^{\otimes m}}}
 \le\|B\|_{p,{\tau_\mu^{\otimes m}}}.
\end{equation}
\end{theorem}

The proof begins with a tensorized qubit logarithmic Sobolev inequality,
uses strong reversibility to obtain qubit hypercontractivity, and transfers the latter to the bosonic semigroup by Theorem~\ref{sandwich:norm-transfer}.
The resulting hypercontractivity inequalities then yield \eqref{eq:unweighted-2lsi}. Displaced thermal
states establish sharpness.

\subsection{Sharp logarithmic Sobolev inequality for qubits}

Our starting point is an improved logarithmic Sobolev inequality for the qubit depolarizing channels. These channels also form a semigroup with the corresponding generator $\mathcal J(B)=\tfrac12\tr(B)I-B$. We denote the corresponding $n$-qubit generator by $\widehat{\mathcal J}=\sum_{k=1}^n\mathcal J^{(k)}$ where $\mathcal J^{(k)}$ acts on the $k$th qubit.
The invariant state of this semigroup is $I/2^n$ for which the corresponding Dirichlet form~\eqref{eq:state-2-form} is simplified to
\begin{align}\label{eq:Dirichlet-dep-exp}
 \mathcal E_{2,\mathrm{dep}}^n(\omega)
 := \mathcal E_{2,\widehat{\mathcal J},I/2^n}(\omega)= -\tr\bigl[\sqrt\omega\,
                    \widehat{\mathcal J}(\sqrt\omega)\bigr].
\end{align}
To describe the improved inequality, let
$h(x)=-x\ln x-(1-x)\ln(1-x)$ be the binary entropy and define
$f:[0,\ln2]\to[0,1/2]$ by
\[
 f(h(x))=\tfrac12-\sqrt{x(1-x)},\qquad 0\le x\le\tfrac12.
\]
This function is well defined because $h$ is strictly increasing on this
interval.

\begin{theorem} \cite[Theorem 2]{Beigi2021}
\label{thm:qubit-2}
For every $n$-qubit state $\omega$,
\begin{equation}\label{eq:qubit-improved-2}
 \mathcal E_{2,\mathrm{dep}}^n(\omega)
 \ge n f\bigl(S(\omega)/n\bigr),
\end{equation}
where as before $S(\omega)$ is the von Neumann entropy.
\end{theorem}

The function $f$ is defined in such a way that the inequality is tight for one-qubit states. The main point of the theorem is that the inequality survives when passing to many qubits. A main point in its proof in~\cite{Beigi2021} is that the function $f$ is convex. To see this explicitly,
put $t=\frac12\ln((1-x)/x)$ for $0<x<1/2$. Differentiation gives
$f'(h(x))=-\sinh(t)/(2t)$. Since $\sinh(t)/t$ is increasing for $t>0$
and $h(x)$ increases as $t$ decreases, this derivative is increasing
as a function of entropy, implying that $f$ is convex. 

Convexity of $f$ means that  any supporting line of $f$ remains below its curve. Moreover, using~\eqref{eq:qubit-improved-2}, any supporting line is a lower bound on $\mathcal E_{2, \mathrm{dep}}^n(\omega)$. We compute this supporting line at our reference qubit state $\sigma_\mu$, corresponding to $\mu=(e^\beta-1)^{-1}$. The values of $f$ and its derivative are
\begin{align}\label{eq:def-e-beta-S-mu}
 e_\beta:=f(S(\sigma_\mu))
 =\tfrac12-\frac1{2\cosh(\beta/2)},\qquad
 f'(S(\sigma_\mu))=-\frac{\sinh(\beta/2)}{\beta}.
\end{align}
Consequently, we have
\begin{equation}\label{eq:qubit-tangent-2}
 \mathcal E_{2,\mathrm{dep}}^n(\omega)-ne_\beta
 +\frac{\sinh(\beta/2)}{\beta}
                       [S(\omega)-nS(\sigma_\mu)]\ge0.
\end{equation}
The following calculation shows that this supporting line provides a logarithmic Sobolev inequality for our qubit generator $\widehat{\mathcal K}_\beta$.

\begin{proposition} \label{prop:thermal-qubit}
For every $n$-qubit state $\omega$,
\begin{equation}\label{eq:thermal-qubit}
 \mathcal E_{2,\widehat{\mathcal K}_\beta,\sigma_\mu^{\otimes n}}(\omega)
 \ge\alpha_\beta D(\omega\Vert\sigma_\mu^{\otimes n}),
 \qquad
 \alpha_\beta=\frac{4\sinh^2(\beta/4)}{\beta}.
\end{equation}
The coefficient is optimal for every $n$.
\end{proposition}

\begin{proof}
We show that~\eqref{eq:thermal-qubit} is equivalent to~\eqref{eq:qubit-tangent-2}. Define
\[
 V=\tfrac12I-\tfrac12\bigl(
 e^{-\beta/2}\ketbra00+e^{\beta/2}\ketbra11\bigr)=\frac{1-\cosh(\beta/2)}2I
   +\frac{\sinh(\beta/2)}2Z,
 \qquad \widehat V=\sum_{k=1}^n V^{(k)},
\]
where $V^{(k)}$ acts on the $k$th qubit. 
Substituting the generator~\eqref{eq:generators-both-2}, direct computation shows that for every one-qubit state $\omega$, we have
\begin{equation}\label{eq:thermal-qubit-generator-conversion}
 \Gamma_{\sigma_\mu}^{-1/2}
 \!\left(\mathcal K_\beta
       \!\left(\Gamma_{\sigma_\mu}^{1/2}(\sqrt\omega)\right)\right)
 =2\tanh(\beta/4)
   \left(\mathcal J(\sqrt\omega)+\tfrac12(V\sqrt\omega+\sqrt\omega V)\right).
\end{equation}
Since square roots of states span the one-qubit operator space, this
is an identity of superoperators. Tensoring it with the identity on
the other qubits and summing over sites gives, for every $n$-qubit state $\omega$,
\[\Gamma_{\sigma_\mu^{\otimes n}}^{-1/2}
 \!\left(\widehat{\mathcal K}_\beta
       \!\left(\Gamma_{\sigma_\mu^{\otimes n}}^{1/2}(\sqrt\omega)\right)\right)
 =2\tanh(\beta/4)
   \left(\widehat{\mathcal J}(\sqrt\omega)+\tfrac12(\widehat V \sqrt\omega+\sqrt\omega\widehat V)\right).
\]
Multiplying by $-\sqrt\omega$, taking the
trace and comparing to~\eqref{eq:Dirichlet-dep-exp} proves
\begin{equation}\label{eq:thermal-qubit-form-conversion}
 \mathcal E_{2,\widehat{\mathcal K}_\beta,\sigma_\mu^{\otimes n}}(\omega)
 =2\tanh(\beta/4)\bigl[
  \mathcal E_{2,\mathrm{dep}}^n(\omega)-\tr(\omega \widehat  V)\bigr].
\end{equation}
Here we use cyclicity to write
$\tr[\sqrt\omega(\widehat V\sqrt\omega+\sqrt\omega \widehat V)]/2
=\tr(\omega \widehat V)$. 

Next, using $\sigma_\mu=\frac{e^{(\beta/2)Z}}{2\cosh(\beta/2)}$ we find that 
\[
 \ln\sigma_\mu
 =\frac{\beta}{2}Z-\ln\!\left(2\cosh(\beta/2)\right)I,
 \qquad
 S(\sigma_\mu)
 =\ln\!\left(2\cosh(\beta/2)\right)
  -\frac{\beta}{2}\tanh(\beta/2).
\]
Consequently,
\[
 \ln\sigma_\mu+S(\sigma_\mu)I
 =\frac{\beta}{2}\bigl(Z-\tanh(\beta/2)I\bigr).
\]
Then, using the definition of $e_\beta$ in~\eqref{eq:def-e-beta-S-mu} we obtain $V=e_\beta I+
 \frac{\sinh(\beta/2)}{\beta}
 \bigl(\ln\sigma_\mu+S(\sigma_\mu)I\bigr)$ and 
\[
 \widehat V=ne_\beta I+
 \frac{\sinh(\beta/2)}{\beta}
 \bigl(\ln(\sigma_\mu^{\otimes n})
       +nS(\sigma_\mu)I\bigr).
\]
Since
$\tr(\omega\ln(\sigma_\mu^{\otimes n}))
=-D(\omega\Vert\sigma_\mu^{\otimes n})-S(\omega)$,
it follows that
\[
 \tr(\omega \widehat V)
 =ne_\beta-
 \frac{\sinh(\beta/2)}{\beta}
 \big(D(\omega\Vert\sigma_\mu^{\otimes n})
       +S(\omega)-nS(\sigma_\mu)\big).
\]
Substituting into \eqref{eq:thermal-qubit-form-conversion}
and using
$\alpha_\beta
=2\tanh(\beta/4)\sinh(\beta/2)/\beta$, we conclude that
\[
\begin{aligned}
 &\mathcal E_{2,\widehat{\mathcal K}_\beta,
                    \sigma_\mu^{\otimes n}}(\omega)
 -\alpha_\beta D(\omega\Vert\sigma_\mu^{\otimes n})\\
 &\quad=
 2\tanh(\beta/4)\left[
 \mathcal E_{2,\mathrm{dep}}^n(\omega)-ne_\beta
 +\frac{\sinh(\beta/2)}{\beta}
       \bigl(S(\omega)-nS(\sigma_\mu)\bigr)
 \right]
 \ge0,
\end{aligned}
\]
where the final inequality is
\eqref{eq:qubit-tangent-2}.

For sharpness, take a one-qubit state $\omega=U\sigma_\mu U^\dagger$
that differs from $\sigma_\mu$. Its entropy is $S(\sigma_\mu)$, and
$\mathcal E_{2,\mathrm{dep}}^1(\omega)=e_\beta$ because this form
is invariant under unitary conjugation. Thus the supporting-line gap
vanishes, and equality holds in \eqref{eq:thermal-qubit} with a
nonzero relative entropy. Tensoring this state with $n-1$ unchanged
references proves optimality for every $n$.
\end{proof}

\subsection{Proof of Theorem~\ref{thm:2-tensor}}

A logarithmic Sobolev inequality involves 
the Dirichlet form that  describes the infinitesimal generator of a semigroup.
Theorem~\ref{sandwich:norm-transfer}, in contrast, applies to a member of the semigroup
at a fixed time. Hypercontractivity connects these two formulations.

\begin{proof}[Proof of Theorem~\ref{thm:2-tensor}]

We use the equivalence between logarithmic Sobolev inequalities
and hypercontractivity recalled earlier in this section.
The qubit semigroup $\Psi_t=e^{t\mathcal K_\beta}$ is primitive.
Its dual generator is symmetric for the thermal inner product and
commutes with the modular action of $\sigma_\mu$, as follows from
\eqref{eq:Kq-eigen}. It is therefore strongly reversible with respect
to $\sigma_\mu$. These properties persist under tensor powers, so
the equivalence applies.
Consequently, Proposition~\ref{prop:thermal-qubit} and~\cite[Corollary~17]{BeigiDattaRouze} give
the estimate
\begin{equation}\label{eq:qubit-hypercontractivity}
 \|(\Psi_t^\dagger)^{\otimes n}(B)\|_{q,\sigma_\mu^{\otimes n}}
 \le \|B\|_{p,\sigma_\mu^{\otimes n}}
\end{equation}
for positive observables $B$ and every $n$, at the time specified in
\eqref{eq:hypercontractive-time}. 

We note that the positivity of $B$ in~\cite[Corollary~17]{BeigiDattaRouze} is not a restriction since by~\cite[Theorem~1]{Watrous2005SchattenNorms},
the Schatten $p\to q$ norm of a completely positive map
is attained on positive semidefinite inputs.
Applying this result to the completely positive map $\Gamma_{\sigma_\mu^{\otimes n}}^{1/q}
 \circ(\Psi_t^\dagger)^{\otimes n}
 \circ\Gamma_{\sigma_\mu^{\otimes n}}^{-1/p}$ extends the weighted norm inequality to arbitrary $B$.

For each fixed $t$, the qubit and bosonic channels
satisfy the correspondence \eqref{eq:qubit-dual-Gaussian-limit}, with
their respective invariant thermal references. By Theorem~\ref{sandwich:norm-transfer}, inequality~\eqref{eq:qubit-hypercontractivity} implies~\eqref{eq:bosonic-hypercontractivity}.

We next obtain the logarithmic Sobolev inequality from the bosonic
hypercontractivity bound. The bosonic product semigroup is positive,
symmetric on $L_2(\tau_\mu^{\otimes m})$, and contractive on the associated
weighted $L_p$ spaces. We may therefore apply the implication of
\cite[Theorem~3.8]{OZ99}, in the form stated in
\cite[Theorem~4.1(a)]{CarboneSasso2008}, to
\eqref{eq:bosonic-hypercontractivity} with $p=2$ and
$q(t)=1+e^{4\alpha_\beta t}$.
For a positive $B$ normalized by $\|B\|_{2,\tau_\mu^{\otimes m}}=1$,
the corresponding entropy functional is
$\tfrac12D(\rho_B\Vert\tau_\mu^{\otimes m})$, where
$\rho_B=((\tau_\mu^{\otimes m})^{1/4}B
(\tau_\mu^{\otimes m})^{1/4})^2$.
Using the closed Dirichlet-form realization recalled above from
\cite{CiprianiFagnolaLindsay2000,CarboneSasso2008}, the resulting
inequality on the form domain is therefore
\[
 \mathcal E_{2,\widehat{\mathcal L}_\beta,\tau_\mu^{\otimes m}}(\rho)
 \ge\alpha_\beta D(\rho\Vert\tau_\mu^{\otimes m}).
\]
In particular, it holds for every finite-energy state, with the
Dirichlet form given by \eqref{eq:bosonic-state-2-form}.

To prove optimality, displace the reference:
$\rho_z=W(z){\tau_\mu^{\otimes m}} W(z)^\dagger$, where $z\in\C^m\setminus\{0\}$.
Unitary conjugation preserves entropy, while displacement raises the
total mean photon number by $|z|^2$. Thus
\[
 D(\rho_z\Vert{\tau_\mu^{\otimes m}})=\beta|z|^2.
\]
The thermal relation
$\ac_j\sqrt{\tau_\mu^{\otimes m}}=e^{-\beta/2}\sqrt{\tau_\mu^{\otimes m}}\,\ac_j$ and Weyl
displacement covariance give
\[
 \ac_j\sqrt{\rho_z}-e^{-\beta/2}\sqrt{\rho_z}\,\ac_j
 =(1-e^{-\beta/2})z_j\sqrt{\rho_z}.
\]
Inserting this into \eqref{eq:bosonic-state-2-form} shows
\[
 \mathcal E_{2,\widehat{\mathcal L}_{\beta},{\tau_\mu^{\otimes m}}}(\rho_z)
 =\frac{2c_\beta(1-e^{-\beta/2})^2}{1-e^{-\beta}}|z|^2
 =2c_\beta\tanh(\beta/4)|z|^2.
\]
The ratio to relative entropy is exactly the coefficient in
\eqref{eq:2-tensor}, in every number of modes. This proves sharpness.
\end{proof}

Our argument also yields the sharp $p=2$ meta logarithmic Sobolev
inequality of~\cite{Beigi-RahimiKeshari2023} in the multimode case.

\section{State lifting and constrained minimum output entropy}
\label{sec:entropy-transfer}

This section transfers relative entropy inequalities from qubit systems to bosonic systems by constructing positive qubit states whose relative entropies converge to their bosonic counterparts. We first state the main theorem, the qubit relative entropy contraction bound needed for the applications, and the resulting bosonic
CMOE theorems. We then prove this theorem using qubit states constructed through Weyl lifting and derive the bosonic applications. The proofs of the qubit bounds
are deferred to Section~\ref{sec:qubit-results}.

\subsection{Qubit contraction and bosonic CMOE results}

For a faithful state $\xi$, the relative entropy is
\[
 D(\rho\Vert\xi)=\tr\rho(\ln\rho-\ln\xi),
\]
with its usual extended-value interpretation. For the bosonic
thermal reference it takes the particularly useful form
\begin{equation}\label{eq:thermal-relative entropy}
 D(\rho\Vert\tau_\mu^{\otimes m})
 =\beta\tr(\rho\Ntot)-S(\rho)-m\ln(1-e^{-\beta}).
\end{equation}
Thus it is finite for finite-energy states. This identity relates
entropy, energy, and the thermal reference, and is the reason that
relative entropy estimates can be used in the CMOE application.

Fix the same admissible channel pair and faithful references as in
Theorem~\ref{sandwich:norm-transfer}: $\Psi(\sigma_\mu)=\sigma_{\mu'}$
and $\Phi(\tau_\mu)=\tau_{\mu'}$, with the parameters related by
\eqref{eq:general-map}. Our result for transferring relative entropy inequalities to bosonic systems is as follows.

\begin{theorem}[Transferring relative entropy inequalities to bosonic systems]
\label{prop:transfer}
Let $c>0$ and suppose that every $n$-qubit state $\omega$, for every
$n\ge1$, satisfies
\begin{equation}\label{eq:abstract-qubitD}
 D(\Psi^{\otimes n}(\omega)\Vert\sigma_{\mu'}^{\otimes n})
 \le cD(\omega\Vert\sigma_\mu^{\otimes n}).
\end{equation}
Then for every $m\ge1$ and every $m$-mode bosonic state $\rho$,
\begin{equation}\label{eq:abstract-bosonD}
 D(\Phi^{\otimes m}(\rho)\Vert\tau_{\mu'}^{\otimes m})
 \le cD(\rho\Vert\tau_\mu^{\otimes m}).
\end{equation}
\end{theorem}

To apply Theorem~\ref{prop:transfer}, we need to establish 
relative entropy contraction bounds for some qubit channels.
To this end, we introduce the family of qubit channels $\Psi_{s, r}$.
For $0\le s\le r\le1$, let
$\Psi_{s,r}$ be the qubit channel defined by its action on the Bloch ball: 
\begin{equation}\label{eq:weighted-bloch}
(x,y,z)\longmapsto(srx,sry,s^2z).
\end{equation}
We use the same references as in Subsection~\ref{sec:qubit-semigroups}
and Section~\ref{sec:norm-lifting}. Fix $\mu>0$ and let
$\beta=\beta(\mu)$. Then
$\sigma_\mu=(I+\kappa_\mu Z)/2$, where
$\kappa_\mu=\tanh(\beta/2)$.
For \(0<s\le r\le1\), define \(\mu'\) and \(\beta'\) by
\begin{equation}\label{eq:pinching-reference}
 \Psi_{s,r}(\sigma_\mu)=\sigma_{\mu'},\qquad
 \tanh(\beta'/2)=s^2\tanh(\beta/2),\qquad
 \mu'=\frac{\mu+1/2}{s^2}-\frac12,
 \qquad\beta'=\beta(\mu').
\end{equation}

\begin{theorem}[Sharp tensorized relative entropy contraction]
\label{thm:weighted-D}
For every \(n\)-qubit state \(\omega\),
\begin{equation}\label{eq:weighted-D}
 D\big(\Psi_{s,r}^{\otimes n}(\omega)\Vert\sigma_{\mu'}^{\otimes n}\big)
 \le r^2\frac{\beta'}{\beta}
       D\big(\omega\Vert\sigma_\mu^{\otimes n}\big).
\end{equation}
The coefficient is optimal for every \(n\). The corresponding
qubit inequality for the maximally mixed reference $\sigma=I/2$
has coefficient $s^2r^2$, obtained continuously as
$\beta\downarrow0$.
For \(s=r=\sqrt{\lambda}\) this inequality gives
\begin{equation}\label{eq:depD}
 D\big(\mathcal D_\lambda^{\otimes n}(\omega)\Vert\sigma_{\mu'}^{\otimes n}\big)
 \le \lambda\frac{\beta'}{\beta}
       D\big(\omega\Vert\sigma_\mu^{\otimes n}\big),
 \qquad \tanh(\beta'/2)=\lambda\tanh(\beta/2).
\end{equation}
\end{theorem}

The proof of this theorem is deferred to Section~\ref{sec:qubit-results}. Combining this qubit inequality with Theorem~\ref{prop:transfer} gives the following two results for bosonic channels. Recall that the additive noise channel $\Phi^{\mathrm{add}}_\gamma$
adds $\gamma/2$ photons per mode.

\begin{theorem}[CMOE for multimode additive-noise channels]\label{thm:heat}
Let $\mu>0$ and $\gamma\ge0$. Every finite-energy $m$-mode state
$\rho$ satisfies
\begin{align}
&D\bigl((\Phi^{\mathrm{add}}_\gamma)^{\otimes m}(\rho)
         \big\Vert\tau_{\mu+\gamma/2}^{\otimes m}\bigr)
 \le\frac{\beta(\mu+\gamma/2)}{\beta(\mu)}
           D(\rho\Vert\tau_\mu^{\otimes m}),
 \label{eq:heat-D}\\
 &S\bigl((\Phi^{\mathrm{add}}_\gamma)^{\otimes m}(\rho)\bigr)
 \ge mg(\mu+\gamma/2)
    +\frac{\beta(\mu+\gamma/2)}{\beta(\mu)}
          [S(\rho)-mg(\mu)].
 \label{eq:heat-affine}
\end{align}
Consequently, we have
\begin{equation}\label{eq:heat-CMOE}
 S\bigl((\Phi^{\mathrm{add}}_\gamma)^{\otimes m}(\rho)\bigr)
 \ge mg\big(g^{-1}(S(\rho)/m)+\gamma/2\big).
\end{equation}
This includes zero input entropy, and the product thermal input
$\tau_{\mu_\rho}^{\otimes m}$, where $\mu_\rho=g^{-1}(S(\rho)/m)$, attains the bound.
\end{theorem}

Our next result concerns $\Phi^{\mathrm{att}}_{\lambda,N}$ which is the attenuator channel with transmissivity $\lambda$ and a
thermal environment of mean photon number $N$. It is obtained by interaction through a beam-splitter unitary when the environment is in a thermal state.

\begin{theorem}[CMOE for multimode attenuator in the heating range]
\label{thm:att-enlarged}
Let $0<\lambda<1$, $N>0$, and choose a faithful reference mean
$0<\mu\le N$. Put $\mu'=\lambda\mu+(1-\lambda)N$. Then every
finite-energy $m$-mode state satisfies
\begin{align}
 &D\bigl((\Phi^{\mathrm{att}}_{\lambda,N})^{\otimes m}(\rho)
        \big\Vert\tau_{\mu'}^{\otimes m}\bigr)
 \le\lambda\frac{\beta(\mu')}{\beta(\mu)}
               D(\rho\Vert\tau_\mu^{\otimes m}),
 \label{eq:att-enlarged-D}\\
 &S\bigl((\Phi^{\mathrm{att}}_{\lambda,N})^{\otimes m}(\rho)\bigr)
 \ge mg(\mu')
    +\lambda\frac{\beta(\mu')}{\beta(\mu)}[S(\rho)-mg(\mu)].
 \label{eq:att-enlarged-affine}
\end{align}
In particular, for $0\le\lambda\le1$ and $N\ge0$, CMOE holds
whenever
\begin{equation}\label{eq:att-enlarged-range}
 S(\rho)\le mg(N).
\end{equation}
That is, under the above condition we have
\begin{equation}\label{eq:att-enlarged-CMOE}
 S\bigl((\Phi^{\mathrm{att}}_{\lambda,N})^{\otimes m}(\rho)\bigr)
 \ge mg\big(\lambda g^{-1}(S(\rho)/m)+(1-\lambda)N\big).
\end{equation}
The product thermal input $\tau_{\mu_\rho}^{\otimes m}$, with
$\mu_\rho=g^{-1}(S(\rho)/m)$, attains this bound.
\end{theorem}

Hirche, Rouz\'e, and Stilck Fran\c{c}a~\cite[Theorem~8.8]{Hirche2022contraction}
identified the exact thermal-reference relative entropy contraction
coefficients of tensor powers of Gaussian channels under the
corresponding multimode CMOE conjecture. For product references with the same thermal mean in every mode,
their additive-noise and attenuation coefficients are
exactly those in \eqref{eq:heat-D} and \eqref{eq:att-enlarged-D}.
Theorems~\ref{thm:heat} and~\ref{thm:att-enlarged} establish these
bounds without assuming CMOE.

\subsection{Convergence of the input relative entropy}
\label{subsec:sandwich-states}

We now develop the state-lifting argument that proves
Theorem~\ref{prop:transfer}. Its input entropy limit and output
lower bound are the two ingredients required to transfer the qubit contraction inequality to bosonic systems.
The elementary linear lift of a Weyl sum need not preserve positivity.
We therefore lift a positive expression of the form
$\varepsilon I+F^\dagger F$, sandwich it by the square root of the
thermal reference, and normalize its trace. The resulting states
supply the input approximations for Theorem~\ref{prop:transfer}.

Let $F=\sum_{\ell=1}^L c_\ell W(z_\ell)$ be a finite Weyl sum, and
write $F_M=\sum_{\ell=1}^L c_\ell W_{M,\mu}(z_\ell)$ for its lift.
For $\varepsilon>0$, form the positive expressions
\begin{equation}\label{sandwich:positive-core}
 X=\varepsilon I+F^\dagger F,\qquad
 X_M=\varepsilon I+F_M^\dagger F_M.
\end{equation}
Here the product is formed after lifting; we do not first reduce
$F^\dagger F$ to a Weyl sum. This choice preserves its positivity
at every $M$. Unitarity of the Weyl and pseudo-Weyl factors gives
\begin{equation}\label{sandwich:bounds}
 \varepsilon I\le X_M\le bI,\qquad
 \varepsilon I\le X\le bI,
 \qquad b=\varepsilon+\Bigl(\sum_{\ell=1}^L|c_\ell|\Bigr)^2.
\end{equation}
Define the normalization constants and states by
\begin{equation}\label{sandwich:state-lift}
 \begin{aligned}
 z_M&=\tr({\sigma_\mu^{\otimes mM}} X_M),&
 \rho_M&=z_M^{-1}\bigl(\sigma_\mu^{\otimes mM}\bigr)^{1/2}X_M\bigl(\sigma_\mu^{\otimes mM}\bigr)^{1/2},\\
 z&=\tr({\tau_\mu^{\otimes m}} X),&
 \rho&=z^{-1}\bigl(\tau_\mu^{\otimes m}\bigr)^{1/2}X\bigl(\tau_\mu^{\otimes m}\bigr)^{1/2}.
 \end{aligned}
\end{equation}
These are density operators because the sandwiches are positive and
have traces $z_M$ and $z$. The bounds in \eqref{sandwich:bounds}
give $\varepsilon\le z_M,z\le b$. Lemma~\ref{norm:thermal-words}
applied with one reference power equal to one gives $z_M\to z$ as $M \to \infty$.
The same order bounds also give
\begin{align}\label{eq:rho-M-rho-UL-bound}
 \frac{\varepsilon} b{\sigma_\mu^{\otimes mM}}\leq \rho_M\le\frac b{\varepsilon}{\sigma_\mu^{\otimes mM}},\qquad
 \frac{\varepsilon} b{\tau_\mu^{\otimes m}}\leq \rho\leq \frac{b}{\varepsilon}{\tau_\mu^{\otimes m}}.
\end{align}
In particular, the bosonic state has finite energy, and both its
relative entropy and those of its qubit lifts are uniformly bounded.

\begin{proposition} \label{sandwich:entropy-endpoint}\label{sandwich:entropy-limit}
The states in \eqref{sandwich:state-lift} satisfy
\begin{equation}\label{sandwich:input-entropy}
 D\big(\rho_M\Vert{\sigma_\mu^{\otimes mM}}\big)\longrightarrow D\big(\rho\Vert{\tau_\mu^{\otimes m}}\big),
\end{equation}
as $M\to\infty$.
The conclusion also holds when $X_M,X$ are replaced by a jointly
admissible family satisfying $aI\le X_M\leq bI$, $aI\leq X\le bI$, where
$0<a\le b<\infty$.
\end{proposition}

\begin{proof}
Relative entropy is the order-one endpoint of the sandwiched
R\'enyi divergences. For
$\alpha>1/2$ with $\alpha\ne1$, the sandwiched R\'enyi divergence is
\[
 \widetilde D_\alpha(\eta\Vert\xi)
 =\frac1{\alpha-1}\ln\tr
 \left(\xi^{(1-\alpha)/(2\alpha)}\eta
                 \xi^{(1-\alpha)/(2\alpha)}\right)^\alpha
\]
on the relevant domains~\cite{MullerLennert,WildeWinterYang,Renyi}.
For the state lifts above, the inverse reference powers cancel
within the sandwich and we obtain
\begin{equation}\label{sandwich:Renyi-power}
 \exp\!\left((\alpha-1)\widetilde D_\alpha(\rho_M\Vert{\sigma_\mu^{\otimes mM}})\right)
 =z_M^{-\alpha}\,\tr\left(\bigl(\sigma_\mu^{\otimes mM}\bigr)^{1/(2\alpha)}X_M\bigl(\sigma_\mu^{\otimes mM}\bigr)^{1/(2\alpha)}\right)^\alpha.
\end{equation}
The analogous identity holds with $\rho,{\tau_\mu^{\otimes m}},X,z$ on the bosonic
side. In that setting the sandwich is first interpreted on the
number-basis core; its bounded positive extension is exactly the
operator on the right-hand side. Thus no uncontrolled inverse
thermal power remains in the trace.

Proposition~\ref{norm:positive-powers} applied to the right-hand side of~\eqref{sandwich:Renyi-power} implies
\[
 \widetilde D_\alpha(\rho_M\Vert{\sigma_\mu^{\otimes mM}})
   \longrightarrow\widetilde D_\alpha(\rho\Vert{\tau_\mu^{\otimes m}}),
\]
for each fixed $\alpha>1/2$, $\alpha\ne1$.
For $0<\delta<1/2$, monotonicity in the R\'enyi parameter $\alpha$ gives
\[
 \widetilde D_{1-\delta}(\rho_M\Vert{\sigma_\mu^{\otimes mM}})
 \le D(\rho_M\Vert{\sigma_\mu^{\otimes mM}})
 \le\widetilde D_{1+\delta}(\rho_M\Vert{\sigma_\mu^{\otimes mM}}).
\]
Taking limits at the two fixed orders therefore yields
\[
 \begin{aligned}
 \widetilde D_{1-\delta}(\rho\Vert{\tau_\mu^{\otimes m}})
 &\le\liminf_{M\to\infty}D(\rho_M\Vert{\sigma_\mu^{\otimes mM}})\\
 &\le\limsup_{M\to\infty}D(\rho_M\Vert{\sigma_\mu^{\otimes mM}})
 \le\widetilde D_{1+\delta}(\rho\Vert{\tau_\mu^{\otimes m}}),
 \end{aligned}
\]
for $0<\delta<1/2$.

Since $X\le bI$, the sandwich construction gives
$\rho\le(b/z)\tau_\mu^{\otimes m}$. By the endpoint continuity
of the sandwiched R\'enyi divergences, in the type-I
specialization of~\cite[Theorem 13]{Renyi}, we have
\[
 \lim_{\delta\downarrow0}
 \widetilde D_{1-\delta}(\rho\Vert\tau_\mu^{\otimes m})
 =
 \lim_{\delta\downarrow0}
 \widetilde D_{1+\delta}(\rho\Vert\tau_\mu^{\otimes m})
 =
 D(\rho\Vert\tau_\mu^{\otimes m}).
\]
Letting $\delta\downarrow0$ in the preceding inequality
therefore proves \eqref{sandwich:input-entropy}.
For the more general assertion, the common bounds
$aI\le X_M\leq bI$, $aI\leq X\le bI$ keep the normalization constants bounded
away from zero and give the same dominations, so the argument
applies unchanged.
\end{proof}

This proposition establishes convergence of the input relative entropies. We next prove the lower bound needed for the output relative entropies.

\subsection{Lower bound for the output relative entropy}
\label{sec:output-pressure}

Our main tool for comparing the output relative entropies is that the qubit channels are well behaved on pseudo-Weyl operators as established in Proposition~\ref{lem:adjoint-limit}. This proposition involves the dual channels. Thus, to apply it we first need to derive a dual representation for relative entropies.

\begin{lemma}\label{lem:quadratic-entropy}
For faithful density operators $\eta$ and $\xi$,
\begin{equation}\label{eq:quadratic-entropy}
 D(\eta\Vert\xi)
 =\int_0^\infty\bigg(\sup_{B\text{ bounded}}\Big\{
 2\operatorname{Re}\tr(\eta B)
 -t\tr(\eta B^\dagger B)-\tr(\xi BB^\dagger)\Big\}-\frac1{1+t}\bigg)\,\dd t.
\end{equation}
The integrand is nonnegative, and the formula is understood in the
extended-value sense. On a finite-mode bosonic system, the supremum
can be restricted to finite Weyl sums.
\end{lemma}

\begin{proof}
The proof is based on
Kosaki's variational formula~\cite{Kosaki1986}, also recalled
in~\cite[Eq.~(8)]{Hollands2021}, that was reformulated in \cite[Proposition~3.5, Eq.~(39)]{BrownFawziFawzi2024} as
\[
 D(\eta\Vert\xi)
 =
 -\int_0^1 \frac1u\,
 \inf_{A\text{ bounded}}
 \left\{
  1+2\operatorname{Re}\tr(\eta A)
  +(1-u)\tr(\eta A^\dagger A)
  +u\tr(\xi AA^\dagger)
 \right\}\,\dd u.
\]
This representation applies to normal states on a von Neumann
algebra, including the infinite-dimensional setting considered here. We can then recover the desired identity \eqref{eq:quadratic-entropy}
under the change of variables $u=\frac1{1+t}$, $A=-(1+t)B$.
Choosing $B=I/(1+t)$ shows that the integrand is nonnegative.

Finally, as in the proof of Theorem~\ref{sandwich:norm-transfer}, the
Kaplansky density theorem supplies uniformly bounded finite Weyl sums
converging strongly-$*$ to any bounded operator. 
All three trace expressions in the supremum are continuous under
such approximations. This proves the last assertion.

\end{proof}

We can now state the desired lower bound for the output relative entropies. The proof is based on Lemma~\ref{norm:thermal-words} applied to the test operators provided by the above lemma.

\begin{lemma}\label{lem:output-entropy-liminf}
Let $\rho$ and $\rho_M$ be the states in
\eqref{sandwich:state-lift}, and let $\Phi$ and $\Psi$ be the
bosonic and qubit channel pair of Theorem~\ref{sandwich:norm-transfer},
with input thermal parameter $\mu$ and output thermal parameter $\mu'$.
Then
\begin{equation}\label{sandwich:output-liminf}
 D\bigl(\Phi^{\otimes m}(\rho)\Vert\tau_{\mu'}^{\otimes m}\bigr)
 \le\liminf_{M\to\infty}
 D\bigl(\Psi^{\otimes mM}(\rho_M)\Vert\sigma_{\mu'}^{\otimes mM}\bigr).
\end{equation}
\end{lemma}

\begin{proof}
We write
$\eta_M=\Psi^{\otimes mM}(\rho_M)$ and
$\eta=\Phi^{\otimes m}(\rho)$. 
Complete
positivity and~\eqref{eq:rho-M-rho-UL-bound} imply
\begin{align}\label{eq:Psi-bound-LU-mu'}
 \frac{\varepsilon}{b}\sigma_{\mu'}^{\otimes mM}\le 
 \eta_M=\Psi^{\otimes mM}(\rho_M)
 \le\frac b\varepsilon\sigma_{\mu'}^{\otimes mM}.
\end{align}
The corresponding bounds hold for $\eta=\Phi^{\otimes m}(\rho)$ with
reference $\tau_{\mu'}^{\otimes m}$. In particular, these outputs
are faithful. Fix $t>0$ and a corresponding finite Weyl sum $B$ in Lemma~\ref{lem:quadratic-entropy}.
Duality gives
\begin{align}
 \tr\bigl[B_{M,\mu'}\Psi^{\otimes mM}(\rho_M)\bigr]
 &=z_M^{-1}\tr\!\left[
 (\Psi^\dagger)^{\otimes mM}(B_{M,\mu'})
 \Gamma_{\sigma_\mu^{\otimes mM}}(X_M)\right]\nonumber\\
 &\longrightarrow z^{-1}\tr\!\left[
 (\Phi^\dagger)^{\otimes m}(B)\Gamma_{\tau_\mu^{\otimes m}}(X)\right]
 =\tr\bigl[B\Phi^{\otimes m}(\rho)\bigr].
 \label{sandwich:output-tests}
\end{align}
Here we use \eqref{eq:adjoint-limit} to replace $(\Psi^\dagger)^{\otimes mM}(B_{M,\mu'})$ by the lift of $(\Phi^\dagger)^{\otimes m}(B)$, and then we apply Lemma~\ref{norm:thermal-words} to establish the limit. The replacement error is at most its operator norm since $\rho_M$ is a state.

 In addition to the above linear expectation,
we need the limits of the two quadratic expectations in \eqref{eq:quadratic-entropy}. Another application of Lemma~\ref{norm:thermal-words} directly gives
\[
 \tr(\sigma_{\mu'}^{\otimes mM}B_{M,\mu'}B_{M,\mu'}^\dagger)
 \longrightarrow\tr(\tau_{\mu'}^{\otimes m}BB^\dagger).
\]
For the other quadratic term, multiply $B^\dagger B$ using Weyl relations to get an equivalent Weyl sum $C=B^\dagger B$. Now consider $E=B^\dagger B -C$ as a Weyl expression with the corresponding Weyl lift $E_{M, \mu'}= B_{M,\mu'}^\dagger B_{M,\mu'} - C_{M, \mu'}$. We note that $E$ is equal to zero as an operator on the bosonic side, but as a specified Weyl expression, $E_{M, \mu'}$ is not necessarily equal to zero. Therefore, by Lemma~\ref{norm:thermal-words} we have
\[
 \tr(\sigma_{\mu'}^{\otimes mM}E_{M, \mu'}^\dagger E_{M, \mu'})\longrightarrow \tr(\tau_{\mu'}^{\otimes m}E^\dagger E)=0.
\]
On the other hand, the  Cauchy--Schwarz inequality and~\eqref{eq:Psi-bound-LU-mu'} give
\[
 \big|\tr(\eta_M E_{M, \mu'})\big|
 \le \Big[\,\tr(\eta_M
    E_{M, \mu'}^\dagger E_{M, \mu'})\Big]^{1/2} \leq \Big[\,\frac{b}{\varepsilon}\tr(\sigma_{\mu'}^{\otimes mM}
    E_{M, \mu'}^\dagger E_{M, \mu'})\Big]^{1/2}
 \longrightarrow0.
\]
Applying the linear expectation limit in \eqref{sandwich:output-tests} to $C$ therefore proves
\[
 \tr(\eta_M B_{M,\mu'}^\dagger B_{M,\mu'})
 \longrightarrow\tr(\eta B^\dagger B).
\]
Therefore, for every fixed $t>0$ and fixed $B$, the entire quadratic
expression in~\eqref{eq:quadratic-entropy} corresponding to $B_{M, \mu'}$ converges. The qubit supremum is at least
its value at $B_{M,\mu'}$. Taking the lower limit and then the
supremum over finite Weyl sums bounds the bosonic supremum from
above by the lower limit of the qubit suprema. The integrands in
\eqref{eq:quadratic-entropy} are nonnegative, so Fatou's lemma proves
\eqref{sandwich:output-liminf}.

\end{proof}

The proof of Lemma~\ref{lem:output-entropy-liminf} uses the channel
structure only to establish faithfulness, uniform domination, and
convergence of lifted Weyl expectations. Consequently, suppose that for a sequence of integers
$M_k\to\infty$, we have faithful $mM_k$-qubit states $\eta_k$, and a faithful
$m$-mode state $\eta$ satisfying $\eta_k\le c\sigma_{\mu'}^{\otimes mM_k}$ for some constant $c>0$ independent of $k$. Moreover, suppose that for every finite Weyl sum $B$ we have
\[
 \tr(\eta_k B_{M_k,\mu'})\longrightarrow\tr(\eta B).
\]
Then, the same
argument as above gives
\begin{equation}\label{eq:output-entropy-liminf}
 D(\eta\Vert\tau_{\mu'}^{\otimes m})
 \le\liminf_{k\to\infty}
 D(\eta_k\Vert\sigma_{\mu'}^{\otimes mM_k}).
\end{equation}

\subsection{Proof of Theorem~\ref{prop:transfer}}
\label{sec:sandwich-approximation}

To extend the relative entropy inequality from the sandwiched states to general finite-energy inputs, we need approximation in relative entropy as well as trace norm. The following lemma supplies this approximation by
controlling energy and entropy separately.

\begin{lemma}[Approximation by sandwiched states]\label{lem:sandwich-core}
Every finite-energy $m$-mode state $\rho$ can be approximated by states of the form 
$$\frac{1}{\tr({\tau_\mu^{\otimes m}} X)}\Gamma_{{\tau_\mu^{\otimes m}}}(X),$$ 
with
$X=\varepsilon I+F^\dagger F$, where $\varepsilon>0$ and $F$ is a
finite Weyl sum. The approximating states $\rho_k$ satisfy
\[
 \|\rho_k-\rho\|_1\to0,\qquad
 \tr(\rho_k\Ntot)\to\tr(\rho\Ntot),\qquad
 S(\rho_k)\to S(\rho),\qquad
 D(\rho_k\Vert\tau_\mu^{\otimes m})
 \to D(\rho\Vert\tau_\mu^{\otimes m}).
\]
Their outputs under $\Phi^{\otimes m}$ satisfy the corresponding
limits, with output relative entropy measured against
$\tau_{\mu'}^{\otimes m}$.
\end{lemma}

The approximations in this lemma are constructed in three steps: truncate the photon
number, add a small thermal component, and approximate the bounded
relative density by finite Weyl sums. The details are given in Appendix~\ref{app:sandwiched-approximation}.
We now combine this approximation with the input entropy limit and the
channel-output lower bound to prove Theorem~\ref{prop:transfer}.

\begin{proof}[Proof of Theorem~\ref{prop:transfer}]
First let $\rho$ be a sandwiched state as in~\eqref{sandwich:state-lift}, and use its lifted states $\rho_M$.
The hypothesis on the qubit channel, the input entropy limit
\eqref{sandwich:input-entropy}, and the channel-output bound in
Lemma~\ref{lem:output-entropy-liminf} give
\[
 \begin{aligned}
 D\bigl(\Phi^{\otimes m}(\rho)\Vert\tau_{\mu'}^{\otimes m}\bigr)
 &\le\liminf_M
 D\bigl(\Psi^{\otimes mM}(\rho_M)\Vert\sigma_{\mu'}^{\otimes mM}\bigr)\\
 &\le c\lim_M D(\rho_M\Vert\sigma_\mu^{\otimes mM})\\
 &=cD(\rho\Vert\tau_\mu^{\otimes m}).
 \end{aligned}
\]
We can infer the same inequality for all finite-energy states $\rho$ once we use Lemma~\ref{lem:sandwich-core}.

\end{proof}

\subsection{Additive Gaussian noise}\label{sec:heat}

We can now apply the qubit inequalities 
of Theorem~\ref{thm:weighted-D} and use Theorem~\ref{prop:transfer} to obtain entropy inequalities for bosonic channels. The CMOE bounds, however, require an additional step as they are in terms of entropies rather than relative entropies. To handle this issue we note that we are free to choose the reference mean $\mu$.
What makes this freedom useful is the
following rewriting of \eqref{eq:thermal-relative entropy}:
\begin{equation}\label{eq:gibbs}
 D(\rho\Vert\tau_\mu^{\otimes m})
 =mg(\mu)-S(\rho)
   +\beta(\mu)\bigl[\tr(\rho\Ntot)-m\mu\bigr].
\end{equation}
Suppose a channel sends $\tau_\mu$ to $\tau_{\mu'}$ and multiplies
the energy excess in the last bracket by $\lambda$. A relative entropy
bound with coefficient $\lambda\beta(\mu')/\beta(\mu)$ then cancels
the energy terms exactly. It leaves an affine lower bound on the
output entropy, valid for every choice of $\mu$. Choosing
$mg(\mu)=S(\rho)$ removes the affine correction and gives the desired CMOE result.

For the additive Gaussian noise we use the qubit channel
$\Psi_{s,1}$ from \eqref{eq:weighted-bloch}. Its action in the Bloch representation is
$(x,y,z)\mapsto(sx,sy,s^2z)$, so the input reference polarization
$\kappa_\mu$ becomes $\kappa_{\mu'}=s^2\kappa_\mu$. Then by
\eqref{eq:general-map}, the limiting Gaussian gain is
$\lambda=1$ and the limit channel is an additive Gaussian noise:
\[
 \chi_{\Phi^{\mathrm{add}}_\gamma(\rho)}(z)
 =e^{-\gamma|z|^2/2}\chi_\rho(z).
\]
More explicitly, the output reference and the Gaussian noise parameter
are
\begin{equation}\label{eq:heat-map}
 \mu'=\frac{\mu+1/2}{s^2}-\frac12=\mu+\frac\gamma2,
 \qquad
 \gamma=(2\mu+1)(s^{-2}-1).
\end{equation}
Thus a prescribed channel $\Phi^{\mathrm{add}}_\gamma$ and a
prescribed faithful reference $\tau_\mu$ are represented by
\[
 s^2=\frac{2\mu+1}{2\mu+\gamma+1},
\]
which belongs to $(0,1]$ for every finite $\gamma\ge0$. Here, the choice of
$s$ depends on the reference, although the bosonic channel $\Phi^{\mathrm{add}}_\gamma$ being
studied is fixed.

\begin{proof}[Proof of Theorem~\ref{thm:heat}]
Choose $s$ as above. With $r=1$, Theorem~\ref{thm:weighted-D}
gives the qubit contraction coefficient
$\beta(\mu')/\beta(\mu)$. Its hypothesis holds for every tensor power
and every qubit input, so Theorem~\ref{prop:transfer} gives
\eqref{eq:heat-D} for arbitrary bosonic inputs with finite energy.

To prove~\eqref{eq:heat-affine} write $E=\tr(\rho\Ntot)$. Additive noise $\Phi_\gamma^{\textup{add}}$ increases the output energy
to $E+m\gamma/2$. Since the reference mean increases by the same
amount,
\[
 \tr\Big((\Phi^{\mathrm{add}}_\gamma)^{\otimes m}(\rho)\Ntot\Big)
       -m(\mu+\gamma/2)=E-m\mu.
\]
Apply \eqref{eq:gibbs} to the two sides of \eqref{eq:heat-D}.
The coefficient multiplying $E-m\mu$ on both sides is
$\beta(\mu+\gamma/2)$, so cancellation gives
\eqref{eq:heat-affine}.

If $S(\rho)>0$, set $\mu=g^{-1}(S(\rho)/m)$ in~\eqref{eq:heat-affine} to obtain~\eqref{eq:heat-CMOE}.
If $S(\rho)=0$ and $\gamma>0$, keep the channel and its input fixed
and let the freely chosen reference mean $\mu$ decrease to zero. The first
term on the right of~\eqref{eq:heat-affine} tends to $mg(\gamma/2)$, while the correction tends
to zero because
\[
 \lim_{\mu\downarrow0}\frac{g(\mu)}{\beta(\mu)}=0.
\]
We note that the channel and the input state $\rho$ are fixed, so the left-hand side is unchanged in this limit. For $\gamma=0$, the channel is the identity and there is nothing to prove.
Finally, the action of the channel on thermal states given in 
\eqref{eq:thermal-mean-map} proves attainment.
\end{proof}

\subsection{Thermal attenuation in the heating range}\label{sec:enlarged}

An attenuator channel both reduces the input energy and adds thermal noise. Our idea to prove Theorem~\ref{thm:att-enlarged}
is to separate these two
effects. The change in relative entropy by the latter effect can be treated by Theorem~\ref{thm:heat} proven above. For the former effect we use the following lemma.

\begin{lemma}[Thermal-reference entropy inequality]\label{lem:thermal-EPI}
Let $\mu>0$ and $0\le\lambda\le1$. Every finite-energy $m$-mode
state $\rho$ satisfies the following equivalent inequalities:
\begin{align}
&S\bigl((\Phi^{\mathrm{att}}_{\lambda,\mu})^{\otimes m}(\rho)\bigr)
 \ge\lambda S(\rho)+(1-\lambda)mg(\mu),
 \label{eq:HanSboson}\\
&D\bigl((\Phi^{\mathrm{att}}_{\lambda,\mu})^{\otimes m}(\rho)
   \Vert\tau_\mu^{\otimes m}\bigr)
 \le\lambda D(\rho\Vert\tau_\mu^{\otimes m}).
 \label{eq:HanDboson}
\end{align}
\end{lemma}

The equivalence of the two inequalities holds because
attenuation towards $\tau_\mu$ sends $\tr(\rho \Ntot)-m\mu$ to
$\lambda(\tr(\rho \Ntot)-m\mu)$, so applying~\eqref{eq:gibbs} on both sides of~\eqref{eq:HanDboson}, the energy terms cancel and we obtain~\eqref{eq:HanSboson}. This observation was first made in~\cite{de2018conditional}. The entropy form~\eqref{eq:HanSboson} is a special case of the linear
quantum entropy power inequality~\cite[Theorem~VII.1]{KS14}. Also, the relative entropy inequality~\eqref{eq:HanDboson} was first established in~\cite[Theorem~8.5]{CarlenMass2}; although the statement there is given only in the one-mode case, the same proof works for multimode states. We give yet another proof of this lemma in
Appendix~\ref{app:Han} based on Weyl lifting. More precisely, transferring quantum Han’s inequality to bosonic systems gives~\eqref{eq:HanDboson}.

\begin{proof}[Proof of Theorem~\ref{thm:att-enlarged}]
The condition $N\ge\mu$ makes available the following decomposition of the channel:
\begin{equation}\label{eq:heat-Han-factor}
 \Phi^{\mathrm{att}}_{\lambda,N}
 =\Phi^{\mathrm{add}}_\gamma\circ
       \Phi^{\mathrm{att}}_{\lambda,\mu},
 \qquad \gamma=2(1-\lambda)(N-\mu).
\end{equation}
Indeed, composing the characteristic functions preserves the gain
$\lambda$ and adds $\gamma$ to the Gaussian noise parameter
$(1-\lambda)(2\mu+1)$; their sum is
$(1-\lambda)(2N+1)$. The first channel preserves $\tau_\mu$ while the
second one sends it to $\tau_{\mu+\gamma/2}=\tau_{\mu'}$.

First apply \eqref{eq:HanDboson} to the attenuation channel $\Phi^{\textup{att}}_{\lambda, \mu}$, and then apply~\eqref{eq:heat-D}, with reference mean $\mu$, to its output.
The two contraction coefficients multiply to
$\lambda\beta(\mu')/\beta(\mu)$, proving
\eqref{eq:att-enlarged-D}. The energy identity for the full channel is
\[
 \tr\bigl((\Phi^{\mathrm{att}}_{\lambda,N})^{\otimes m}(\rho)
             \Ntot\bigr)-m\mu'
 =\lambda\bigl[\tr(\rho\Ntot)-m\mu\bigr].
\]
Substituting \eqref{eq:gibbs} in the relative entropy inequality
therefore gives \eqref{eq:att-enlarged-affine}.

For positive input entropy, the choice $\mu=g^{-1}(S(\rho)/m)$ is allowed by
\eqref{eq:att-enlarged-range} and gives
\eqref{eq:att-enlarged-CMOE}. If $S(\rho)=0$ and $N>0$, let
$\mu\downarrow0$ through $0<\mu\le N$ in the affine inequality,
with the channel fixed. Its correction tends to zero and its first
term tends to $mg((1-\lambda)N)$. If $N=0$ and $S(\rho)=0$,
the assertion is entropy nonnegativity. Finally, $\lambda=0$ gives
thermal replacement and $\lambda=1$ gives the identity. The action
on thermal inputs proves attainment in every case.

\end{proof}

\section{Qubit entropy inequalities}
\label{sec:qubit-results}

This section establishes the qubit entropy inequalities underlying
Section~\ref{sec:entropy-transfer}. We first state a sharp tensor-power
entropy bound for the channel $\Psi_{s,r}$ defined in
\eqref{eq:weighted-bloch}, together with its depolarizing-channel
consequence. We then prove this bound and derive the relative entropy
contraction of Theorem~\ref{thm:weighted-D}. These results are interesting on their own and may have other applications.

We use the binary entropy function $h(x) = -x\ln x-(1-x)\ln(1-x)$ and write
\[
 \widetilde h(z)=h((1-z)/2),\qquad -1\le z\le1.
\]
The function $\widetilde h$ is even and strictly decreasing on $[0,1]$.
Its inverse $\widetilde h^{-1}:[0,\ln2]\to[0,1]$ always refers to
this decreasing branch. Thus the entropy of a qubit with Bloch vector of length
$r$ is $\widetilde h(r)$.

For a Hermitian unitary $U$, define the pinching maps
by
\begin{equation*}\label{eq:pinching-def}
\Delta_U(\omega)=\frac{\omega+U\omega U}{2}.
\end{equation*}
This pinching map replaces $\omega$ with its block diagonal when written in the eigenbasis of $U$. We also define the partial pinching map by
$$
 \mathcal P_{U,r}(\omega)=r\omega+(1-r)\Delta_U(\omega),
 \qquad 0\le r\le1.$$
For the Pauli matrices $X,Y,Z$ from Section~\ref{sec:qubit-semigroups},
consider
\begin{equation}\label{eq:weighted-pinching}
 \Psi_{s,r}=\mathcal P_{Z,r}\circ\mathcal P_{Y,s}\circ\mathcal P_{X,s},
 \qquad 0\le s\le r\le1.
\end{equation}
This realizes the Bloch action in \eqref{eq:weighted-bloch}.
Here $s$ and $r$ parametrize the partial pinchings: the transverse
coefficient in the convention of \eqref{eq:qubit-standard-Bloch} is
$sr$, and the longitudinal coefficient is $s^2$. In particular,
$\Psi_{s,1}=\mathcal P_{Y,s}\circ\mathcal P_{X,s}$ consists of two
transverse pinchings, whereas $\Psi_{s,s}=\mathcal D_{s^2}$ is the
depolarizing channel. The condition $s\le r$ is used in the
convexity argument below.

For $0\le a,b\le1$, introduce the function
\begin{align}\label{eq:Fab} F_{a,b}(u,z)
 =\widetilde h\!\left(\sqrt{a^2\,\widetilde h^{-1}(u)^2+(b^2-a^2)z^2}\right),
\end{align} 
on the domain
\begin{align}\label{eq:Fab-D}
 \mathscr D=\big\{(u,z): |z|\le1,\ 0\le u\le\widetilde h(z)\big\}.
\end{align}
We note that, since $\widetilde h$ is decreasing, for any $(u, z)\in \mathscr D$ we have 
\begin{align}\label{eq:Fab-bound-tildeh}
F_{a, b}(u, z) \leq \widetilde h(bz). 
\end{align}

The motivation for this function is the following observation: suppose that a one-qubit state $\omega$ has entropy $u=S(\omega)$ and polarization $z=\tr(\omega Z)$, and let $\omega'$ be the output of the channel 
with Bloch representation $(x,y,z)\mapsto(ax,ay,bz)$ applied to $\omega$. Then, 
$$S(\omega')=F_{a,b}(u,z).$$ 
The next theorem says that the same profile remains a
lower bound for arbitrary tensor powers of \eqref{eq:weighted-pinching}.

For an $n$-qubit state $\omega$, we denote its marginal on qubit $i$ by $\omega_i$. We also use
$A_i$ to denote a one-qubit operator $A$ that acts on qubit $i$.

\begin{theorem}[Sharp output entropy bound]
\label{thm:composed-tensor}
Let \(0\le s\le r\le1\). For every \(n\)-qubit state \(\omega\),
put \(z=n^{-1}\sum_{i=1}^n\tr(\omega Z_i)\). Then
\begin{equation}\label{eq:composed-tensor}
 S(\Psi_{s,r}^{\otimes n}(\omega))
 \ge nF_{sr,s^2}\!\left(S(\omega)/n,z\right).
\end{equation}
Equality holds for \(\omega=\gamma^{\otimes n}\), for every
qubit state \(\gamma\).
\end{theorem}

We note that the pair on the right-hand side of~\eqref{eq:composed-tensor} lies in the domain $\mathscr D$ because by subadditivity and
concavity of $\widetilde h$ we have
\[
 S(\omega)\le\sum_{i=1}^nS(\omega_i)
 \le\sum_{i=1}^n\widetilde h(\tr(\omega Z_i))
 \le n\widetilde h\!\left(\frac1n\sum_{i=1}^n\tr(\omega Z_i)\right).
\]

Setting \(s=r=\sqrt{\lambda}\), we recover the depolarizing channel: 
\(\Psi_{s,r}=\mathcal D_\lambda\). Therefore, Theorem~\ref{thm:composed-tensor} 
directly yields the following corollary.

\begin{corollary}[Quantum Mrs.~Gerber's lemma]\label{thm:QMGL}
For every \(n\)-qubit state $\omega$ and \(0\le \lambda\le1\),
\begin{equation}\label{eq:QMGL}
 S\big(\mathcal D_\lambda^{\otimes n}(\omega)\big)
 \ge nF_{\lambda,\lambda}(S(\omega)/n,0).
\end{equation}
Here \(F_{\lambda,\lambda}\) is independent of its second argument and
\[
 F_{\lambda,\lambda}(u,z)=\widetilde h(\lambda\widetilde h^{-1}(u))
 =h\!\left(\frac{1-\lambda}{2}+\lambda h^{-1}(u)\right),
\]
where \(h^{-1}\) takes values in \([0,1/2]\).
Products of qubit states with the same entropy attain equality.
\end{corollary}

The rest of this section is devoted to the proof of Theorem~\ref{thm:composed-tensor} as well as the proof of Theorem~\ref{thm:weighted-D} stated in the previous section.

\subsection{Separating one qubit from the remaining system}

As mentioned above, the one-qubit bound of~\eqref{eq:composed-tensor} is indeed an equality. To obtain the corresponding bound for $n >1$, we proceed by induction. We single out one qubit $Q$ and denote the remaining qubits by $R$. 
To apply induction, we need to assign some entropy value to subsystem $Q$ to be used as the first argument of the function on the right-hand side of~\eqref{eq:composed-tensor}. A first choice for this entropy value is $S(Q)_{\omega}$, but this ignores the correlations between $Q$ and $R$ and does not work. A second natural choice is $S(Q|R)_\omega$, but this quantity would not work either simply because it can take negative values.

We will use the decomposition of $\Psi_{s, r}$ into the partial pinching maps in~\eqref{eq:weighted-pinching}, so this entropy value should be consistent with these successive operations. Since full pinching is a projective measurement
with its outcome unrecorded, it is natural to consider the
conditional states of $R$ induced by a measurement on $Q$.
For a measurement basis $\mathcal B$, let $\widehat Q$ record
the outcome and let $\omega^{\mathcal B}$ denote the resulting
classical--quantum state. We assign to $Q$ the quantity
\[
 S(QR)_\omega-S(R|\widehat Q)_{\omega^{\mathcal B}}.
\]
As we show below, this quantity lies between zero and
$S(Q)_\omega$. For a product state, it equals $S(Q)_\omega$
for every measurement basis. We retain the original joint
entropy $S(QR)_\omega$, rather than the post-measurement
entropy, so that the entropy assigned to $Q$ and the averaged
conditional entropy of $R$ sum to the input entropy required
in~\eqref{eq:composed-tensor}.

Finally, we minimize $S(R|\widehat Q)_{\omega^{\mathcal B}}$
over the measurement basis, thereby maximizing the entropy
assigned to $Q$. The resulting optimized conditional entropy
is invariant under unitaries on $Q$ and cannot decrease under
random-unitary channels on $Q$. These properties allow us to
use it consistently through all three partial pinchings,
even though their measurement bases differ.

Let \(Q\) be a qubit and \(R\) an arbitrary finite-dimensional
system. We use the subsystem entropy notation $S(A)_\rho:=S(\rho_A)$.
For an orthonormal basis $\mathcal B=\{\ket{e_0},\ket{e_1}\}$ of $Q$, define
\begin{align*}
 p_x\omega_R^x&=\bra{e_x}\omega_{QR}\ket{e_x},\qquad
 \omega^{\mathcal B}_{\widehat Q R}
 =\sum_{x=0}^1p_x\ketbra{e_x}{e_x}_{\widehat Q}\otimes\omega_R^x.
\end{align*}
Here $\widehat Q$ is the classical register recording the measurement outcome.
The superscript $\mathcal B$ specifies the measurement basis. We note that 
$\omega_{\widehat QR}^{\mathcal B}=(\Delta_U\otimes\id_R)(\omega)$, where
$U=\ketbra{e_0}{e_0}-\ketbra{e_1}{e_1}$.
Now define
\begin{equation}\label{eq:cstar-definition}
 S_*(R|\widehat Q)_\omega
 :=\min_{\mathcal B}S(R|\widehat Q)_{\omega^{\mathcal B}},\qquad
 S(R|\widehat Q)_{\omega^{\mathcal B}}
 =\sum_{x=0}^1p_xS(R)_{\omega_R^x}.
\end{equation}
In the optimized quantity, the subscript $\omega$ denotes the original
state on $QR$, and the star indicates minimization over measurement bases. Compactness of the space of bases and continuity of the averaged conditional entropy ensure that the minimum exists. This is the optimized conditional entropy associated with projective
measurements in quantum discord~\cite{OllivierZurek2002}.

\begin{lemma}\label{lem:cstar}
Let $\omega=\omega_{QR}$ be a bipartite state with $Q$ being a qubit. 
For every given orthonormal basis \(\mathcal B=\{\ket{e_0}, \ket{e_1}\}\) on $Q$,
\begin{equation}\label{eq:c-domain}
 0\le S(QR)_{\omega}-S(R|\widehat Q)_{\omega^{\mathcal B}}\le S(\widehat Q)_{\omega^{\mathcal B}} = h(p_0),
\end{equation}
where $p_0= \bra{e_0}\omega_Q\ket{e_0}$.
Moreover, for $z=\tr(\omega_QZ)$, we have
\begin{equation}\label{eq:cstar-strong-domain}
 0\le S(QR)_{\omega}-S_*(R|\widehat Q)_{\omega}
 \le S(Q)_{\omega}\le \widetilde h(z),
\end{equation}
and \(S_*(R|\widehat Q)_{\omega}\le S(R)_{\omega}\).
The function \(\omega\mapsto S_*(R|\widehat Q)_{\omega}\) is concave, invariant under unitaries on \(Q\),
and nondecreasing under random-unitary channels on \(Q\).
Also, for every channel \(\mathcal T\) on the auxiliary system~\(R\) it holds that
\begin{equation}\label{eq:cstar-auxiliary}
S(QR)_{(\id\otimes\mathcal T)(\omega)}
       -S_*(R|\widehat Q)_{(\id\otimes\mathcal T)(\omega)}
 \ge S(QR)_{\omega}-S_*(R|\widehat Q)_{\omega}.
\end{equation}
\end{lemma}

\begin{proof}
Let $\omega_{QRE}$ be a purification of $\omega_{QR}$ and write
$\omega^{\mathcal B}_{\widehat QRE}$ for its post-measurement state. Each conditional
state on $RE$ is pure, so its $R$ and $E$ marginals have the same
entropy. Thus, we have
\begin{align}\label{eq:proof-S-QR-E-I}
 S(QR)_{\omega}-S(R|\widehat Q)_{\omega^{\mathcal B}}
 =S(E)_\omega-S(E|\widehat Q)_{\omega^{\mathcal B}}=I(E;\widehat Q)_{\omega^{\mathcal B}}.
\end{align}
This lies between zero and $S(\widehat Q)_{\omega^{\mathcal B}}=h(p_0)$, proving
\eqref{eq:c-domain}. Moreover, data processing under the partial trace
over $R$ gives
\[
 I(E;\widehat Q)_{\omega^{\mathcal B}}\le I(RE;\widehat Q)_{\omega^{\mathcal B}}=S(RE)_\omega=S(Q)_{\omega},
\]
where the first equality uses purity of the conditional $RE$ states.
Maximizing over $\mathcal B$ gives the first upper bound in
\eqref{eq:cstar-strong-domain}, and its second bound $S(Q)_{\omega}\le\widetilde h(z)$ is a consequence of the concavity of entropy.
Concavity of entropy also gives the concavity of $\omega\mapsto S_*(R|\widehat Q)_{\omega}$ and its monotonicity under random-unitary channels.

Finally,~\eqref{eq:cstar-auxiliary} is a consequence of the data processing inequality once we use~\eqref{eq:proof-S-QR-E-I} and consider a Stinespring dilation of $\mathcal T$.
\end{proof}

\subsection{Biased binary pinching}

We next estimate the entropy increase caused by a partial pinching relative to a full pinching. We bound this entropy difference with that of a particular single-qubit state having the same measurement bias
and the same entropy deficit relative to complete pinching. For that qubit state, the output entropy can be computed exactly, giving rise to the following formulation of our bound.

\begin{proposition}
[Biased binary pinching]\label{prop:BP}
Let $U=U^\dagger=U^{-1}$ be a self-adjoint unitary on a qubit $Q$,
and let $\omega=\omega_{QR}$ be a joint state with a finite-dimensional
auxiliary system $R$. Set
\[
 z=\tr\big((U\otimes I_R)\omega\big),\qquad
 u=S(QR)_{\omega}-S(QR)_{(\Delta_U\otimes \id_R)(\omega)}+\widetilde h(z).
\]
Then $0\le u\le\widetilde h(z)$, and for every $0\le r\le1$,
\begin{equation}\label{eq:BP}
 0\leq S(QR)_{(\Delta_U\otimes \id_R)(\omega)}-S(QR)_{(\mathcal P_{U,r}\otimes \id_R)(\omega)}
 \le \widetilde h(z)-\widetilde h\!\left(\sqrt{r^2\,\widetilde h^{-1}(u)^2+(1-r^2)z^2}\right).
\end{equation}
Equality holds for every product state $\omega_Q\otimes\omega_R$.
\end{proposition}

\begin{proof}
If $U=\pm I$, then $(\Delta_U\otimes \id_R)(\omega)=\omega$, $z=\pm1$, and $u=0$,
so the asserted inequality is an equality. Otherwise, we can assume without loss of generality that $U$ is the Pauli $Z$ operator. In this case, letting $\mathcal B$ be the computational basis, we have
$S(QR)_{(\Delta_Z\otimes \id_R)(\omega)}-\widetilde h(z)=S(R|\widehat Q)_{\omega^{\mathcal B}}$,
so $u=S(QR)_{\omega}-S(R|\widehat Q)_{\omega^{\mathcal B}}$. Then,
the bounds $0\le u\le\widetilde h(z)$ follow from
\eqref{eq:c-domain}.

To prove~\eqref{eq:BP}, let \(\omega_0=(\Delta_Z\otimes \id_R)(\omega)\) and \(A=\omega-\omega_0\). With this notation
the partial-pinching path is written as
\[
 (\mathcal P_{Z,r}\otimes \id_R)(\omega)=r\omega+(1-r)\Delta_Z\omega
 =\omega_0+rA.
\]
We restrict to the support of $\omega_0$, which is invariant under $Z$.
On this support $\omega_0>0$, and
\[
Z\omega_0Z^\dagger=\omega_0,\qquad ZAZ^\dagger=-A,\qquad
 -\omega_0\le A\le\omega_0.
\]
In particular, \(\tr (A\ln\omega_0)=0\), and hence
 \begin{align*}
 J(r)&:=S(QR)_{(\Delta_Z\otimes \id_R)(\omega)}-S(QR)_{(\mathcal P_{Z,r}\otimes \id_R)(\omega)}\\
 &=S(QR)_{\omega_0}-S(QR)_{\omega_0+rA}     =D(\omega_0+rA\Vert\omega_0)\geq 0.
 \end{align*}
Thus $J(r)$ is the deficit relative to full pinching, and the
entropy increase caused by partial pinching is $J(1)-J(r)$.
We compare this entropy deficit with that of the auxiliary single-qubit state
\[
 \gamma=\frac{I+aX+zZ}{2},
 \qquad
 a=\sqrt{\widetilde h^{-1}(u)^2-z^2}.
\]
The bounds $0\le u\le\widetilde h(z)$ ensure that \(\gamma\) is a valid qubit state.
Its entropy is $S(Q)_{\gamma}=u$ and its $Z$ bias is $\tr(\gamma Z)=z$. On this auxiliary qubit,
\[
 \Delta_Z(\gamma)=\frac{I+zZ}{2},\qquad
 \mathcal P_{Z,r}(\gamma)=\frac{I+raX+zZ}{2}.
\]
The latter state has eigenvalues $(1\pm\sqrt{z^2+r^2a^2})/2$.
Its entropy is therefore $\widetilde h(\sqrt{z^2+r^2a^2})$, and
its deficit relative to full pinching is
\begin{equation}\label{eq:BP-compare-J}
 J_\gamma(r):=S(Q)_{\Delta_Z(\gamma)}-S(Q)_{\mathcal P_{Z,r}(\gamma)}
 =\widetilde h(z)
  -\widetilde h\!\left(\sqrt{z^2+r^2a^2}\right).
\end{equation}
Thus \(J_\gamma(1)=\widetilde h(z)-u=J(1)\), and the desired inequality~\eqref{eq:BP}
is equivalent to \(J(r)\le J_\gamma(r)\).

Define
\[
 f(x)=\|A-x\omega_0\|_1,\qquad
 K(x)=f(x)-x=2\tr(A-x\omega_0)_+.
\]
Then \(K\ge0\), and \(K(x)=0\) for \(x\ge1\).
Frenkel's integral representation
\cite[Theorem~6]{Frenkel}, in the form given in
\cite[Eq.~(1.6)]{HircheTomamichel}, yields
\begin{equation}\label{eq:Jintegral}
 J(r)=r^2\int_0^1\frac{K(x)}{1-r^2x^2}\,dx,
 \qquad 0\le r\le1.
\end{equation}
For the comparison qubit, the same formula
holds with
\begin{equation}\label{eq:BP-compare-K}
 K_\gamma(x)=\left(\sqrt{a^2+z^2x^2}-x\right)_+.
\end{equation}
Now our key observation is that
\begin{equation}\label{eq:BP-monotone}
 f(x)^2-z^2x^2
 \quad\text{is nondecreasing for }x\ge0.
\end{equation}
To prove this, we note that $A-x\omega_0$ is invertible on the support of $\omega_0$, except at most finitely many values of \(x\). For every nonexceptional value $x>0$, put
\[
 V_x=\operatorname{sign}(A-x\omega_0),\qquad L_x=\frac{ZV_x+V_xZ}{2}.
\]
We have $f(x) = \tr((A-x\omega_0)V_x)$ and \(f'(x)=-\tr(\omega_0V_x)\).
Using \((A-x\omega_0)+Z(A-x\omega_0)Z=-2x\omega_0\), \(A-x\omega_0=V_x|A-x\omega_0|\), and \(V_x^2=I\),
cyclicity of trace gives
\[
 \tr(|A-x\omega_0|L_x)=\tr((A-x\omega_0)Z)=-xz,
\]
and
$$\tr(|A-x\omega_0|L_x^2)
 =\frac12\tr\big[\big((A-x\omega_0)+Z(A-x\omega_0)Z\big)V_x\big]=xf'(x).$$
Therefore, the Cauchy--Schwarz inequality implies
\[
 x^2z^2
 \le \tr(|A-x\omega_0|)\,\tr\big(|A-x\omega_0|L_x^2\big)
 =xf(x)f'(x).
\]
This proves \eqref{eq:BP-monotone} away from the finitely many
exceptional points, and continuity gives the assertion everywhere.

When \(K_\gamma(x)>0\), the sign of \(K(x)-K_\gamma(x)\)
is the same as the sign of \(f(x)^2-z^2x^2-a^2\).
If $a=0$, then $J(1)=J_\gamma(1)=0$, so both deficits vanish
for all $r$ and the assertion follows. Assume henceforth that $a>0$.
Then $K_\gamma$ is positive on an interval containing $x=0$, and since
\(K\ge0\), \eqref{eq:BP-monotone} shows that
\(K(x)-K_\gamma(x)\) changes sign at most once, from negative to positive.
Moreover, \(J(1)=J_\gamma(1)\) established above, gives
\[
 \int_0^1\frac{K(x)-K_\gamma(x)}{1-x^2}\,dx=0.
\]

If $K=K_\gamma$, the desired inequality is an equality. Otherwise,
the sign-change property and the vanishing integral give a threshold
$x_0\in(0,1)$ such that $K-K_\gamma\le0$ below $x_0$ and
$K-K_\gamma\ge0$ above it. The threshold need not be unique.
For $0<r<1$, the function
\[
 q_r(x)=\frac{r^2(1-x^2)}{1-r^2x^2}
\]
is nonincreasing. Therefore,
\[
 J(r)-J_\gamma(r)
 =\int_0^1
   \frac{\big(q_r(x)-q_r(x_0)\big)\big(K(x)-K_\gamma(x)\big)}{1-x^2}\,dx
 \le0.
\]
The endpoints $r=0$ and $r=1$ are equalities, completing the proof.
\end{proof}

\subsection{Partial pinchings}

We now combine the single-pinching bound of
Proposition~\ref{prop:BP} along the decomposition
of $\Psi_{s,r}$ in~\eqref{eq:weighted-pinching}. We first record the
monotonicity and convexity properties of the entropy profile $F_{a,b}$:
monotonicity allows us to compose the local bounds, while joint convexity
will be used in the subsequent tensorization argument.
The proof of the following lemma is given in
Appendix~\ref{app:joint-convex}.

\begin{lemma}\label{lem:joint-convex}
The domain \(\mathscr D\) defined in~\eqref{eq:Fab-D} is convex.
For \(0\le b\le a\le1\), the function \(F_{a,b}\) is jointly convex on \(\mathscr D\).
Moreover, for all \(0\le a,b\le1\), without any ordering restriction,
\[
 0\le\partial_u F_{a,b}(u,z)\le a^2
\]
in the interior of \(\mathscr D\). Consequently, for every fixed $z$ and
$0\le u_1\le u_2\le\widetilde h(z)$,
\[
 0\le F_{a,b}(u_2,z)-F_{a,b}(u_1,z)\le a^2(u_2-u_1).
\]
In particular, for fixed $c$ and $z$, the function
\(x\mapsto x+F_{a,b}(c-x,z)\) is nondecreasing wherever it is defined.
\end{lemma}

The last term on the right-hand side of~\eqref{eq:BP} is the function
\(F_{r,1}(u,z)\) defined in~\eqref{eq:Fab}. Thus, the inequality of Proposition~\ref{prop:BP} can be written as
$$S(QR)_{(\mathcal P_{U,r}\otimes \id_R)(\omega)}\geq S(R|\widehat Q)_{\omega^{\mathcal B}}+ F_{r, 1}\big(S(QR)_{\omega} - S(R|\widehat Q)_{\omega^{\mathcal B}}, z\big) ,$$
Here $U$ is a nontrivial self-adjoint qubit unitary, $\mathcal B$ is its
eigenbasis, and $\widehat Q$ records the measurement outcome. To apply this bound successively, we use the monotonicity of
$x\mapsto x+F_{r,1}(c-x,z)$ from Lemma~\ref{lem:joint-convex}.
To this end, since by definition
\(S_*(R|\widehat Q)_{\omega}\le S(R|\widehat Q)_{\omega^{\mathcal B}}\) and \eqref{eq:cstar-strong-domain} controls the
enlarged domain, we find that
\begin{equation}\label{eq:BP-cstar}
 S(QR)_{(\mathcal P_{U,r}\otimes \id_R)(\omega)}
 \ge S_*(R|\widehat Q)_{\omega}+F_{r,1}(S(QR)_{\omega}-S_*(R|\widehat Q)_{\omega},z).
\end{equation}

We now return to the three partial pinchings in~\eqref{eq:weighted-pinching}.
Their composition changes the preferred Pauli axis at each step.
Nevertheless, the optimized conditional entropy is invariant under that change and
cannot decrease under any of these random-unitary channels. This lets
us use the original optimized conditional entropy $S_*(R|\widehat Q)_{\omega}$ through all three steps. This is the main idea behind the use of $S_*(R|\widehat Q)_{\omega}$ in our proofs.

\begin{lemma}\label{lem:composed-local}\label{lem:XY-local}\label{lem:local}
For every state \(\omega_{QR}\), put
\(z=\tr(\omega_Q Z)\). Then
\begin{equation}\label{eq:composed-local}
 S(QR)_{(\Psi_{s,r}\otimes\id_R)(\omega)}
 \ge S_*(R|\widehat Q)_{\omega}+F_{sr,s^2}\big(S(QR)_{\omega}-S_*(R|\widehat Q)_{\omega},\, z\big).
\end{equation}
\end{lemma}

\begin{proof}
Letting $(x,y,z)$ be the Bloch vector of $\omega_Q$, Lemma~\ref{lem:cstar} gives
\begin{equation}\label{eq:composed-local-domain}
 \widetilde h^{-1}(u)^2\ge x^2+y^2+z^2, \qquad u=S(QR)_{\omega}-S_*(R|\widehat Q)_{\omega}.
\end{equation}
We apply the three partial pinchings in
$\Psi_{s,r}=\mathcal P_{Z,r}\circ\mathcal P_{Y,s}\circ\mathcal P_{X,s}$ one by one, keeping the original quantity $S_*(R|\widehat Q)_{\omega}$ throughout.
This is possible because $S_*$ cannot decrease under these random-unitary
channels, and Lemma~\ref{lem:joint-convex} allows us to lower the
optimized conditional entropy in the bound~\eqref{eq:BP-cstar} whenever the resulting
argument remains in the domain of $F$.

Write
\[
 \omega'=(\mathcal P_{X,s}\otimes\id_R)(\omega),\qquad
 \omega''=(\mathcal P_{Y,s}\otimes\id_R)(\omega').
\]
The first application of~\eqref{eq:BP-cstar} gives
\begin{align}
\label{eq:S-omega'-bound}
 S(QR)_{\omega'}\ge S_*(R|\widehat Q)_{\omega}+F_{s,1}(u,x).
\end{align}
The Bloch vector of $\omega'_Q$ is $(x, sy, sz)$, and
$S_*(R|\widehat Q)_{\omega'}\ge S_*(R|\widehat Q)_{\omega}$ by Lemma~\ref{lem:cstar}. Thus, if
$S(QR)_{\omega'}-S_*(R|\widehat Q)_{\omega}\le\widetilde h(sy)$, we can apply
\eqref{eq:BP-cstar} to $\omega'$, lower its optimized conditional entropy to
$S_*(R|\widehat Q)_{\omega}$, and use monotonicity of $F_{s, 1}$ in the first argument to obtain
\begin{align}
 S(QR)_{\omega''}& \ge S_*(R|\widehat Q)_{\omega'}
+F_{s,1}\big(S(QR)_{\omega'} - S_*(R|\widehat Q)_{\omega'},sy\big)\\
& \ge S_*(R|\widehat Q)_{\omega}
+F_{s,1}\big(S(QR)_{\omega'} - S_*(R|\widehat Q)_{\omega},sy\big)\\
   &\ge S_*(R|\widehat Q)_{\omega}
   +F_{s,1}\big(F_{s,1}(u,x),sy\big),
   \label{eq:composed-local-two}
\end{align}
where the last inequality uses~\eqref{eq:S-omega'-bound} to apply the monotonicity of $F_{s, 1}$ in its first argument. 

If instead $S(QR)_{\omega'}-S_*(R|\widehat Q)_{\omega}>\widetilde h(sy)$, the final bound in
\eqref{eq:composed-local-two} still holds:
$$S(QR)_{\omega''}\geq S(QR)_{\omega'}\geq S_*(R|\widehat Q)_{\omega}+\widetilde h(sy)\geq S_*(R|\widehat Q)_{\omega}+F_{s,1}\big(F_{s,1}(u,x),sy\big).$$ Here, we use the fact that partial pinching cannot decrease
entropy and apply~\eqref{eq:Fab-bound-tildeh} at the last step.
For the latter bound, we need to check the domain condition
$F_{s, 1}(u, x)\leq \widetilde h(sy)$ which holds since by~\eqref{eq:composed-local-domain},
\[
 \widetilde h^{-1}(F_{s,1}(u,x))^2
 =s^2\,\widetilde h^{-1}(u)^2+(1-s^2)x^2\ge s^2y^2.
\]

Next, simplifying the last term in~\eqref{eq:composed-local-two}, by the definition of $F_{a, b}$ we have
\[
 F_{s,1}\big(F_{s,1}(u,x),sy\big)
 =\widetilde h\!\left(
 \sqrt{s^4\, \widetilde h^{-1}(u)^2+s^2(1-s^2)(x^2+y^2)}\right)
 \ge F_{s,s^2}(u,z),
\]
where the inequality uses the fact that $\widetilde h$ is strictly decreasing on $[0,1]$ and~\eqref{eq:composed-local-domain}.  Consequently,
\begin{equation}\label{eq:composed-local-XY}
 S(QR)_{\omega''}\ge S_*(R|\widehat Q)_{\omega}+F_{s,s^2}(u,z).
\end{equation}

The Bloch vector of $\omega''_Q$ is $(sx, sy, s^2z)$, and
$S_*(R|\widehat Q)_{\omega''}\ge S_*(R|\widehat Q)_{\omega}$. Moreover,
\[
 \widetilde h^{-1}(F_{s,s^2}(u,z))^2
 =s^2\Big(\widetilde h^{-1}(u)^2-z^2\Big)+s^4z^2
 \ge s^4z^2,
\]
so $F_{s,s^2}(u,z)\le\widetilde h(s^2z)$ and the following nested
evaluation is in its domain. If
$S(QR)_{\omega''}-S_*(R|\widehat Q)_{\omega}\le\widetilde h(s^2z)$, applying
\eqref{eq:BP-cstar} and Lemma~\ref{lem:joint-convex} exactly as above gives
\begin{align*}
 S(QR)_{(\Psi_{s,r}\otimes\id_R)(\omega)}
 &\ge S_*(R|\widehat Q)_{\omega}+F_{r,1}\big(F_{s,s^2}(u,z),s^2z\big)\\
 &=S_*(R|\widehat Q)_{\omega}+F_{sr,s^2}(u,z).
\end{align*}
The equality follows directly from the definition of $F_{a, b}$.
If $S(QR)_{\omega''}-S_*(R|\widehat Q)_{\omega}>\widetilde h(s^2z)$, the same bound follows
from entropy monotonicity and
$F_{r,1}(F_{s,s^2}(u,z),s^2z)\le\widetilde h(s^2z)$.
This proves~\eqref{eq:composed-local}, including the parameter endpoints
by continuity.
\end{proof}

\subsection{Tensorization and relative entropy contraction}

We now have all the ingredients to prove Theorems~\ref{thm:composed-tensor} and~\ref{thm:weighted-D}.

\begin{proof}[Proof of Theorem~\ref{thm:composed-tensor}]
Considering the action of $\Psi_{s, r}$, the inequality turns into an equality if $n=1$. Suppose the theorem holds for \(n-1\). Split an $n$-qubit state $\omega = \omega_{Q_1, \dots, Q_n}$ into a qubit \(Q=Q_1\) and $n-1$ qubits \(R=Q_2,\cdots ,Q_n\). Put
\[
\widetilde\omega=\big(\id_Q\otimes\Psi_{s,r}^{\otimes(n-1)}\big)(\omega),
 \qquad z_R=\frac1{n-1}\sum_{i=2}^n\tr(\omega Z_i).
\]
Fix a measurement basis $\mathcal B$ on $Q$, with outcome
probabilities $p_x$ and conditional states of the remaining qubits $\omega_R^x$.
The channel on $R$ commutes with this measurement, so the conditional
output is $\Psi_{s,r}^{\otimes(n-1)}(\omega_R^x)$ and the
probabilities are unchanged. For each conditional state, induction
gives its output entropy in terms of its input entropy and average $Z$ bias. Joint convexity of $F_{sr, s^2}$ established in Lemma~\ref{lem:joint-convex} then allows us to average over $x$. Since
$\sum_xp_x\omega_R^x=\omega_R$, the averaged bias is $z_R$,
independently of the measurement basis.
Consequently,
\[
 S(R|\widehat Q)_{\widetilde\omega^{\mathcal B}}
 \ge(n-1)F_{sr,s^2}
       \big(S(R|\widehat Q)_{\omega^{\mathcal B}}/(n-1),z_R\big).\]
As $F_{sr,s^2}$ is
nondecreasing in its first argument, letting $\mathcal B$ be the optimal basis for the state $\widetilde \omega$ we obtain
\begin{equation}\label{eq:c-induction}
 S_*(R|\widehat Q)_{\widetilde\omega}\ge(n-1)F_{sr,s^2}
       \left(S_*(R|\widehat Q)_{\omega}/(n-1),z_R\right).
\end{equation}
We note that these evaluations are at pairs belonging to \(\mathscr D\) because
\[
 S_*(R|\widehat Q)_{\omega}\le S(R|\widehat Q)_{\omega^{\mathcal B}}\le S(R)_{\omega}
 \le\sum_{i=2}^n \widetilde h\big(\tr(\omega Z_i)\big)
 \le(n-1)\widetilde h(z_R).
\]
The monotonicity under channels on the auxiliary system in \eqref{eq:cstar-auxiliary} of Lemma~\ref{lem:cstar} gives
\(S(QR)_{\widetilde\omega}-S_*(R|\widehat Q)_{\widetilde\omega}\ge S(QR)_{\omega}-S_*(R|\widehat Q)_{\omega}\), while
\(z_Q=\tr(\omega Z_Q)\) is unchanged. Apply
Lemma~\ref{lem:composed-local} to $\widetilde \omega$. Then, the above bounds and joint
convexity imply
\begin{align*}
 S(QR)_{\Psi_{s,r}^{\otimes n}(\omega)}
 &\ge S_*(R|\widehat Q)_{\widetilde\omega}+
 F_{sr,s^2}(S(QR)_{\widetilde\omega}-S_*(R|\widehat Q)_{\widetilde\omega},z_Q)\\
 &\ge(n-1)F_{sr,s^2}(S_*(R|\widehat Q)_{\omega}/(n-1),z_R)
       +F_{sr,s^2}(S(QR)_{\omega}-S_*(R|\widehat Q)_{\omega},z_Q)\\
 &\ge nF_{sr,s^2}(S(QR)_{\omega}/n,z).
\end{align*}
The one-qubit identity and entropy additivity give equality for
$\omega=\gamma^{\otimes n}$.
\end{proof}

We finish this section by proving Theorem~\ref{thm:weighted-D}. Recall the parameters for the reference states from the previous section:
\begin{equation*}
 \Psi_{s,r}(\sigma_\mu)=\sigma_{\mu'},\qquad
 \tanh(\beta'/2)=s^2\tanh(\beta/2),\qquad
 \mu'=\frac{\mu+1/2}{s^2}-\frac12,
 \qquad\beta'=\beta(\mu').
\end{equation*}

\begin{proof}[Proof of Theorem~\ref{thm:weighted-D}]
We provide a supporting plane of the entropy bound of Theorem~\ref{thm:composed-tensor} to convert the bound just
proved into the required relative entropy contraction.

Differentiating \(F_{sr,s^2}\) at the point 
\[
 (u_0,z_0)=\big(S(\sigma_\mu),\tr(\sigma_\mu Z)\big)=\Big(\widetilde h(\tr(\sigma_\mu Z)),  \tr(\sigma_\mu Z)\Big)=\Big(\widetilde h(\tanh(\beta/2)),\tanh(\beta/2)\Big)
\]
corresponding to $\sigma_\mu$ gives
\[
 \partial_u F_{sr,s^2}(u, z)\Big|_{(u_0, z_0)}=r^2\frac{\beta'}{\beta},\qquad
 \partial_zF_{sr,s^2} (u, z)\Big|_{(u_0, z_0)}
 =\frac{r^2-s^2}{2}\beta'.
\]
This point lies on the boundary $u=\widetilde h(z)$, but the profile
and these derivatives extend smoothly to a neighborhood of it for
$0<s\le r\le1$. Then, applying the supporting-plane inequality from joint convexity of $F_{sr, s^2}(u, z)$ at $(u_0, z_0)$ and using the bound of Theorem~\ref{thm:composed-tensor} we find that
\begin{align}
 S\big(\Psi_{s,r}^{\otimes n}(\omega)\big)
 &\ge nF_{sr,s^2}\!\left(S(\omega)/n,z\right)\nonumber\\ &\geq n F_{sr, s^2}(u_0, z_0)+r^2\frac{\beta'}{\beta}\big[S(\omega)-n S(\sigma_\mu)\big]\notag +\frac{r^2-s^2}{2}\beta'\Big[\sum_i\tr(\omega Z_i)-n \tr(\sigma_\mu Z)\Big]\nonumber \\ & = n S(\sigma_{\mu'})+r^2\frac{\beta'}{\beta}\big[S(\omega)-n S(\sigma_\mu)\big]\notag +\frac{r^2-s^2}{2}\beta'\Big[\sum_i\tr(\omega Z_i)-n \tr(\sigma_\mu Z)\Big].
 \label{eq:weighted-tangent}
\end{align}
Next, we note that
\[ D\big(\omega\Vert\sigma_\mu^{\otimes n}\big)
 =nS(\sigma_\mu)-S(\omega)-\frac 12\beta \Big[\sum_i\tr(\omega Z_i)-n \tr(\sigma_\mu Z)\Big],
\]
and similarly
\begin{align*} D\big(\Psi_{s,r}^{\otimes n}(\omega)\Vert\sigma_{\mu'}^{\otimes n}\big)
 & =nS(\sigma_{\mu'})-S\big(\Psi_{s,r}^{\otimes n}(\omega)\big)-\frac 12\beta' \Big[\sum_i\tr\big(\Psi_{s,r}^{\otimes n}(\omega) Z_i\big)-n \tr(\sigma_{\mu'} Z)\Big]\\
 &=nS(\sigma_{\mu'})-S\big(\Psi_{s,r}^{\otimes n}(\omega)\big)-\frac {s^2}2\beta' \Big[\sum_i\tr(\omega Z_i)-n \tr(\sigma_{\mu} Z)\Big]
\end{align*}
Using these in the previous inequality we arrive at the desired bound~\eqref{eq:weighted-D}.

For sharpness, take \(\omega_\varepsilon=\sigma_\mu+\varepsilon X\)
for the Pauli $X$ operator and sufficiently small real \(\varepsilon\). The reference $\sigma_\mu$ and
perturbation $\omega_{\varepsilon}$ have the same diagonal, so their relative entropy is
the entropy difference, as also computed above. Expansion of the entropy function therefore gives
\[
 D(\omega_\varepsilon\Vert\sigma_\mu)
 =\frac{\beta}{\tanh(\beta/2)}\varepsilon^2+O(\varepsilon^4),\qquad
 D(\Psi_{s,r}(\omega_\varepsilon)\Vert\sigma_{\mu'})
 =\frac{\beta'}{\tanh(\beta'/2)}s^2r^2\varepsilon^2+O(\varepsilon^4).
\]
Their ratio tends to \(r^2\beta'/\beta\). Tensoring the
perturbed qubit with reference states proves optimality for every
\(n\). Letting \(\beta\downarrow0\) gives the stated maximally
mixed coefficient, which is again attained to second order by the same perturbation.
\end{proof}

\section{Amplitude damping and Gaussian CMOE}\label{sec:amp-damp}

Our CMOE bounds in the previous sections cover the additive noise channel as well as attenuators in the heating range. In particular, they do not cover pure-loss or amplifier channels. 
In this section, we formulate a conjecture regarding the contraction coefficients of the qubit amplitude-damping channel and show that, if true, it would extend
the same method to pure loss and quantum-limited amplification,
and hence to every phase-covariant Gaussian channel considered here. 
We state the conjecture and its full CMOE consequence first, and then prove
the conditional implication.

\subsection{Relative entropy contraction for amplitude damping}
\label{sec:AD}

The qubit channel corresponding to pure loss is amplitude damping. Theorem~\ref{prop:transfer} also applies to this channel. Nevertheless, 
a sharp entropy inequality for the tensor powers
of amplitude damping is missing and cannot be covered by the methods of the previous section. 

With survival parameter $0\leq r\leq 1$, the amplitude-damping channel is
\begin{equation}\label{eq:AD}
 \mathcal B_r\!\begin{pmatrix}1-x&c\\\overline c&x\end{pmatrix}
 =\begin{pmatrix}1-rx&\sqrt r\,c\\
                    \sqrt r\,\overline c&rx\end{pmatrix}.
\end{equation}
To include the polarization needed for amplification as well, for every $\beta\in\R$, define qubit references by
\[
 \sigma=\frac{I+\tanh(\beta/2)Z}{2},\qquad
 \sigma'=\mathcal B_r(\sigma)
          =\frac{I+\tanh(\beta'/2)Z}{2}.
\]
For finite $\beta$ and $0<r<1$, both references are faithful and
\begin{equation}\label{eq:AD-reference}
 e^{-\beta'}=\frac{r}{1+e^\beta-r}.
\end{equation}
When $\beta>0$, $\sigma$ is the reference $\sigma_\mu$ of
Section~\ref{sec:qubit-semigroups}, with
$\mu=(e^\beta-1)^{-1}$. Negative $\beta$ reverses the qubit
polarization. 

\begin{conjecture}[Relative entropy contraction for amplitude damping]
\label{conj:AD}
Let $\beta\in\R$ and $0<r<1$ and define $\sigma, \sigma'$ as above.
Then for every $n\ge1$ and every $n$-qubit state $\omega$,\footnote{In the definition of $c_r(\beta)$,
each quotient of the form $x/\tanh(x/2)$ has its continuous value $2$ at $x=0$.}
\begin{equation}\label{eq:ADtarget}
 D\bigl(\mathcal B_r^{\otimes n}(\omega)
       \Vert(\sigma')^{\otimes n}\bigr)
 \le c_r(\beta)D(\omega\Vert\sigma^{\otimes n}),
 \qquad
 c_r(\beta)=r\,
 \frac{\beta'/\tanh(\beta'/2)}{\beta/\tanh(\beta/2)}.
\end{equation}
\end{conjecture}

The coefficient $c_r(\beta)$ is dictated by a
perturbation of the reference state. To see this, let
$\omega_\varepsilon=\sigma+\varepsilon X$ for the Pauli $X$ operator and sufficiently small
real $\varepsilon$. Expanding the entropy terms gives
\begin{equation}\label{eq:ADnecessity}
 D(\omega_\varepsilon\Vert\sigma)
   =\frac{\beta}{\tanh(\beta/2)}\varepsilon^2+O(\varepsilon^4),\qquad
 D\bigl(\mathcal B_r(\omega_\varepsilon)\Vert\sigma'\bigr)
   =r\frac{\beta'}{\tanh(\beta'/2)}
                  \varepsilon^2+O(\varepsilon^4).
\end{equation}
The ratio of these quantities tends to $c_r(\beta)$, so any valid contraction coefficient must be at least
this large even for one qubit. Tensoring the perturbed qubit with
copies of $\sigma$ proves the same necessity for every $n$.

The following implication identifies the two polarization regimes
needed to obtain the full bosonic conclusion.

\begin{proposition}\label{prop:full}
Assume Conjecture~\ref{conj:AD} in the regimes
$0<\beta<\beta'$ and $\beta<\beta'<0$. Then every channel of
\eqref{eq:gaussian-general} and every finite-energy $m$-mode state
satisfy
\begin{equation}\label{eq:general-CMOE}
 S(\Phi_{\lambda,\gamma}^{\otimes m}(\rho))
 \ge mg\Big(\lambda g^{-1}(S(\rho)/m)+ \frac{\gamma+\lambda-1}{2}\Big).
\end{equation}
Products of identical thermal states attain this bound.
\end{proposition}

\subsection{Proof of Proposition~\ref{prop:full}}
\label{sec:full-CMOE}

This section is devoted to the proof of Proposition~\ref{prop:full}. We start with the pure loss channel. Using~\eqref{eq:general-map}, for $\beta>0$, the amplitude damping channel has the limiting channel corresponding to the parameters
\begin{equation}\label{eq:ADmap}
 \lambda
        =\frac{r}{1+2\mu(1-r)},
 \qquad \gamma=1-\lambda,
 \qquad \mu'=\lambda\mu.
\end{equation}
The limiting channel is therefore pure loss $\Phi^{\textup{att}}_{\lambda, 0}$. Conversely, for each
$\mu>0$ and $0<\lambda<1$, the choice $r=\frac{\lambda(2\mu+1)}{1+2\lambda\mu}$ lies in $(0,1)$ and realizes this channel.

The contraction 
coefficient in Conjecture~\ref{conj:AD} can be written as $c_r(\beta)=\lambda\frac{\beta(\lambda\mu)}{\beta(\mu)}$. Then, assuming the conjecture, Theorem~\ref{prop:transfer} would
give
\begin{equation}\label{eq:lossD}
D\bigl((\Phi^{\mathrm{att}}_{\lambda,0})^{\otimes m}(\rho)
  \Vert\tau_{\lambda\mu}^{\otimes m}\bigr)
\le\lambda\frac{\beta(\lambda\mu)}{\beta(\mu)}  D(\rho\Vert\tau_\mu^{\otimes m}).
\end{equation}
Expanding the relative entropies in terms of entropies and energy terms as in \eqref{eq:gibbs} then gives
\begin{equation}\label{eq:loss-affine}
 S\bigl((\Phi^{\mathrm{att}}_{\lambda,0})^{\otimes m}(\rho)\bigr)
 \ge mg(\lambda\mu)
       +\lambda\frac{\beta(\lambda\mu)}{\beta(\mu)}
      \big[S(\rho)-mg(\mu)\big].
\end{equation}
Choosing $\mu=g^{-1}(S(\rho)/m)$ proves pure-loss CMOE for positive input entropy; at zero input entropy, the bound is entropy nonnegativity.

\medskip

The opposite polarization direction is needed for amplification.
Define amplitude raising by bit-flip conjugation,
\[
 \mathcal B_r^\uparrow(\omega)=X\mathcal B_r(X\omega X)X.
\]
Using~\eqref{eq:general-map}, the limiting channel has parameters
\begin{equation}\label{eq:raising-map}
 \lambda=\frac{r}{2r(\mu+1)-(2\mu+1)}>1,
 \qquad \gamma=\lambda-1,
 \qquad \mu'=\lambda\mu+\lambda-1,
\end{equation}
provided $r>(2\mu+1)/(2\mu+2)$. It is therefore the quantum-limited
amplifier $\Phi^{\mathrm{amp}}_{\lambda,0}$. Conversely, for each
$\mu>0$ and $\lambda>1$, the choice
$r=\frac{\lambda(2\mu+1)}{2\lambda(\mu+1)-1}$ lies in this range
and realizes this channel.

We note that relative entropy is invariant under
conjugation. Then, Conjecture~\ref{conj:AD} for the negative range of parameter $\beta$ would imply
\begin{equation}\label{eq:amplifier-D}
 D\bigl((\Phi^{\mathrm{amp}}_{\lambda,0})^{\otimes m}(\rho)
        \Vert\tau_{\lambda\mu+\lambda-1}^{\otimes m}\bigr)
 \le \lambda\frac{\beta(\lambda\mu+\lambda-1)}{\beta(\mu)}
          D(\rho\Vert\tau_\mu^{\otimes m}),
\end{equation}
which is equivalent to 
\begin{equation}\label{eq:amplifier-affine}
 S\bigl((\Phi^{\mathrm{amp}}_{\lambda,0})^{\otimes m}(\rho)\bigr)
 \ge{} mg(\lambda\mu+\lambda-1)
       +\lambda\frac{\beta(\lambda\mu+\lambda-1)}{\beta(\mu)}        \big[S(\rho)-mg(\mu)\big].
\end{equation}
Matching the reference entropies as above gives CMOE for the quantum-limited amplifier.

Pure loss and quantum-limited amplification are sufficient for the
whole phase-covariant Gaussian family considered here, using factorization. 
Such a factorization is standard in the study of Gaussian optimizers for bosonic systems and has been used, e.g., in~\cite{GiovannettiHolevoGarciaPatron} and~\cite{DePalmaCMOE}.

\begin{proof}[Proof of Proposition~\ref{prop:full}]
The positive-polarization regime gives pure-loss CMOE by
\eqref{eq:loss-affine}. The negative-polarization regime, after
bit-flip conjugation, gives quantum-limited amplifier CMOE by
\eqref{eq:amplifier-affine}. For zero input entropy, the pure-loss bound is entropy nonnegativity.
For amplification, let $\mu\downarrow0$ in
\eqref{eq:amplifier-affine}: its correction tends to zero because
$g(\mu)/\beta(\mu)\to0$, and the first term tends to
$mg(\lambda-1)$. Thus both bounds hold at every finite input entropy.

Put $w=(\gamma+\lambda-1)/2$. Complete positivity of $\Phi_{\lambda,\gamma}$ is equivalent to
$w\ge\max\{0,\lambda-1\}$. Consequently $w+1\ge1$ and
$\eta=\lambda/(w+1)$ belongs to $[0,1]$. Then the channel $\Phi_{\lambda, \gamma}$ factors as
\begin{equation}\label{eq:factorization}
 \Phi_{\lambda,\gamma}
 =\Phi^{\mathrm{amp}}_{w+1,0}
       \circ\Phi^{\mathrm{att}}_{\eta,0}.
\end{equation}
Put $\rho'=(\Phi^{\mathrm{att}}_{\eta,0})^{\otimes m}(\rho)$. Our bound for 
pure loss gives $S(\rho')\ge mg\big(\eta g^{-1}(S(\rho)/m)\big)$. Since $g$ and
$g^{-1}$ are increasing, our quantum-limited amplifier bound then gives
\[
 \begin{aligned}
 S\bigl((\Phi_{\lambda, \gamma})^{\otimes m}(\rho)\bigr)&=S\bigl((\Phi^{\mathrm{amp}}_{w+1,0})^{\otimes m}(\rho')\bigr)\\
 &\ge mg\Big((w+1)g^{-1}\big(S(\rho')/m\big)+w\Big)\\
 &\ge mg\Big((w+1)\eta g^{-1}\big(S(\rho)/m\big)+w\Big)\\
 & =mg\Big(\lambda g^{-1}\big(S(\rho)/m\big)+w\Big).
 \end{aligned}
\]
All intermediate states have finite energy.  Also, thermal inputs
remain thermal throughout the factorization, proving attainment.
\end{proof}

\section*{Conclusion}\label{sec:conclusion}

Gross's proof of the Gaussian logarithmic Sobolev inequality
provides a guiding principle: establish an inequality on a
two-point space, tensorize it, and pass to a Gaussian limit.
In this work, we developed a method for quantizing this argument.
Qubit systems took the role of the classical two-point spaces,
and Weyl lifting represented bosonic Weyl expressions by collective qubit observables while retaining the noncommutative structure
needed for quantum inequalities.

We established two theorems for transferring superoperator norm inequalities and relative entropy inequalities from tensor powers of qubit channels to bosonic Gaussian channels. These results provide a route to Gaussian inequalities through finite-dimensional tensorization problems. We applied this approach to hypercontractivity, logarithmic
Sobolev inequalities, and constrained minimum output entropy
bounds. The qubit inequalities obtained along the way are
also of independent interest.

Two directions deserve further investigation. The first is
to extend our method beyond the channel families
considered here. In particular, extending the method to quantum convolution channels could yield bosonic entropy inequalities involving two inputs. The second is to establish further qubit
tensorization results whose bosonic counterparts remain
open. The amplitude-damping conjecture formulated here
provides one concrete target: through Theorem~\ref{prop:transfer},
it would yield the full constrained minimum output entropy
inequality for the phase-covariant Gaussian channels.

\paragraph*{Use of artificial intelligence.} OpenAI Codex with ChatGPT~6 Astra assisted with exploring and checking proof strategies, and with preparing the manuscript. The current exposition of the proofs is due to the authors. The authors are responsible for the mathematical claims, the final presentation, and any remaining errors.

{\small
\bibliographystyle{abbrvurl} 
\bibliography{CLTBIB}
}

\appendix

\section{Bounded modular functional calculus}\label{app:modular-FC}

We give the proof of Lemma~\ref{norm:bounded-modular-calculus} in a few steps. We use the notation  
$$\mathcal U_M(t)=e^{it\ln\Delta_M}, \qquad \mathcal U(t)=e^{it\ln\Delta}.$$

\paragraph{Step 1: Paired vectors built from admissible words.}
Call $(V_M,V)$ a \emph{canonical pair} if $V_M$ is a finite linear
combination of operators obtained by applying finite words in
$\mathcal L_{A_{j,M}}, \mathcal L_{W_M(z)}$ and $\mathcal U_M(t)$ to $\xi_M^{1/2}$ (with all coefficients and
factors being fixed independently of $M$), and $V$ is obtained by the same expression computed in terms of bosonic operators.

We note that $\mathcal U_M(t)\xi_M^{1/2}=\xi_M^{1/2}$, $\mathcal U(t)\xi^{1/2}=\xi^{1/2}$ and 
\[
 \mathcal U_M(t)\mathcal L_{A_{j,M}}
 =\mathcal L_{\xi_M^{it}A_{j, M}\xi_M^{-it}}\mathcal U_M(t), \qquad \mathcal U(t)\mathcal L_{A_{j}}
 =\mathcal L_{\xi^{it}A_{j}\xi^{-it}}\mathcal U(t).
\]
The latter equations can be verified by computing the action of both sides on operators of the form $X_M\xi_M^{1/2}$ and $X\xi^{1/2}$, respectively.  
These relations allow us to move every modular unitary $\mathcal U_M(t), \mathcal U(t)$ to the right at the cost of extending the jointly admissible family by including pairs of the form $(\xi_M^{it} A_{j, M} \xi_M^{-it}, \xi^{it}A_j\xi^{-it})$. Then, by the definition of jointly admissible families and Lemma~\ref{lem:admissible-closure}, for canonical pairs $(V_M, V)$ and $(W_M, W)$ we have $\langle V_M,W_M\rangle_{\mathrm{HS}}
 \rightarrow\langle V, W\rangle_{\mathrm{HS}}$. More generally, by the same reasoning, it holds that   
\begin{equation}\label{bmc:canonical}
 \big\langle V_M,\mathcal U_M(t)W_M\big\rangle_{\mathrm{HS}}
 \longrightarrow\big\langle V,\mathcal U(t)W\big\rangle_{\mathrm{HS}}
 \qquad \forall t\in\mathbb R.
\end{equation}
In particular, $\|V_M\|_{\mathrm{HS}}\to\|V\|_{\mathrm{HS}}$.

\paragraph{Step 2: Density on the bosonic side.}
The bosonic operators of canonical pairs are dense in Hilbert--Schmidt space.
Indeed, irreducibility of the Weyl representation and the Kaplansky
density theorem provide, for any bounded $X$, a uniformly bounded sequence
of finite Weyl sums $F_k$ converging strongly to $X$. To obtain finite
sums, we first approximate in the $C^*$-algebra generated by the Weyl
operators and then approximate each element in operator norm by a
finite sum. Then, considering the spectral decomposition of $\xi$ and using the dominated convergence theorem, we conclude that 
\[
 \big\|(F_k-X)\xi^{1/2}\big\|_{\mathrm{HS}}^2
 \longrightarrow0.
\]
Each $F_k\xi^{1/2}$ is canonical.
Finally, operators $X\xi^{1/2}$, where $X$ runs over bounded operators,  are dense in the Hilbert--Schmidt space.

\paragraph{Step 3: Spectral measures and one continuous factor.}
For a canonical pair $(V_M, V)$, define 
\[
 \nu_M(E)=\langle V_M,\mathbf 1_E(\ln \Delta_M)V_M\rangle_{\mathrm{HS}},
 \qquad
 \nu(E)=\langle V,\mathbf 1_E(\ln \Delta)V\rangle_{\mathrm{HS}},
\]
where $\mathbf 1_E$ is the indicator function of the set $E$.
Then, $\nu_M, \nu$ are finite positive measures on $\mathbb R$.  By~\eqref{bmc:canonical},
their characteristic functions satisfy
\[
 \int e^{itx}\,d\nu_M(x)
 =\big\langle V_M,\mathcal U_M(t)V_M\big\rangle_{\mathrm{HS}}
 \longrightarrow
 \big\langle V,\mathcal U(t)V\big\rangle_{\mathrm{HS}}
 =\int e^{itx}\,d\nu(x).
\]
We note that by the strong continuity
of $\mathcal U(t)$, the limiting function on the right is continuous at zero. If $\|V\|_{\mathrm{HS}}>0$, we can normalize the measures by their
masses and apply L\'evy's continuity theorem to obtain
\[
 \big\langle V_M,f(\ln \Delta_M)V_M\big\rangle_{\mathrm{HS}}
 \longrightarrow\big\langle V, f(\ln\Delta)V\big\rangle_{\mathrm{HS}}
 \qquad \forall f\in C_b(\mathbb R).
\]
The same holds even if $V=0$. Since this holds for every canonical pair, we also obtain
\begin{equation}\label{bmc:cross}
 \big\langle W_M,f(\ln \Delta_M)V_M\big\rangle_{\mathrm{HS}}
 \longrightarrow\big\langle W,f(\ln \Delta)V\big\rangle_{\mathrm{HS}}.
\end{equation}
Moreover, using the test function $|f|^2$ yields
\begin{equation}\label{bmc:norm}
 \big\|f(\ln \Delta_M)V_M\big\|_{\mathrm{HS}}^2
 \longrightarrow\big\|f(\ln \Delta)V\big\|_{\mathrm{HS}}^2.
\end{equation}

By Step 2, for any canonical pair $(V_M, V)$, $f\in C_b(\mathbb R)$ and $\varepsilon>0$ there is a canonical pair $(W_M, W)$
such that $\big\|f(\ln \Delta)V-W\big\|_{\mathrm{HS}}<\varepsilon$.
Expanding the squared distances and using
\eqref{bmc:canonical}--\eqref{bmc:norm}, we find that
\begin{equation}\label{bmc:distance}
 \lim_{M\to\infty}\big\|f(\ln \Delta_M)V_M-W_M\big\|_{\mathrm{HS}}^2
 =\big\|f(\ln \Delta)V-W\big\|_{\mathrm{HS}}^2<\varepsilon^2.
\end{equation}
This is the key approximation: a canonical approximation on the
bosonic side has the same limiting error on the qubit side. 

\paragraph{Step 4: Successively inserting all the factors.}
For clarity, say that a pair $(X_M, X)$ can be approximated by
canonical pairs if, for every $\varepsilon>0$, there is a canonical
pair $(V_M, V)$ satisfying
\begin{equation}\label{bmc:approximation}
 \limsup_{M\to\infty}{\|X_M-V_M\|}_{\mathrm{HS}}<\varepsilon,
 \qquad {\|X-V\|}_{\mathrm{HS}}<\varepsilon.
\end{equation}
We note that 
canonical pairs themselves have this property. Then to establish the desired limit in the statement of the lemma, it suffices to show that the application of any factor in~\eqref{eq:modular-FC-limit} preserves approximability by canonical pairs. For a continuous factor, first approximate $(X_M,X)$ by a canonical
pair $(V_M,V)$. Applying $f(\ln\Delta_M)$ and $f(\ln\Delta)$
multiplies the approximation errors by at most $\|f\|_\infty$.
Equation~\eqref{bmc:distance} then approximates the resulting pair
$(f(\ln\Delta_M)V_M,f(\ln\Delta)V)$ by a canonical pair.
The triangle inequality proves the required preservation property. 
For left multiplications, put
$C=\max\{\sup_M\|A_{j,M}\|_\infty,\|A_j\|_\infty\}$.
Then
\[
 {\|\mathcal L_{A_{j,M}}(X_M-V_M)\|}_\mathrm{HS}
 \le C{\|X_M-V_M\|}_\mathrm{HS},
\]
with the same estimate on the bosonic side. On the other hand, if $(V_M, V)$ is canonical, then
$(\mathcal L_{A_{j,M}}V_M,\mathcal L_{A_j}V)$ is also canonical. Thus, the above bound shows that left multiplications also preserve approximability by canonical pairs. 
This finishes the proof of Lemma~\ref{norm:bounded-modular-calculus}.

\section{Approximation by thermally sandwiched states}
\label{app:sandwiched-approximation}

We prove the approximation statement used to extend the relative entropy inequality from sandwiched states to arbitrary finite-energy states.

\begin{proof}[Proof of Lemma~\ref{lem:sandwich-core}]
We construct the approximation in three steps: truncate the photon
number, add a small thermal component, and approximate the bounded
relative density by finite Weyl sums.

Let $P_L=\mathbf1_{\{\Ntot\le L\}}$ be the projection on the space of states with total photon number at most $L$. Put
\[
 \rho^{(L)}=\frac{1}{\tr(P_L\rho)}P_L\rho P_L,
\]
using the cutoffs for which the denominator is nonzero. These
states converge to $\rho$ in trace norm.
Since $P_L$ commutes with $\Ntot$, the monotone convergence theorem also gives
\[
 \tr\Big(\rho^{(L)}\Ntot\Big)
 =\frac{\tr\big(\rho\Ntot P_L\big)}{\tr(P_L\rho)}
 \longrightarrow\tr\big(\rho\Ntot\big).
\]
Entropy is known to be continuous on sets of states
with bounded mean energy~\cite{winter_tight_2016}. Hence
$S(\rho^{(L)})\to S(\rho)$. The thermal identity
\begin{equation}\label{sandwich:thermal-relative entropy}
D(\rho\Vert{\tau_\mu^{\otimes m}})
 =-S(\rho)+\beta(\mu)\tr(\rho\Ntot)+m\ln(\mu+1),
\end{equation}
then proves convergence of the input relative entropies. Moreover, the Gaussian channel $\Phi$ obeys the affine energy transformation
\[
 \tr\bigl[\Phi^{\otimes m}(\rho)\Ntot\bigr]
 =\lambda\tr\big(\rho\Ntot\big)+mw,
 \qquad w=\frac{\gamma+\lambda-1}{2}.
\]
Thus trace-norm contractivity and energy convergence give trace
norm and energy convergence for the outputs of the channel. Their energies are
uniformly bounded, so the same entropy continuity theorem and
the thermal identity at $\mu'$ also give convergence of output
entropy and output relative entropy. This observation applies to
each of the following two approximation steps as well.

For fixed $L$, define the relative density
\[
 X^{(L)}=\bigl(\tau_\mu^{\otimes m}\bigr)^{-1/2}\rho^{(L)}\bigl(\tau_\mu^{\otimes m}\bigr)^{-1/2}.
\]
We note that $X^{(L)}$ is a bounded positive finite-rank operator, because the inverse
thermal factors act on the finite-dimensional range of $P_L$.
For $\varepsilon>0$, its regularization gives the state
\[
 \frac{1}{1+\varepsilon}\Gamma_{{\tau_\mu^{\otimes m}}}\big(X^{(L)}+\varepsilon I\big)
 =\frac{1}{1+\varepsilon}\big(\rho^{(L)}+\varepsilon{\tau_\mu^{\otimes m}}\big).
\]
Letting $\varepsilon\downarrow0$ gives convergence in trace norm
and energy. The input and output entropy conclusions then follow
from the preceding argument.

For fixed $L$ and $\varepsilon$, choose uniformly bounded finite
Weyl sums $F_k$ converging strongly-$*$ to $\sqrt{X^{(L)}}$.
The density of the Weyl algebra gives these approximations as in
the proofs in Section~\ref{sec:norm-lifting}. Now let
\[
 X_k=\varepsilon I+F_k^\dagger F_k.
\]
These are precisely the positive expressions required by the
sandwich construction. They are uniformly bounded and converge
strongly to $X^{(L)}+\varepsilon I$. A finite-rank approximation
of $\bigl(\tau_\mu^{\otimes m}\bigr)^{1/2}$ shows that
$\Gamma_{{\tau_\mu^{\otimes m}}}(X_k)\to\Gamma_{{\tau_\mu^{\otimes m}}}(X^{(L)}+\varepsilon I)$ in
trace norm. Their unnormalized energies also converge since
\[
\tr\Big[\Ntot\Gamma_{{\tau_\mu^{\otimes m}}}(X_k)\Big]
 =\tr\big(\Ntot{\tau_\mu^{\otimes m}} X_k\big)
\longrightarrow\tr\Big[\Ntot{\tau_\mu^{\otimes m}}(X^{(L)}+\varepsilon I)\Big].
\]
The equality is understood first with finite photon-number
cutoffs and then by the monotone convergence theorem as $\Ntot{\tau_\mu^{\otimes m}}$ is trace
class. Normalizing by $\tr({\tau_\mu^{\otimes m}} X_k)\to1+\varepsilon$ preserves
the trace and energy limits. Once again, bounded-energy entropy
continuity supplies all the remaining input and output limits.

Taking these approximations successively, and then choosing a
diagonal sequence in $k$, $\varepsilon$, and $L$, proves the
lemma.
\end{proof}

\section{A Han-inequality proof of Lemma~\ref{lem:thermal-EPI}}
\label{app:Han}

The attenuation inequality of Lemma~\ref{lem:thermal-EPI} has a
finite-dimensional explanation based on Weyl lifting. Consider a permutation-invariant state on many qubits, and retain a fixed fraction of
those qubits, tracing out the rest. Quantum Han's inequality says that its relative entropy
to a product reference state is at most that fraction of the full-state
relative entropy. Now the point is that in the
fluctuation limit, retaining a fraction of qubits turns into attenuation with a thermal
environment. We prove this correspondence for the sandwiched states of
Section~\ref{sec:entropy-transfer}, and then use their approximation property
to obtain the inequality for arbitrary finite-energy bosonic states.

\begin{proof}[Proof of Lemma~\ref{lem:thermal-EPI}]
The endpoints $\lambda=0$ and $\lambda=1$ are immediate. At the first
endpoint the output is always $\tau_\mu^{\otimes m}$, while at the second endpoint the
channel is the identity. In the following we assume that $\lambda=p/q\in(0,1)$ is the ratio of positive integers $p,q$ with $p<q$. 
Having established the desired inequality for rational transmissivity parameters, the inequality for arbitrary $\lambda$ is obtained by a continuity argument; note that the underlying states have finite energy, so the entropy continuity of~\cite{winter_tight_2016} is available.

\paragraph{Quantum Han's inequality.}
Arrange the $mM$ qubits in $M$ groups $C_1,\ldots,C_M$, each containing
one qubit for each of the $m$ modes. As before, the corresponding reference state on these qubits is $\sigma_\mu^{\otimes mM}$. 

For an arbitrary state $\omega$ and any subset
$J=\{j_1<\cdots<j_K\}\subseteq\{1,\ldots,M\}$ of size $K$, the chain rule
and strong subadditivity give
\[
 \begin{aligned}
 S(\omega_J)
 =\sum_{r=1}^K
     S(C_{j_r}| C_{j_1}\cdots C_{j_{r-1}})_\omega\ge\sum_{j\in J}S(C_j| C_1\cdots C_{j-1})_\omega.
 \end{aligned}
\]
Averaging these inequalities over all choices of $K$-subsets includes each group
with probability $K/M$. Then, a second application of the chain rule yields
\begin{equation}\label{eq:Han-subset-average}
 \binom MK^{-1}\sum_{|J|=K} S(\omega_J)
 \ge\frac KM S(\omega).
\end{equation}
This turns into an equality for product states. Moreover, since
$\ln(\sigma_\mu^{\otimes mK})$ is the sum of one-qubit operators, we have
\[
 \binom MK^{-1}\sum_{|J|=K}
    -\tr\bigl(\omega_J\ln\sigma_\mu^{\otimes mK}\bigr)
 =-\frac KM\tr\bigl(\omega\ln{\sigma_\mu^{\otimes mM}}\bigr).
\]
Combining this identity with \eqref{eq:Han-subset-average} implies
\[
 \binom MK^{-1}\sum_{|J|=K}
   D\big(\omega_J\big\Vert\sigma_\mu^{\otimes mK}\big)
 \le\frac KM D\big(\omega\big\Vert{\sigma_\mu^{\otimes mM}}\big).
\]
If $\omega$ is invariant under permutations of the
groups, all the terms in this average coincide. With $\omega_{[K]}$
denoting the marginal state on the first $K$ groups, we obtain
\begin{equation}\label{eq:HanDqubit}
D\big(\omega_{[K]}\big\Vert{\sigma_\mu^{\otimes mK}}\big)
 \le\frac KM D\big(\omega\big\Vert{\sigma_\mu^{\otimes mM}}\big).
\end{equation}

\paragraph{Symmetric sandwiched states.}
Let $F=\sum_{r=1}^L c_rW(z_r)$ be a finite Weyl sum and choose
$\varepsilon>0$. As in \eqref{sandwich:positive-core}, form the positive
expressions before normalizing their thermal sandwiches:
\[
 \begin{aligned}
 X&=\varepsilon I+F^\dagger F,&
 X_M&=\varepsilon I+F_M^\dagger F_M,\\
 z&=\tr({\tau_\mu^{\otimes m}} X),& z_{M}&=\tr({\sigma_\mu^{\otimes mM}} X_M),\\
 \rho&=z^{-1}\Gamma_{{\tau_\mu^{\otimes m}}}(X),&
 \rho_{M}&=z_{M}^{-1}\Gamma_{{\sigma_\mu^{\otimes mM}}}(X_M).
 \end{aligned}
\]
We emphasize that the product
$F_M^\dagger F_M$ is formed after lifting. We note that, by construction, $\rho_{M}$ is invariant under permutations as above. Therefore, letting
$\rho_{M,[K]}$ be the marginal state on 
the first $K$ groups, \eqref{eq:HanDqubit} gives
\begin{equation}\label{eq:Han-core-qubit}
 D(\rho_{M, [K]}\Vert{\sigma_\mu^{\otimes mK}})
 \le\frac KM D(\rho_{M}\Vert{\sigma_\mu^{\otimes mM}}).
\end{equation}
The results of Section~\ref{sec:entropy-transfer}, in particular Proposition~\ref{sandwich:entropy-limit}, already give
\begin{equation}\label{eq:Han-input-limit}
 z_{M}\longrightarrow z,\qquad
 D(\rho_{M}\Vert{\sigma_\mu^{\otimes mM}})
     \longrightarrow D(\rho\Vert{\tau_\mu^{\otimes m}}).
\end{equation}

\paragraph{Computing the fluctuation limit.} Recall that $\lambda=p/q$. Then, we can restrict to $M=qM'$ and $K=pM'$ as $M'\to\infty$, in which case
$K/M=\lambda$ and both $K$ and $M-K$ tend to infinity. Moreover, the tensor-product definition of the
pseudo-Weyl operators gives the exact factorization
\begin{equation}\label{eq:Gibbs-two-block-Weyl}
 W_{M,\mu}(z)
 =W_{K,\mu}(\sqrt{\lambda}\,z)
       \otimes W_{M-K,\mu}(\sqrt{1-\lambda}\,z),
\end{equation}
where the tensor factors correspond to the retained and
discarded groups.

Let $B$ be a finite Weyl sum, and let $B_{K,\mu}$ be its Weyl lift on the retained groups.
Expand the trace
\[ \tr\bigl[(B_{K,\mu}\otimes I)          \bigl(\sigma_\mu^{\otimes mM}\bigr)^{1/2}X_M\bigl(\sigma_\mu^{\otimes mM}\bigr)^{1/2}\bigr]\]
using the fixed expression $F_M=\sum_r c_r W_{M, \mu}(z_r)$. Next, using the factorization~\eqref{eq:Gibbs-two-block-Weyl}, any term can be factored as a product of two traces. For instance, the term corresponding to \(\overline{c_r}c_s\) is equal to  
\begin{align*}
&\operatorname{tr}\!\left[
B_{K,\mu}\bigl(\sigma_\mu^{\otimes mK}\bigr)^{1/2}
W_{K,\mu}(\sqrt\lambda z_r)^\dagger
W_{K,\mu}(\sqrt\lambda z_s)\bigl(\sigma_\mu^{\otimes mK}\bigr)^{1/2}
\right]\\
&\quad\times
\operatorname{tr}\!\left[
\bigl(\sigma_\mu^{\otimes m(M-K)}\bigr)^{1/2}
W_{M-K,\mu}(\sqrt{1-\lambda}z_r)^\dagger
W_{M-K,\mu}(\sqrt{1-\lambda}z_s)\bigl(\sigma_\mu^{\otimes m(M-K)}\bigr)^{1/2}
\right].
\end{align*}
We can then apply Lemma~\ref{norm:thermal-words} to each of these factors to obtain the limit
\begin{align*}
&\operatorname{tr}\!\left[
B\bigl(\tau_\mu^{\otimes m}\bigr)^{1/2}
W(\sqrt\lambda z_r)^\dagger
W(\sqrt\lambda z_s)\bigl(\tau_\mu^{\otimes m}\bigr)^{1/2}
\right]\\
&\quad\times
\operatorname{tr}\!\left[
\bigl(\tau_\mu^{\otimes m}\bigr)^{1/2}
W(\sqrt{1-\lambda}z_r)^\dagger
W(\sqrt{1-\lambda}z_s)\bigl(\tau_\mu^{\otimes m}\bigr)^{1/2}
\right].
\end{align*}
Putting the limit of these terms together, we find that 
\[ \tr\bigl[(B_{K,\mu}\otimes I)          \bigl(\sigma_\mu^{\otimes mM}\bigr)^{1/2}X_M\bigl(\sigma_\mu^{\otimes mM}\bigr)^{1/2}\bigr]\longrightarrow \tr\bigl[(B\otimes I)          \bigl(\tau_\mu^{\otimes m} \otimes \tau_\mu^{\otimes m}\bigr)^{1/2}\widetilde X\bigl(\tau_\mu^{\otimes m}\otimes \tau_\mu^{\otimes m}\bigr)^{1/2}\bigr],\]
where
\[
\widetilde X=\varepsilon I+\widetilde F^\dagger\widetilde F, \qquad \widetilde F
=\sum_r c_r\,W(\sqrt\lambda z_r)
                 \otimes W(\sqrt{1-\lambda}z_r).
\]
In fact, letting $U_{\lambda}$ be the beam-splitter unitary that satisfies
\[
 U_\lambda(W(z)\otimes I)U_\lambda^\dagger
 =W(\sqrt\lambda\,z)\otimes
       W(\sqrt{1-\lambda}\,z),
\]
we have $\widetilde X = U_\lambda (X\otimes I)U_\lambda^\dagger$. Therefore, writing the above limit in terms of the states, we obtain
\begin{align*}\label{eq:Gibbs-Han-moments}
 \tr\big(B_{K,\mu}\, \rho_{M, [K]}\big) &= \tr\big((B_{K,\mu}\otimes I)\, \rho_{M}\big)\\ &\longrightarrow  z^{-1}\, \tr\left( B \,\tr_{2}\left[ \Gamma_{\tau_\mu^{\otimes m}\otimes \tau_\mu^{\otimes m}} \left(  U_\lambda (X\otimes I)  U_{\lambda}^{\dagger} \right)\right]    \right)\\
 & \qquad = z^{-1}\, \tr\left( B \,\tr_{2}\left[ U_\lambda\Gamma_{\tau_\mu^{\otimes m}\otimes \tau_\mu^{\otimes m}} \left(   (X\otimes I)   \right)U_{\lambda}^{\dagger}\right]    \right)\\
 & \qquad =  \tr\left( B \,\tr_{2}\left[ U_\lambda\left(\rho\otimes \tau_\mu^{\otimes m} \right) U_{\lambda}^{\dagger}\right]    \right)\\
 & \qquad =  \tr\Big( B \,(\Phi^{\mathrm{att}}_{\lambda,\mu})^{\otimes m}(\rho)    \Big).
\end{align*}
The second equality uses that the beam-splitter unitary commutes with
the product of thermal states at the same temperature.

\paragraph{Passing the relative entropy inequality to the limit.}
With $b=\varepsilon+(\sum_{r=1}^L|c_r|)^2$, partial trace gives
\[
 \frac{\varepsilon}{b}\sigma_\mu^{\otimes mK}
 \le\rho_{M, [K]}\le\frac b\varepsilon\sigma_\mu^{\otimes mK},
\]
and the limiting state $(\Phi^{\mathrm{att}}_{\lambda,\mu})^{\otimes m}(\rho) $ satisfies the analogous bounds relative to
$\tau_\mu^{\otimes m}$.  Since $K=pM'$ goes to infinity, the general-sequence
extension \eqref{eq:output-entropy-liminf} following
Lemma~\ref{lem:output-entropy-liminf} applies to $\rho_{M,[K]}$,
with block size $K$, limiting state $\eta=(\Phi^{\mathrm{att}}_{\lambda,\mu})^{\otimes m}(\rho)$, and reference parameter
$\mu'=\mu$.
Together with \eqref{eq:Han-core-qubit} and the input entropy limit,
this yields
\[
 \begin{aligned}
 D\big((\Phi^{\mathrm{att}}_{\lambda,\mu})^{\otimes m}(\rho) \Vert\tau_\mu^{\otimes m}\big)
 &\le\liminf_{M\to\infty}
 D(\rho_{M, [K]}\Vert\sigma_\mu^{\otimes mK})\\
 &\le\lambda D(\rho\Vert\tau_\mu^{\otimes m}).
 \end{aligned}
\]
This proves \eqref{eq:HanDboson} for every sandwiched state.

For a general input $\rho$, which has finite energy, Lemma~\ref{lem:sandwich-core} supplies sandwiched-state
approximations whose input and output relative entropies converge.
Passing to the limit proves \eqref{eq:HanDboson} for every such input.

\end{proof}

\section{Joint convexity and monotonicity of \texorpdfstring{$F_{a,b}$}{F(a,b)}}\label{app:joint-convex}

\begin{proof}[Proof of Lemma~\ref{lem:joint-convex}]
The convexity of the domain $\mathscr D$ is a consequence of the concavity of \(\widetilde h\). For the entropy monotonicity, let $0\le a,b\le1$ and put
\[w=\sqrt{a^2\Big(\widetilde h^{-1}(u)^2-z^2\Big)+b^2z^2}.
\]
Since $|z|\le \widetilde h^{-1}(u)\le1$ on $\mathscr D$, we have $0\le w\le \widetilde h^{-1}(u)$. Define
\(\varphi(x)=\operatorname{artanh}(x)/x\), with $\varphi(0)=1$.
Differentiating in the interior gives
\begin{equation}\label{eq:Fab-first}
 \partial_uF_{a,b}(u,z)=a^2\frac{\varphi(w)}{\varphi\big(\,\widetilde h^{-1}(u)\big)}\in[0,a^2],
\end{equation}
because $\varphi$ is positive and increasing. Integration and continuity
give the finite-difference bound on the whole domain. Also, the monotonicity of $x\mapsto x+F_{a, b}(c-x, z)$ is a consequence of the above derivative bound and $0\leq a^2\leq 1$.

For joint convexity, first assume $0<b<a<1$ and write
$c=a^2-b^2$, so that $w=\sqrt{a^2\widetilde h^{-1}(u)^2-cz^2}$.
Using the function \(\varphi\) above, for $x\in [0,1)$ define
\[
 K_0(x)=\frac{\varphi'(x)}{x\varphi(x)},\qquad
 R_0(x)=x^2+\frac1{K_0(x)}
       =\frac{x^3}{x-(1-x^2)\operatorname{artanh}x}.
\]
These functions take their values at $x=0$ by continuity. The series
\[
 \frac1{R_0(x)}
 =\sum_{j\ge1}\frac{2x^{2j-2}}{(2j-1)(2j+1)},
 \qquad 0\le x<1,
\]
shows that \(R_0(x)\) is decreasing. Consequently \(1/K_0(x)=R_0(x)-x^2\)
is decreasing and \(K_0(x)\) is increasing. 
Now computing the second derivatives, we obtain
\begin{align*}
 \partial_{uu}F_{a,b}(u, z)
 =\frac{a^2\varphi(w)}{\varphi\big(\,\widetilde h^{-1}(u)\big)^2}
    \Big[K_0\big(\,\widetilde h^{-1}(u)\big)-a^2K_0(w)\Big]\ge0,\\
       \partial_{zz}F_{a,b}(u, z)
 =c\varphi(w)\bigl[1-cz^2K_0(w)\bigr],\qquad
 \partial_{uz}F_{a,b}(u, z)
 =-\frac{ca^2z\varphi(w)K_0(w)}{\varphi(\widetilde h^{-1}(u))}.
 \end{align*}
Here we use $w\leq \widetilde h^{-1}(u)$, $0\leq a^2\leq 1$ and the monotonicity of $K_0(x)$ for the first line. 
Computing the determinant of the Hessian, we also have 
\begin{align*}
\det\nabla^2F_{a,b}(u, z)
=\frac{ca^2\varphi(w)^2}{\varphi\big(\,\widetilde h^{-1}(u)\big)^2}
       K_0\big(\,\widetilde h^{-1}(u)\big)K_0(w)\Big[R_0(w)-a^2R_0\big(\,\widetilde h^{-1}(u)\big)\Big]\ge0,
\end{align*}
where the bound follows from the monotonicity of $R_0(x)$ and $0\leq a^2\leq 1$. In this parameter range $\partial_{uu}F_{a,b}>0$, so the
nonnegative determinant makes the Hessian positive semidefinite.
Continuity extends convexity to $0\le b\le a\le1$ and to the
boundary of $\mathscr D$.
\end{proof}

\end{document}